\documentclass[%
 reprint,
 amsmath,amssymb,
 aps,
]{revtex4-2}

\usepackage{graphicx}% Include figure files
\usepackage{dcolumn}% Align table columns on decimal point
\usepackage{bm}% bold math
\usepackage{subcaption}
\usepackage{float}

\begin{document}

\preprint{APS/123-QED}

\title{Neutron Stars In $f(R,T)$ Theory: Slow Rotation Approximation}
\author{Masum Murshid}
\email{masum.murshid@wbscte.ac.in}
\author{Mehedi Kalam}
 \email{kalam@associates.iucaa.in}
\affiliation{Department of Physics, Aliah University, II-A/27, Action Area II, Newtown, Kolkata-700156, India.}%


\begin{abstract}
\textbf{In this paper, we study the slowly rotating neutron stars in $f(R, T)$ gravity based on Hartle-Thorne formalism. We first consider the simplest matter-geometry coupled modified gravity, namely $f(R, T)=R+2\chi T$. We compute the mass, radius, moment of inertia, change in radius, and binding energy due to rotation, eccentricity, quadrupole moment, and the tidal love number. The quantities, which are of the second order in angular velocity, like change in radius and binding energy due to rotation, eccentricity, and quadrupole moment, deviate more from their corresponding general relativistic counterparts in lighter neutron stars than heavier ones. Whereas the moment of inertia, which is of the first order in angular velocity, in  $f(R, T)=R+2\chi T$ modified gravity, barely diverges from the general relativistic one. The Equation of state-independent I-Love-Q relation retains in this $f(R, T) $ modified gravity, and it coincides with the general relativistic ones within less than one percent even for the maximum allowed coupling parameters. We also study the slowly rotating neutron star in $f(R, T)=R+\alpha R^{2}+2\chi T$ up to first order their angular velocity. We calculate the mass, radius, and moment of inertia of neutron stars in this modified gravity. The results show that the impact of the matter-geometric coupling parameter is greater on lighter neutron stars in both of these modified gravity models.}
\end{abstract}

\maketitle

\section{Introduction}
Neutron stars are the corpse of massive stars. Black holes, the end state of more massive stars, are completely described by mass, charge and spin as suggested by the No hair and uniqueness theorem \cite{Chrusciel2012}. No such theorem restrict the parameters required to characterize a neutron star to the mass, charge and spin to make it as simple as black holes. Neutron stars are ultra-high dense objects, surpassing the nuclear mass density. The equation of state (EOS) that is the pressure and energy relation uniquely determine the mass-radius relationship $M(R)$ of neutron stars \cite{Lattimer&Prakash2004}. But the problem is that the EOS at this very high density is not well understood. Nonetheless, some astrophysical observations allow us to make certain guesses about EOS \cite{Lattimer2004,Lattimer2007,Ozel2010,Steiner2010,Psaltis2013,Guver2013,Miller2020,Hu2020}. Neutron stars act as a natural laboratory to study the behaviour of the fundamental particles at such immense density. They also produce Gravitational Waves (GWs) during the inspiral and merger of Neutron star binaries. Ground-based detectors like  LIGO \cite{LIGO}, VIRGO \cite{VIRGO} and KAGRA \cite{KAGRA} can detect these GWs produced by neutron star binaries during their last minutes of the orbit before coalescence. These Gravitational Waves could decode information about dense matter, gravitation and cosmology \cite{Abbott2017}. Besides, neutron stars produce strong gravitational fields where we can test general relativity in these Strong fields regime which has passed all the tests only in the weak fields regime \cite{Will2005}.

 NS gains rotation from its progenitor star due to the conservation of angular momentum. Some NSs may rotate at an extreme rotation rate, called the Masss-Shedding rotation limit; above this rotation limit, matters of the NS get loosened from its surface. When NS rotates, it gets deformed and takes the form of an oblate because of its non-rigidness, which snatches away the spherical symmetric property of a rotating NS. The oblateness of rotating NS introduces different mass multipoles and shows their effect on spacetime. The rotation also causes a change in the gravitational mass of the rotating NS as measured by a distant observer. Though the rotating NSs have axisymmetric properties, the loss of spherical symmetry makes the study of RNSs cumbersome. But if NS rotates slowly, the rotation can be treated as perturbation over a non-rotating NS, which makes life easy. Hartle and Thorne, in their seminal papers \cite{Hartle1967,Hartle1968}, studied all these rotational effects in slow rotation approximation. Hartle and Thorne metric is an approximate  Einstein's equations solution of the slowly and rigidly rotating stellar objects with an accuracy up to the second-order in stellar angular velocity. One can get the Kerr metric in the Bayer-Lindquist coordinate from this metric after a suitable coordinate transformation \cite{Abramowicz2003}. Manko et al. \cite{Manko2000b,Manko2000a}  found an exact vacuum solution of Einstein-Maxwell's equations analytically using Sibgatullin devises \cite{Sibgatullin1991}. This solution involves only five parameters expressed by simple rational functions and describes the exterior field of both slowly and rapidly rotating neutron stars with a poloidal magnetic field. Berti et al. \cite{Berti2005} tested the accuracy of Hartle-Thorne slow rotation approximation by comparing it with Manko's exact solution and with the numerical solution of full Einstein's equations. They computed relative errors in the quadrupole moment and the radii of innermost stable circular orbits (ISCOs); and found  Hartle-Thorne approximation predicts ISCO radii with an accuracy of 1 per cent even for the fasted millisecond pulsars.

 The inspiral merging of NS binary produces gravitational waves. These gravitational waves, in the early stage, are governed by the point-mass dynamics of the NS. As the GWs carry away energy and angular momentum, the inspiring orbital radius decreases and frequency increases. In the late time of binary merging, the NSs are perturbed by the tidal field of the companion star and causing the NSs to deform . This tidal deformation affects the gravitational field through higher multipoles and hence produces noises in GWs signal in the late time. Measuring these noises in terms of the Signal to noise ratio (SNR) from GWs can give us ample information about the NS. Hinderer \cite{Hinderer2008} found the tidal love number by considering the tidal deformation as a small perturbation in metric and combined all the Einstein equations for this perturbed metric into a single second-order differential equation. Flanagan and Hinderer \cite{Flanagan2008} showed that for the inspiring two neutron stars having mass $1.4M_{\odot}$ each 50kpc away from us, one constrains the parameter $\lambda$, related to tidal love no, to $\lambda \leq 2\times 10^{37}$ $gcm^{2}s^{2}$ with  $90$ confidence.  Binnington and Poisson \cite{Binnington2009} distinguished electric-type and magnetic-type love numbers. Later, following the Hinderer paper, Yagi and Yunes \cite{Yagi2013} found that the relationship between the moment of inertia, the love numbers and the quadrupole is insensitive to the NS's internal structure for slowly rotating neutron stars. This insensitivity comes from mainly two reasons. i) the I-Love-Q relation depends sensitively on the outer region of the NS where all the EOS approach each other. ii) The more the NS compactness, the more the I-Love-Q relation becomes insensitive to the internal structure, approaching the black hole's I-Love-Q relation. The I-Love-Q relation can be used to determine the average spin of the NS from the SNR of the detected Gravitational waves. If these I-Love-Q relations remain independent of EOS in alternative theories of gravity, then the deviation of their relationship from that of the GR can be used to discriminate between GR and possible extensions of the theory. This I-Love-Q universality breaks down for rapidly rotating Stars.

 Einstein's theory of general relativity (GR) has passed all the tests in weak field gravity regimes or intermediate energy scales with magnificent precision. But the failure of GR in small and large energies indicates a need for modification in GR. Many scientists have proposed different types of modified theories of gravity \cite{Clifton2011,Nojiri:2010wj,Nojiri2017,Berti2015,Silva2016,Wu2018,DeFelice2010,Sotiriou2008,Lobo2008}. Astashenok et al.  studied neutron stars in perturbative $f(R)$ gravity with realistic EoS \cite{Astashenok:2013vza}. They found that the stars with very high central density ($\rho > 10\rho_{ns}$ where $\rho_{ns}=2.7\times 10^{14}$ $g/cm^{3}$) are also stable in $R^{2}$ gravity with cubic corrections. The maximum mass of such stars is approximately $1.9M_{\odot}$ with a minimum radius of around $9$km. The neutron stars having dense matter in strong mean magnetic fields can be as massive as $M>4M_{\odot}$ in $f(R)$ gravity with cubic corrections. In $f(\mathcal{G})$ gravity with quadratic corrections (where $\mathcal{G}$ is Gauss-Bonnet invariant ), neutron stars with high strangeness fraction are found to be stable \cite{Astashenok:2014nua}. Astashenok et al., in their study of the nonperturbative model of quark stars in $f(R)$ gravity, showed that the fine-tuning for the central scalar curvature $R$ is equivalent to the fine-tuning of the scalar field $\phi$ in the corresponding scalar-tensor field theory \cite{Astashenok:2014dja}. They showed that the thermal spectrum emerging from the surface of the star can be used to discriminate modified theories of gravity from GR.  Astashenok et al. showed that, in astrophysical context, in $f(R)=R+\alpha R^{2}$ gravity, only the positive value of $\alpha$ is allowable as the gravitational mass of an astronomical object increases indefinitely with radial distance for a negative value of $\alpha$ \cite{Astashenok:2017dpo}. The coupling between curvature and axion field gives rise to supermassive compact stars with mass $M \sim 2.2 - 2.3 M_{\odot}$ for realist EoS \cite{Astashenok:2020cfv}. This coupling also increases the maximal frequency of the rotation of the star in comparison to GR \cite{Astashenok:2020cqq}. Astashenok et al.  explained the GW190814 supermassive neutron star with consistent mass-radius relation in the extended theories of gravity using EoS consistent with LIGO observational constraints \cite{Astashenok:2020qds}. The elevated role of extensions of GR towards the successful description of the GW190814 events was discussed by \cite{Astashenok:2021xpm}. In GR as well as $f(R)$ gravity, the maximum gravitational mass of a neutron star cannot be larger than $3M_{\odot}$ \cite{Astashenok:2021xpm,Astashenok:2021btj,Astashenok:2021peo}. Thus, to explain neutron stars having mass greater than $3M_{\odot}$, we need alternative extensions of GR other than $f(R)$.
 
 Harko et al. proposed the modified theory of gravity by introducing an arbitrary coupling between matter and geometry \cite{Harko2011}. They considered the Lagrangian as an arbitrary function of $T$, the trace of the energy-momentum tensor, and Ricci scalar $R$. Though they proposed many forms of the $ f(R, T)$ function, the minimal matter-geometry coupling, that is, $f(R, T)=R+2\chi T$, because of its simple linear structure might have mathematical advantages in specific contexts, becomes more popular in the study of the astrophysical and cosmological aspects in $ f(R, T)$ modified gravity. P. S.  Debnath presented the bulk viscous cosmological model of the universe in this $f(R, T)$  modified gravity. They showed that both in the contracting and expanding universe, the null energy condition is violated \cite{Debnath:2018wct}. S. Bhattacharjee et al. studied non-singular bouncing cosmology \cite{Bhattacharjee:2020eec} and presented modelling of inflation scenario \cite{Bhattacharjee:2020jsf}, and M. Gamonal investigated the slow-roll approximation of the cosmic inflation \cite{Gamonal:2020itt} in this modified theory of gravity. H Shabani et al. investigate the classical bouncing solution in $f(R, T)=R+h(T)$ gravity, where $h(T)$ is a function of $T$ \cite{Shabani:2017kis}. Moraes et al. studied the compact stars in $f(R, T)=R+2\chi T$ gravity using polytropic and MIT bag EoS \cite{Moraes:2015uxq}. Lobato et al. and Silva et al. constrained the coupling parameter for $f(R, T)=R+2\chi T$ gravity to $\chi \leq 0.02$  in the light of massive and GW107817 \cite{Lobato2020,daSilva2022}. Pretel et al. investigated the radial stability of compact stars \cite{Pretel:2020oae} and studied charged quark stars for this modified gravity \cite{Pretel:2022dbx}. Bora et al. studied the gravitational wave echoes from the static spherically symmetric compact stars in this modified gravity \cite{Bora:2022dnu}. In this modified gravity, the exterior solution of the field equation of the spherically symmetric objects is Schwarzschild as the outside of the star $R=0$. Besides that, some other people studied compact stars in this $f(R, T)$ modified gravity \cite{Das:2016mxq,Rej:2021ngp,Deb:2018gzt,Deb:2018sgt,PhysRevD.100.044014,Biswas:2020gzd,Maurya:2019iup,Biswas:2021wfn}.
 
 In modified theories of gravity, the blackhole solution is the Kerr metric, the blackhole solution in GR, or is deviated from the Kerr solution by a small margin such that these modified theories cannot be distinguished from GR using present and future astrophysics observations. Unlike the blackholes, the coupling of matter with gravity in the NS provides alternative ways to test modified theories of gravity in the strong field regime \cite{Stairs2003}. The cost of the coupling of matter with gravity is paid off by the fact that the mass-radius relation is sensitive to both the EoS and the underlying theory of gravity. However, various universal relations, independent of EoS, can be found for NSs which opens the windows of possibilities of the comparative studies of NSs in different modified theories of gravity without prior details knowledge of EoS. The relation among I-Love-Q,  Quasi Normal modes (QNMs) frequencies and the combination of mass and radius,  moment of inertia and QNMs frequencies, and moment of inertia and compactness of NSs are insensitive to EoS \cite{Yagi2017}. Pappas and Apostolatos found three hair theorems for NSs in GR, which is an approximate generalization of the BH no-hair theorem \cite{Pappas2014} . Among all the universal relations, I-Love-Q relations are more insensitive to EoS than others. Therefore, studying these I-Love-Q relations in the different modified theories of gravity is of utmost interest. Pani and Berti \cite{Pani2014} calculated the I-Love-Q relations in scalar-tensor theory, Sham et al. \cite{Sham2013} in  EiBI gravity and Yagi et al.  \cite{Yagi2013b} in dCS gravity. Kleihaus et al. \cite{Kleihaus2014} studied the I-Q relation for rapidly rotating NSs in EdGB gravity.

In recent years, the study of rotating NSs has gained attention. Berti et al. studied rotating NSs using Hartle-Thorne and Manko formalism and the numerical solution \cite{Berti2005}. Pani and Berti \cite{Pani2014} constructed the models of slowly rotating NSs in scalar-tensor theories of gravity and Pani et al. \cite{pani2011} in Einstein-Dilaton-Gauss-Bonnet gravity. Yazadjiev et al. \cite{Yazadjiev2016} analysed the same in the scalar theories with massive scalar fields. Boumaza \cite{Boumaza2021} investigated the rotating NSs in the shift symmetric scalar torsion theory, and Staykov et al. \cite{Staykov2014} do the same in $R^{2}$ gravity. Pattersons and Sulaksono \cite{Pattersons2021} studied the mass correction and deformation produced by slowly rotating anisotropic NSs. Pappas et al. studied supermassive neutron stars in  $f(R, T)$ gravity \cite{Pappas:2022gtt}. Pretel \cite{Pretel2022}  computed the moment of inertia of anisotropic NSs in $f(R, T)$ gravity. \textbf{In this article, we aim to explore slowly rotating neutron stars in modified gravity theories. Specifically, we will investigate two modified gravity models: $f(R, T)=R+2\chi T$ up to the second order in angular velocity using Hartle-Thorne formalism and $f(R, T)=R+\alpha R^{2}+2\chi T$ up to the first order in angular velocity. Previous studies have inspired these models, and we hope to contribute further to the understanding of neutron stars in modified gravity. The structure of this article is the following: In section II, we discuss $f(R, T)$ gravity. In sections III and IV, we establish the mathematical formalism for rotating stars in $f(R, T)=R+2\chi T$ and $f(R, T)=R+\alpha R^{2}+2\chi T$ modified gravity, respectively. In section V, we discuss tidal deformability. In Section VI, we provide some information about the equation of state. After that, we discuss the results found in our study in section VII and section VIII. In the last sections, we draw our conclusions based on our findings.}

\section{f(R,T) Modified gravity}
Harko et al. proposed a modified theory of gravity by considering the gravitational Lagrangian to be an arbitrary function of the Ricci scalar $R$ and of the trace of stress-energy tensor $T$ \cite{Harko2011}. The field equations of this proposed modified theory can be followed from the action given below
\begin{equation}\label{EQ:1}
S=\frac{1}{2\kappa}\int d^{4}x\sqrt{-g}f(R,T)+\int d^{4}x\sqrt{-g} \mathcal{L}_{m}
\end{equation}  
where $\mathcal{L}_{m}$ is the matter Lagrangian density and $f(R,T)$ is any arbitrary function of $R$ and $T$.

 Varying the action given by Eq.\ref{EQ:1} w.r.t $g_{\mu\nu}$, we get modified Einstein's field equations as follows
\begin{align}\label{EQ:2}
f_{R}(R,T)R_{\mu\nu}-\frac{1}{2}g_{\mu\nu}
f(R,T)+
(g_{\mu\nu}\Box-\nabla_{\mu}\nabla_{\nu})f_{R}(R,T)\\ \nonumber
=\kappa T_{\mu\nu}-f_{T}(R,T) T_{\mu\nu}-f_{T}(R,T)\Theta_{\mu\nu}
\end{align}
where $f_{R}(R,T)\equiv \dfrac{\delta f(R,T)}{\delta R}$ , $f_{T}(R,T)\equiv \dfrac{\delta f(R,T)}{\delta T}$, $\Box \equiv \frac{1}{\sqrt{-g}}\delta_{\mu}(\sqrt{-g}g^{\mu \nu}\delta_{r})$ and the tensor $\Theta_{\mu\nu}$ is defined as
\begin{align}\label{EQ:3}
\Theta_{\mu\nu}\equiv g^{\alpha\beta}\dfrac{\delta T_{\alpha\beta}}{\delta g^{\mu\nu}}=-2T_{\mu\nu}+g_{\mu\nu}\mathcal{L}_{m}-2g^{\alpha\beta}\dfrac{\delta^{2}\mathcal{L}_{m}}{\delta g^{\mu\nu}\delta g^{\alpha\beta}}
\end{align}
The covariant divergence of Eq.\ref{EQ:2} will give us \citep{Barrientos2014}
\begin{widetext}
\begin{equation}\label{EQ:4}
\nabla^{\mu}T_{\mu\nu}=\frac{f_{T}(R,T)}{\kappa-f_{T}(R,T)}\left[\left(\Theta_{\mu\nu}+T_{\mu\nu}\right)\nabla^{\mu}ln f_{T}(R,T)-\frac{1}{2}g_{\mu\nu}\nabla^{\mu}T+\nabla^{\mu}\Theta_{\mu\nu}\right]
\end{equation}
\end{widetext}

\section{$f(R, T)=R+2\chi T$ modified gravity }
In this section, we consider $f(R, T)=R+2\chi T$, where $\chi$ is the matter-gravity coupling constant and Lagrangian density to be $\mathcal{L}_{m}=P$, where $P$ are the pressure. For this considerations, the Eqs.\ref{EQ:2}-\ref{EQ:4} become
\begin{equation}\label{EQ:5}
\Theta_{\mu\nu}=-2T_{\mu\nu}+P g_{\mu\nu}
\end{equation}
\begin{equation}\label{EQ:6}
G_{\mu\nu}=\kappa T_{\mu\nu}+\chi T g_{\mu\nu}+2\chi \left(T_{\mu\nu}-P g_{\mu\nu}\right)
\end{equation}
where $G_{\mu\nu}$ is the Einstein's tensor, and
\begin{equation}\label{EQ:7}
\nabla^{\mu}T_{\mu\nu}=\frac{2\chi}{\kappa+2\chi}\left[\nabla^{\mu}(P g_{\mu\nu}-\frac{1}{2}g_{\mu\nu} \nabla^{\mu}T)\right]
\end{equation}

%\subsection{Slow Rotation Effect}
We choose Boyer-Lindquist type coordinate $(t,r,\theta,\phi)$ to represent the most general stationary axisymmetric spacetime for rotaion which can be written as
\begin{align} \label{EQ:8}
ds^{2}&= -e^{\nu}[1-2 \epsilon^2 (h_{0}+h_{2}P_{2}]dt^{2} \nonumber \\  &+\frac{1+2\epsilon^2(m_{0}+m_{2}P_{2})/(r-2m)}{1-\frac{2m}{r}}dr^2  \nonumber\\  &+r^2 [1+2\epsilon^2 (v_{2}-h_{_2})P_{2}][d\theta^2+\sin^{2}\theta (d\phi-\epsilon \omega dt)^2]
\end{align}
where $\epsilon$ is a bookkeeping parameter for slow rotation and $P_{2}$ is the 2nd order Legendre polynomial. $P_{2}=P_{2}(\cos\theta)=(3\cos^{2}\theta-1)/2$. The function $r$ and $m$ are zeroth order, $\omega$ is first order, and $m_{0}$,$h_{0}$, $h_{2}$, $v_{2}$ and $m_{2}$ are second order in rotation. $\omega$ gives the angular velocity of local inertial frames.

 Since the perturbation causes displacement of matters of the star, one cannot carry out calculation in $(t,r,\theta,\phi)$ coordinates having old pressure and energy relation. To overcome this problem, one has to transform the radial coordinate in such a way that the displaced density in the new radial coordinate has same density as the unperturb non rotating one. As pointed out in Refs \citep{Hartle1967,Hartle1968}, one can bypass this coordinate transformation by expanding the pressure and the density in the old $(t,r,\theta,\phi)$ coordinate as
\begin{align}\label{EQ:9}
P&=P_{0}+\epsilon^2 (\rho_{0}+P_{0})(p_{0}+p_{2} P_{2})\\
\rho&=\rho_{0}+\epsilon^2 (\rho_{0}+P_{0})\dfrac{d\rho_{0}}{dP_{0}}(p_{0}+p_{2} P_{2})
\end{align}
 subjected to assumption of the barotropic Equation of State (EOS) $P=P(\rho)$. Where $\rho_{0}$ and $P_{0}$ are unperturbed matter density and pressure, and $p_{0}$ and $p_{2}$ are the pressure perturbation terms for second order slow rotation perturbation.
 The energy-momentum tensor and the four velocity of an uniformly rotating star is given by
 \begin{equation}\label{EQ:11}
T_{\mu \nu}=(P+\rho)u_{\mu} u_{\nu}+P g_{\mu \nu} 
 \end{equation}
and 
\begin{equation}\label{EQ:12}
u^{\mu}=(u^{0},0,0,\epsilon \Omega u^{0})
\end{equation}
where $\Omega$ represents the fluid angular velocity as measured by the far away distance observer the normalised condition $u^{\nu}u_{\nu}=1$ gives 
\begin{equation}\label{EQ:13}
u^{0}=[g_{\mu \nu}+2 \epsilon \Omega g_{t \phi} +\epsilon^2 \Omega^2  g_{\phi \phi}]
\end{equation}

\begin{figure*}[!ht]
\centering
\begin{subfigure}{0.49\textwidth}
    \includegraphics[width=\textwidth]{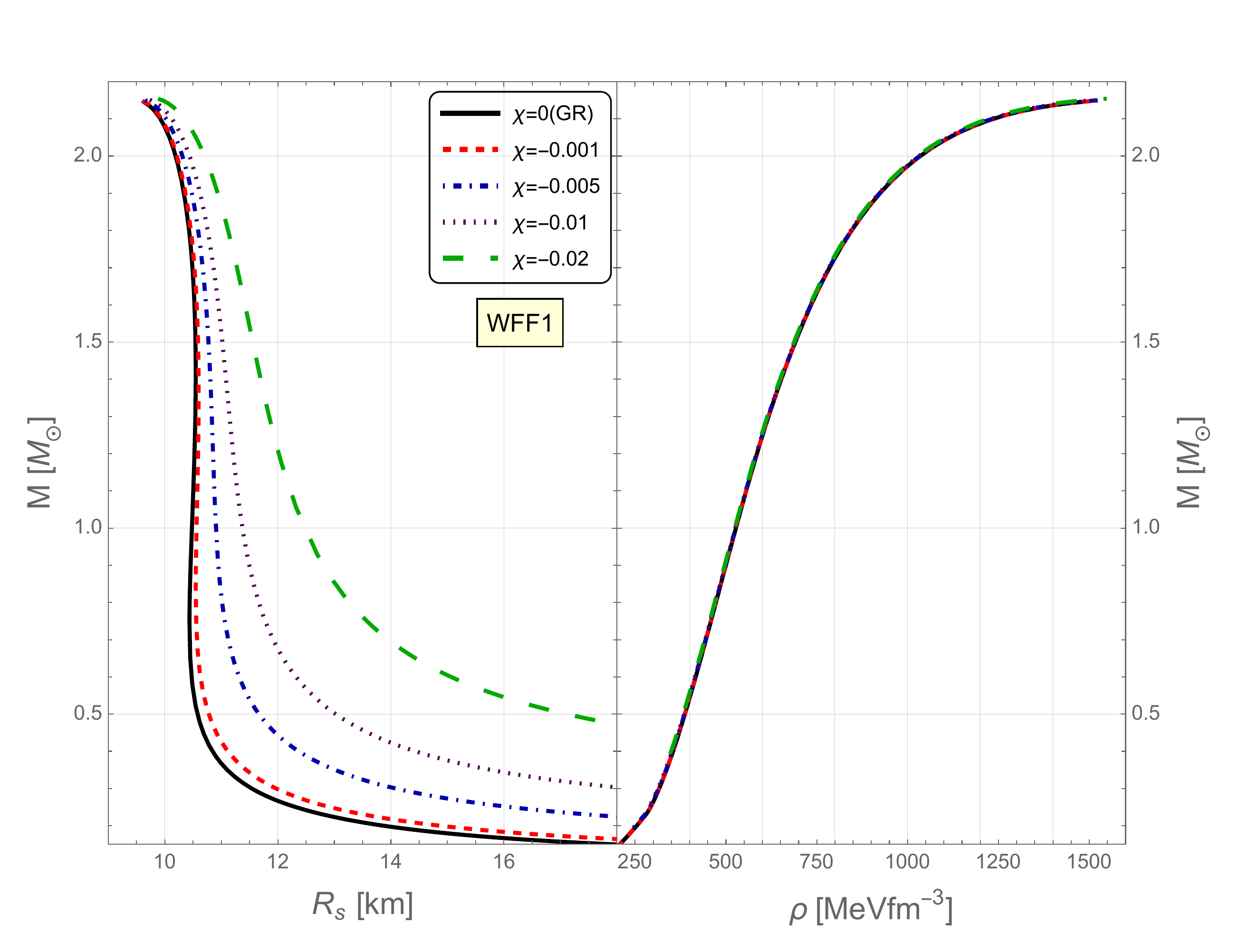}
%    \caption{sly}
    \label{fig:1a}
\end{subfigure}
\hfill
\begin{subfigure}{0.49\textwidth}
    \includegraphics[width=\textwidth]{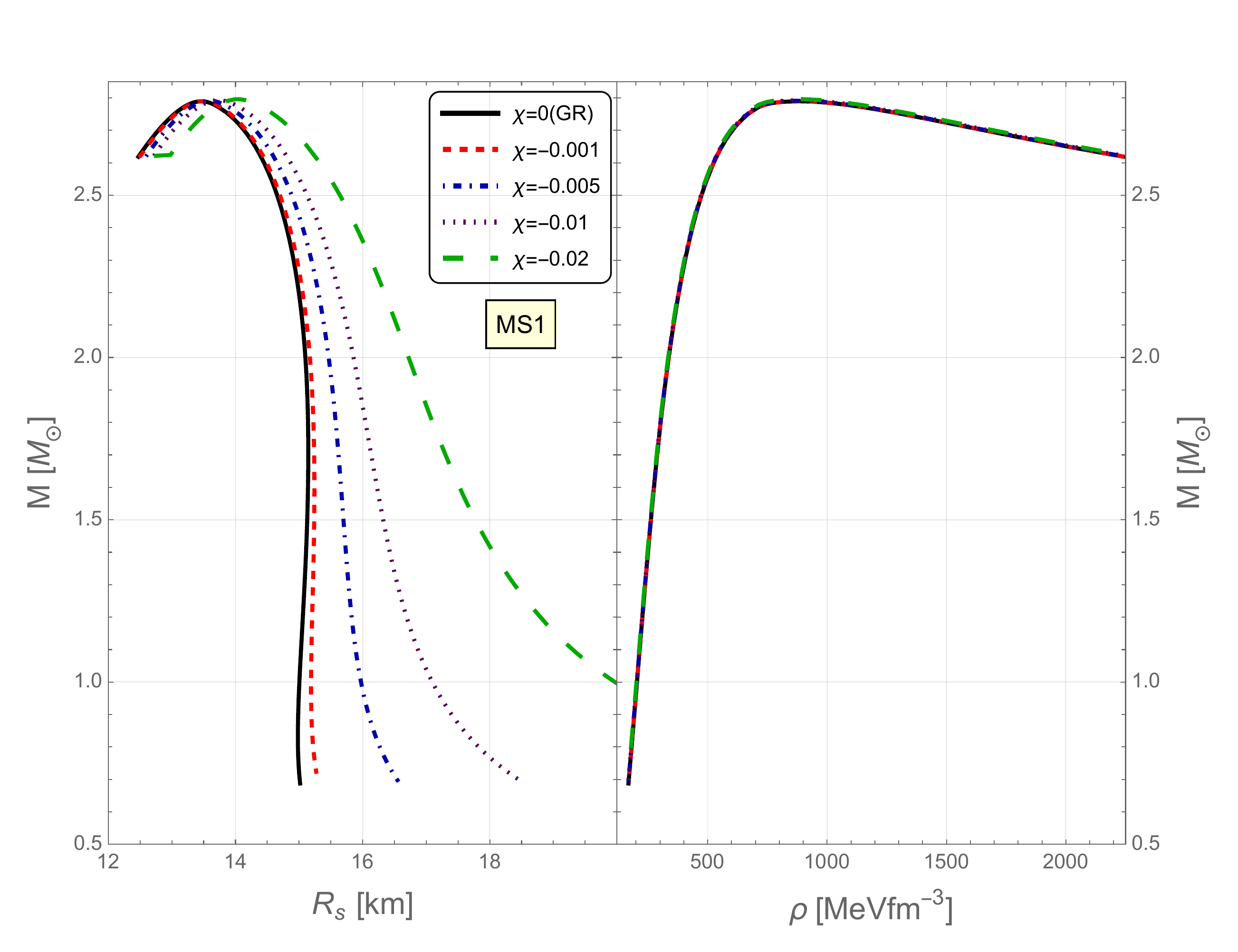}
 %   \caption{bbb2}
    \label{fig:1b}
\end{subfigure}
        \hfill
\begin{subfigure}{0.49\textwidth}
    \includegraphics[width=\textwidth]{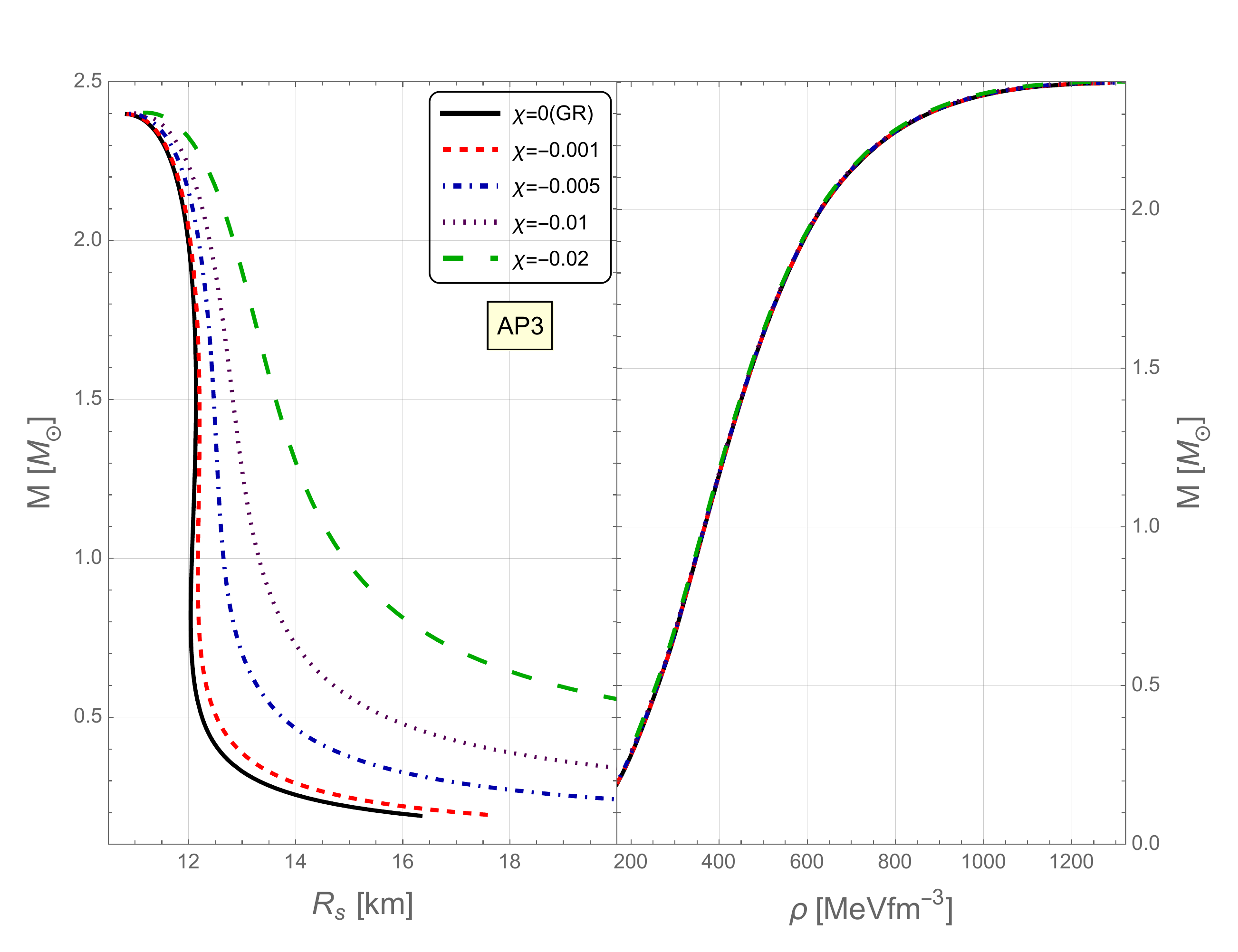}
 %   \caption{alf4}
    \label{fig:1c}
\end{subfigure}
\hfill
\begin{subfigure}{0.49\textwidth}
    \includegraphics[width=\textwidth]{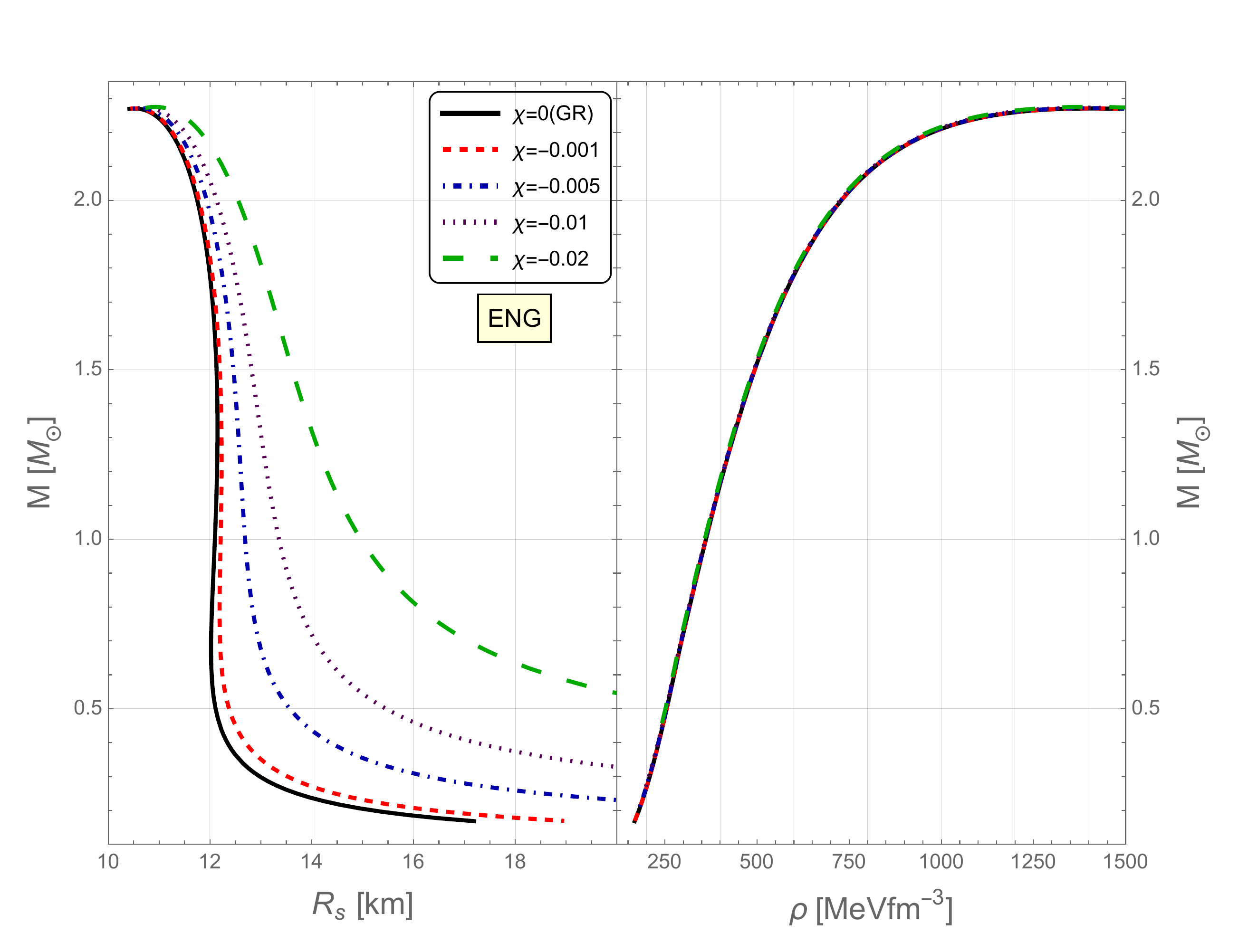}
%    \caption{pal6}
    \label{fig:1d}
\end{subfigure}
\hfill
\begin{subfigure}{0.49\textwidth}
    \includegraphics[width=\textwidth]{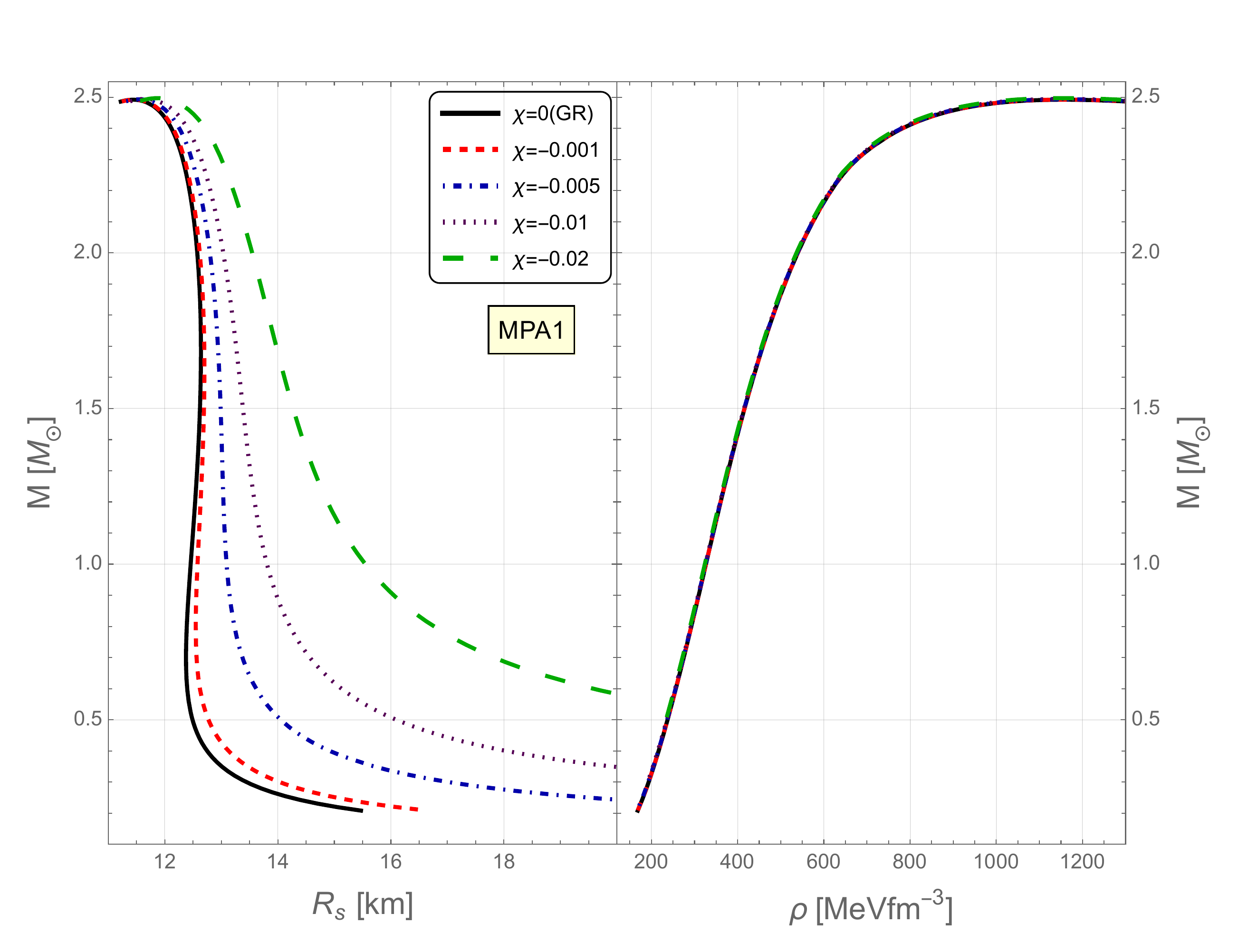}
 %   \caption{alf4}
    \label{fig:1e}
\end{subfigure}
\hfill
\begin{subfigure}{0.49\textwidth}
    \includegraphics[width=\textwidth]{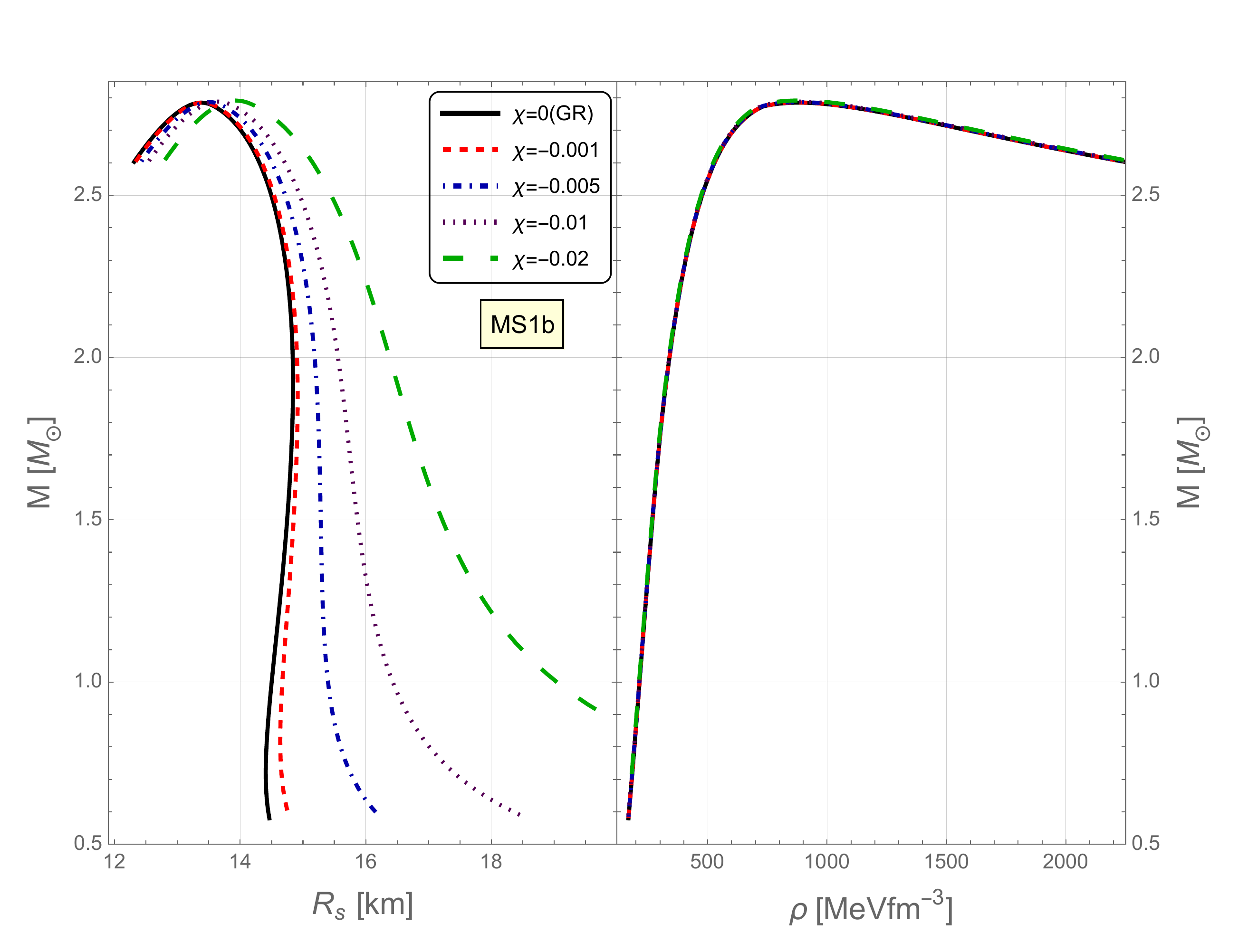}
%    \caption{pal6}
    \label{fig:1f}
\end{subfigure}
\caption{Mass-Radius relation  and Mass-Density relation for six EoSs in $f(R, T)=R+2\chi T$ modified gravity }
\label{fig:1}
\end{figure*}

\subsection{Zeroth Order Effect}
Since we consider the slow rotational effect as a perturbation on the non rotating star, the zeroth order in rotation gives the same equations as the non-rotating relativistic equations. The $(t,t)$ and $(r,r)$ components of $f(R,T)$ field equations yield
\begin{equation}\label{EQ:14}
\dfrac{dM}{dr}-\frac{1}{2} r^2 \left[\left(\kappa +3 \chi \right) \rho - \chi  P\right]=0
\end{equation}
\begin{equation}\label{EQ:15}
\dfrac{d\nu}{dr}-\frac{2 M+r^3 \left[ (\kappa +3 \chi ) P-\chi  \rho\right]}{r (r-2 M)}=0
\end{equation}
The covariant divergence equation $(i.e. \nabla^{\mu}T_{\mu r} =0)$ gives
\begin{equation}\label{EQ:16}
\dfrac{d P}{dr}+\frac{\gamma  (\kappa +2 \chi ) (P+\rho ) \left(2 M+r^3 \left[
   (\kappa +3 \chi )P-\rho  \chi \right]\right)}{2 r (r-2 M) \left[ 
   (\kappa +3 \chi )\gamma-\chi \right]}=0
\end{equation}
Given the EOS $P=P(\rho)$, one can find radius $(R_{s})$ of the star by doing integration from the centre to outward, and imposing the condition $\rho(R_{s})=P(R_{s})=0$ at the surface.
\subsection{First Order Effect}
The only nonvanishing component $(t,\phi)$ of the field equations at linear order of rotation gives rise to a second order differential equation of $\bar{\omega }$ and which is written as
\begin{widetext}
\begin{equation}\label{EQ:17}
\dfrac{d^2 \bar{\omega } }{dr^2}-\frac{\dfrac{d\bar{\omega }}{dr} \left(16 M+r^3 (\kappa +2 \chi ) (P+\rho )-8 r\right)+4 r^2 (\kappa +2 \chi ) (P+\rho ) \bar{\omega }}{2 r (r-2 M)}=0
\end{equation}
\end{widetext}
The exterior solution (i.e. where $\rho=P=0$) of the above equation takes the form written as below
\begin{equation}\label{EQ:18}
\bar{\omega }=\Omega-\frac{2J}{r^3}
\end{equation}
where $J$ is the total angular momentum of the star. The moment of Inertia gives how first the NS rotates given the total angular momentum is defined as
\begin{equation}\label{EQ:19}
I=\frac{J}{\Omega}
\end{equation}
At first we do Taylor series expansion of Eq.\ref{EQ:17} about the star's centre (i.e. at $r \to 0$), and then solving that expanded equation we obtain the asymptotic solution of $\bar{\omega }$ about $r \to 0$ as
\begin{equation}\label{EQ:20}
\bar{\omega }=\bar{\omega }_{c}+ \frac{1}{5} r^2 \bar{\omega }_{c} (\kappa +2 \chi )
   \left(\rho(P_{c})+P_{c}\right)
\end{equation}  
  The regularity condition of $\bar{\omega }$ at the centre  and the homogeneity of Eq.\ref{EQ:17} gives us freedom to set $\bar{\omega }_{c}$ to unity.

 To ensure the continuity and differentiability of the solution, one has to match the exterior solution to interior solution at the NS surface (i.e. at $r=R_{s}$) using the matching conditions
  \begin{equation}\label{EQ:21}
\bar{\omega }_{int}(R_{s})=  \bar{\omega }_{ext}(R_{s})
  \end{equation}
  \begin{equation}\label{EQ:22}
  \dfrac{d \bar{\omega }_{int}}{d r}\bigg|_{r=R_{s}} =\dfrac{d \bar{\omega }_{ext}}{d r}\bigg|_{r=R_{s}}
  \end{equation}

\subsection{Second Order Effect}  
  The $(r,\theta)$, $(\theta,\theta)$ and $(\phi,\phi)$ components of the field equation and $\nabla^{\mu}T_{\mu \theta}$ component of covariant derivative become nonvanishing at quadratic order in spin. Solving these four nonvanshing equations alongside the equations we get for zeroth and first order rotational effect, we obtain five first order linear differential equations and two non-differential equations. In the second order correction, only two modes survive for axisymmetric perturbation which are $(l,m)=(0,0)$ and $(l,m)=(2,0)$. 
  \subsubsection{Spherical Deformation}
  The $(l,m)=(0,0)$ mode produces spherical deformation in a star. This spherical deformation is mathematically represented by three first order differential equations of the mass perturbation factor $m_{0}$ , the pressure perturbation factor $p_{0}$
 and $h_{0}$ as 
 \begin{widetext}
 \begin{equation}\label{EQ:23}
 \dfrac{dm_{0}}{dr}-\frac{p_0 r^2 (P+\rho ) (-\gamma  \chi +\kappa +3 \chi )}{2 \gamma }-\frac{1}{12} e^{-\nu } r^3 \left[(r-2 M) \left(\bar{\omega }'\right)^2+4 r \bar{\omega }^2 (\kappa +2\chi ) (P+\rho )\right]=0
 \end{equation}
  \begin{align}\label{EQ:24}
\dfrac{d p_{0}}{dr}&+m_0 \left(\frac{\gamma  P r^2 (\kappa +2 \chi ) (\kappa +3 \chi )}{(r-2 M)^2
   (\gamma  (\kappa +3 \chi )-\chi )}+\frac{\gamma  (-\kappa -2 \chi ) \left(\rho  r^2 \chi
   -1\right)}{(r-2 M)^2 (\gamma  (\kappa +3 \chi )-\chi )}\right) \nonumber \\
	& +p_0 \left(\frac{\chi  \gamma '}{\gamma 
   (\gamma  (\kappa +3 \chi )-\chi )}+\frac{P r^2 (\kappa +2 \chi )}{2 (r-2 M)}-\frac{\rho  r^2 (\kappa +2
   \chi ) (\chi -\gamma  (\kappa +3 \chi ))}{2 (r-2 M) (\gamma  (\kappa +3 \chi )-\chi )}\right)  \nonumber \\
	&-\frac{\gamma  e^{-\nu } r \bar{\omega } (\kappa +2 \chi ) \left(\bar{\omega } \left(-6 M+r^3 (\rho  \chi
   -P (\kappa +3 \chi ))+2 r\right)+2 r (r-2 M) \bar{\omega }'\right)}{3 (r-2 M) (\gamma  (\kappa +3 \chi
   )-\chi )}-\frac{\gamma  e^{-\nu } r^3 (\kappa +2 \chi ) \left(\bar{\omega }'\right)^2}{12 (\gamma 
   (\kappa +3 \chi )-\chi )}=0
 \end{align}
  \begin{align}\label{EQ:25}
  \dfrac{d h_{0}}{dr}+m_0 \left(\frac{\rho  r^2 \chi -1}{(r-2
   M)^2}-\frac{P r^2 (\kappa +3 \chi )}{(r-2 M)^2}\right)+p_0 \left(\frac{\rho  r^2 (\chi -\gamma  (\kappa
   +3 \chi ))}{2 \gamma  (r-2 M)}-\frac{P r^2 (\gamma  (\kappa +3 \chi )-\chi )}{2 \gamma  (r-2 M)}\right)+\frac{1}{12} e^{-\nu } r^3 \left(\bar{\omega }'\right)^2=0
  \end{align}
 \end{widetext}

 These equation are to be integrated outward from the centre. $m_{0}$ and $p_{0}$ vanish at the surface of the star. Outside the star, the asymptotic solutions (for $r \to \infty$) of Eq.\ref{EQ:23} \& Eq.\ref{EQ:25} give
  \begin{equation}\label{EQ:26}
  m_{0}=\delta M-\frac{J^{2}}{r^{3}}
\end{equation}
and
\begin{equation}\label{EQ:27}
h_{0}=-\frac{\delta M}{r}-\frac{2 \delta M~
   M}{r^2}-\frac{4 \delta
   M~ M^2}{r^3}+\frac{J^2-8 \delta M ~M^3}{r^4}
\end{equation}
where $\delta M$  is the change in mass due the rotation. Therefore, the total mass-energy of the star is then given by
\begin{equation}\label{EQ:28}
M_{tot}=M+\delta M=M+m_{0}+\frac{J^2}{R^{3}}
\end{equation}

\subsubsection{Quadrupole Deformation}
The quadrupole deformation of a rotating star is represented by two linear differential equation of $h_{2}$ and $v_{2}$, and they are expresses as
\begin{widetext}
\begin{align}\label{EQ:29}
\dfrac{d v_{2}}{dr}&+\frac{h_2 \left(2 M+r^3 (P (\kappa +3 \chi )-\rho  \chi )\right)}{r (r-2 M)} \nonumber \\
&-\frac{e^{-\nu } r^2 \left(-2 M+r^3 (P (\kappa +3 \chi )-\rho  \chi )+2 r\right) \left((r-2 M)
   \left(\bar{\omega }'\right)^2+2 r \bar{\omega }^2 (\kappa +2 \chi ) (P+\rho )\right)}{12 (r-2
   M)}=0
\end{align}
and
\begin{align}\label{EQ:30}
\dfrac{d h_{2}}{dr}  &+\frac{4 v_{2}}{2 M+r^3 (P (\kappa +3 \chi )-\rho 
   \chi )}  +\frac{h_{2} \left(-4 M^2+2 M r \left(r^2 (\kappa  \rho +P (3 \kappa +8 \chi
   ))+2\right)\right)}{r (r-2 M) \left(2
   M+r^3 (P (\kappa +3 \chi )-\rho  \chi )\right)}\nonumber \\
&+ \frac{r^4 h_{2}\left(P^2 r^2 (\kappa +3 \chi )^2-P \left(\kappa +2 \rho  r^2 \chi  (\kappa +3 \chi )+2
   \chi \right)+\rho  \left(-\kappa +\rho  r^2 \chi ^2-2 \chi \right)\right)}{r (r-2 M) \left(2
   M+r^3 (P (\kappa +3 \chi )-\rho  \chi )\right)}\nonumber \\
&
   -\frac{e^{-\nu } r^2 (r-2 M)
   \left(\bar{\omega }'\right)^2 \left(4 M^2+4 M \left(r^3 (P (\kappa +3 \chi )-\rho  \chi )+r\right)+r^6
   (P (\kappa +3 \chi )-\rho  \chi )^2-2 r^2\right)}{12 (r-2 M) \left(2 M+r^3 (P (\kappa +3 \chi
   )-\rho  \chi )\right)}\nonumber \\
   &- \frac{ r^3 \bar{\omega }^2 e^{-\nu }  (\kappa +2 \chi ) (P+\rho ) \left(4 M^2+4 M r \left(r^2 (P
   (\kappa +3 \chi )-\rho  \chi )-1\right)+r^6 (P (\kappa +3 \chi )-\rho  \chi )^2+2 r^2\right)}{6 (r-2 M) \left(2 M+r^3 (P (\kappa +3 \chi
   )-\rho  \chi )\right)}=0
  \end{align}
\end{widetext}

\begin{figure*}[!ht]
\centering
\begin{subfigure}{0.32\textwidth}
    \includegraphics[width=\textwidth]{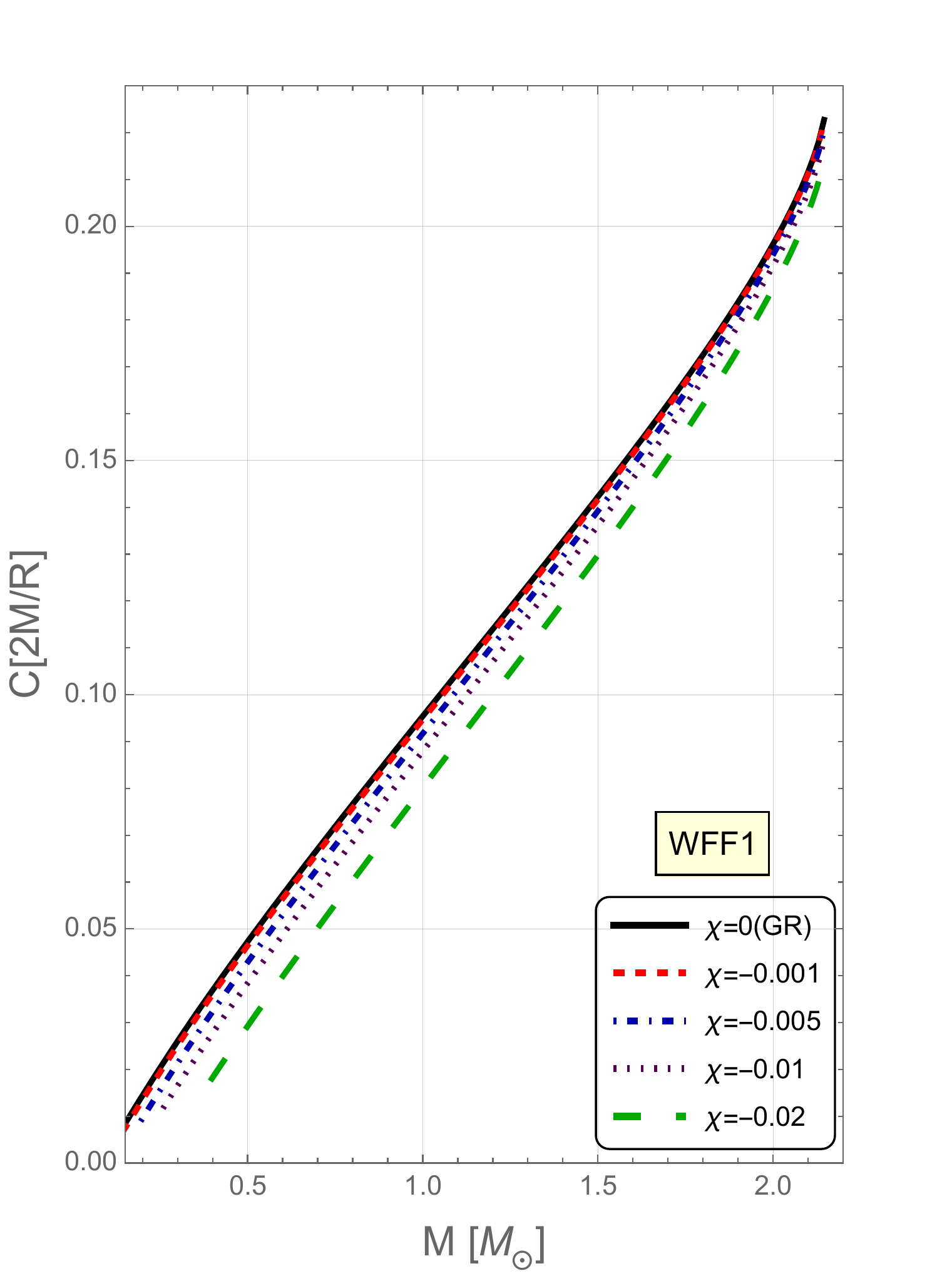}
%    \caption{sly}
    \label{fig:2a}
\end{subfigure}
\hfill
\begin{subfigure}{0.32\textwidth}
    \includegraphics[width=\textwidth]{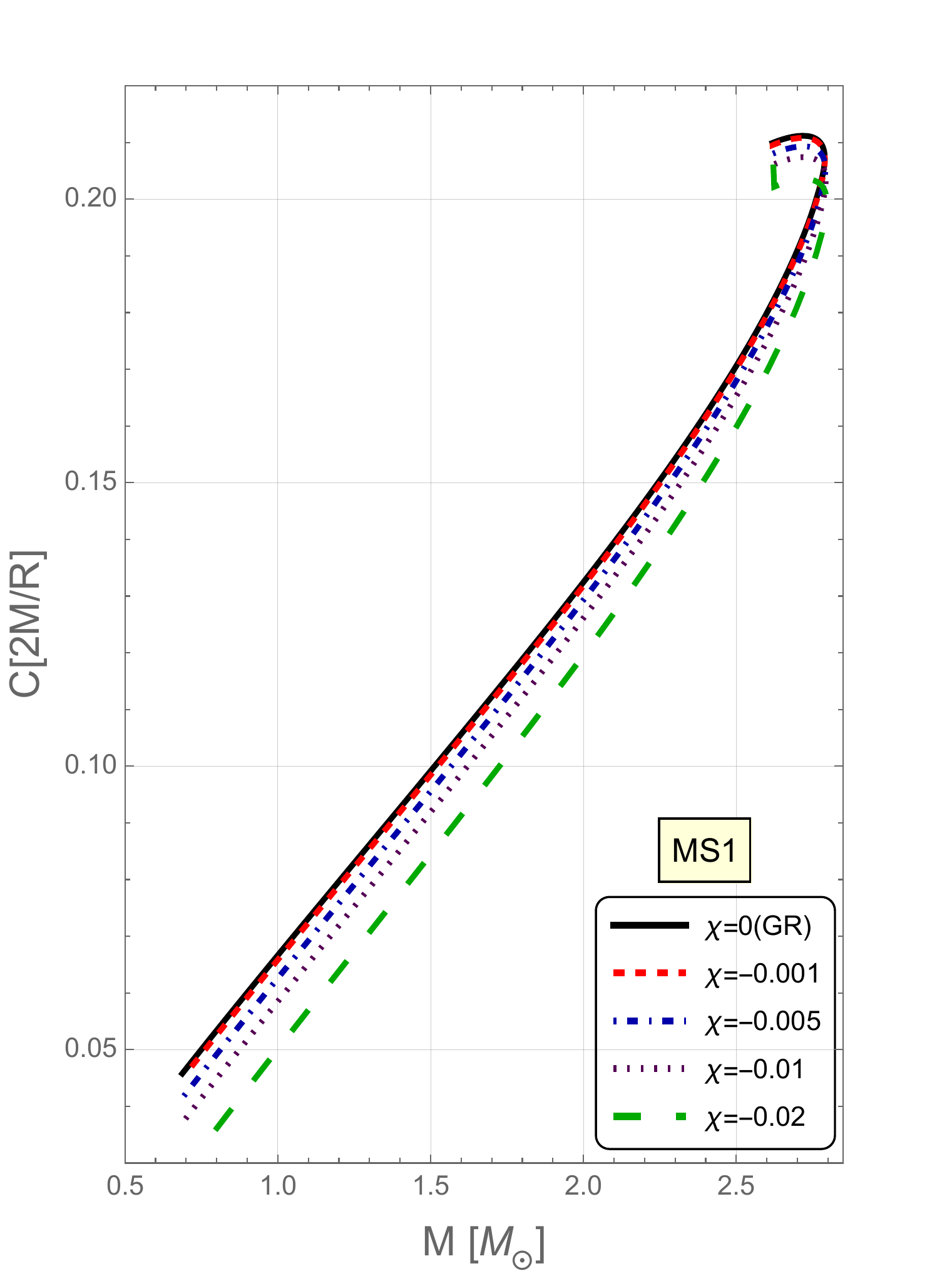}
 %   \caption{bbb2}
    \label{fig:2b}
\end{subfigure}
        \hfill
\begin{subfigure}{0.32\textwidth}
    \includegraphics[width=\textwidth]{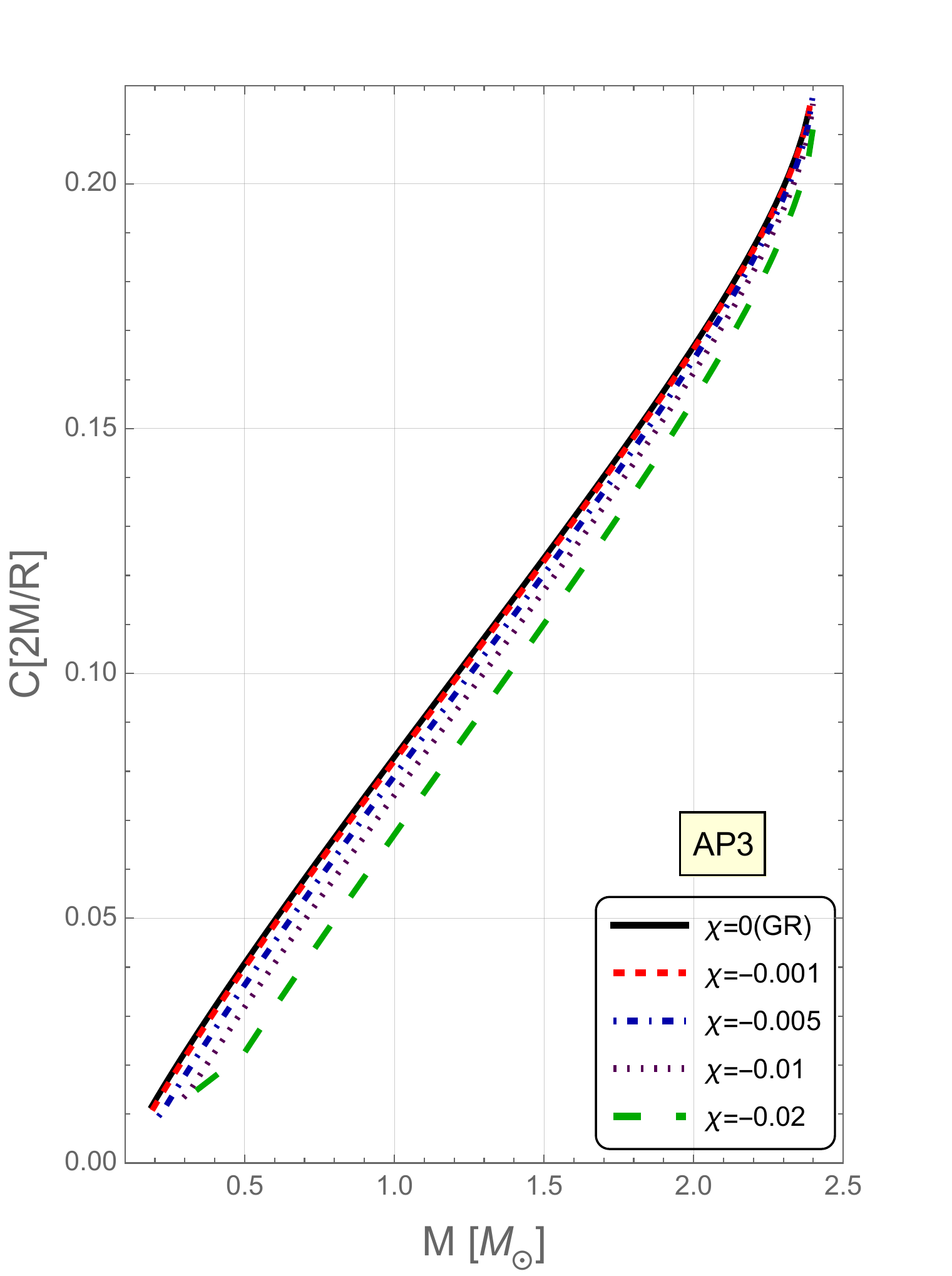}
 %   \caption{alf4}
    \label{fig:2c}
\end{subfigure}
\hfill
\begin{subfigure}{0.32\textwidth}
    \includegraphics[width=\textwidth]{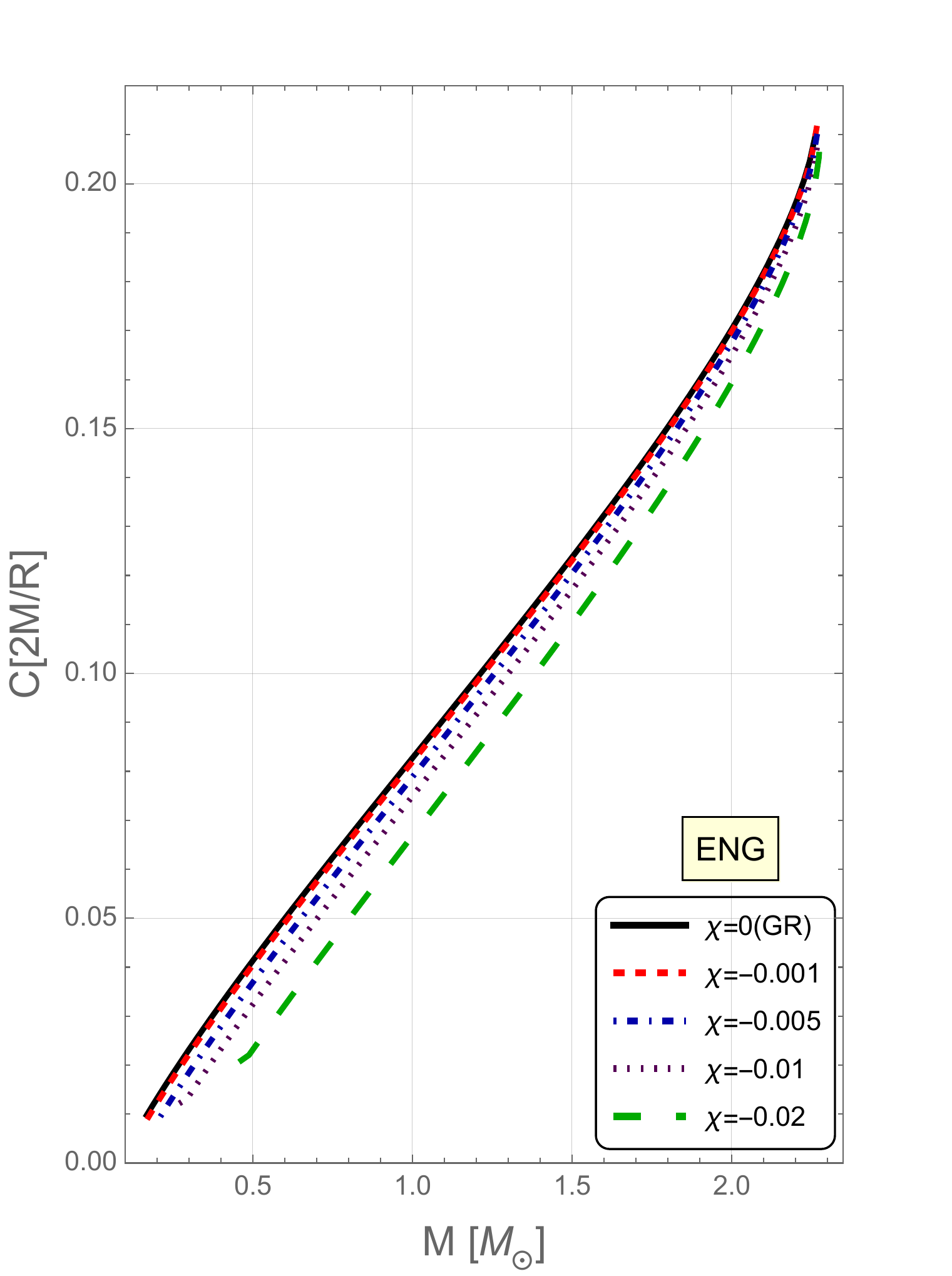}
%    \caption{pal6}
    \label{fig:2d}
\end{subfigure}
\hfill
\begin{subfigure}{0.32\textwidth}
    \includegraphics[width=\textwidth]{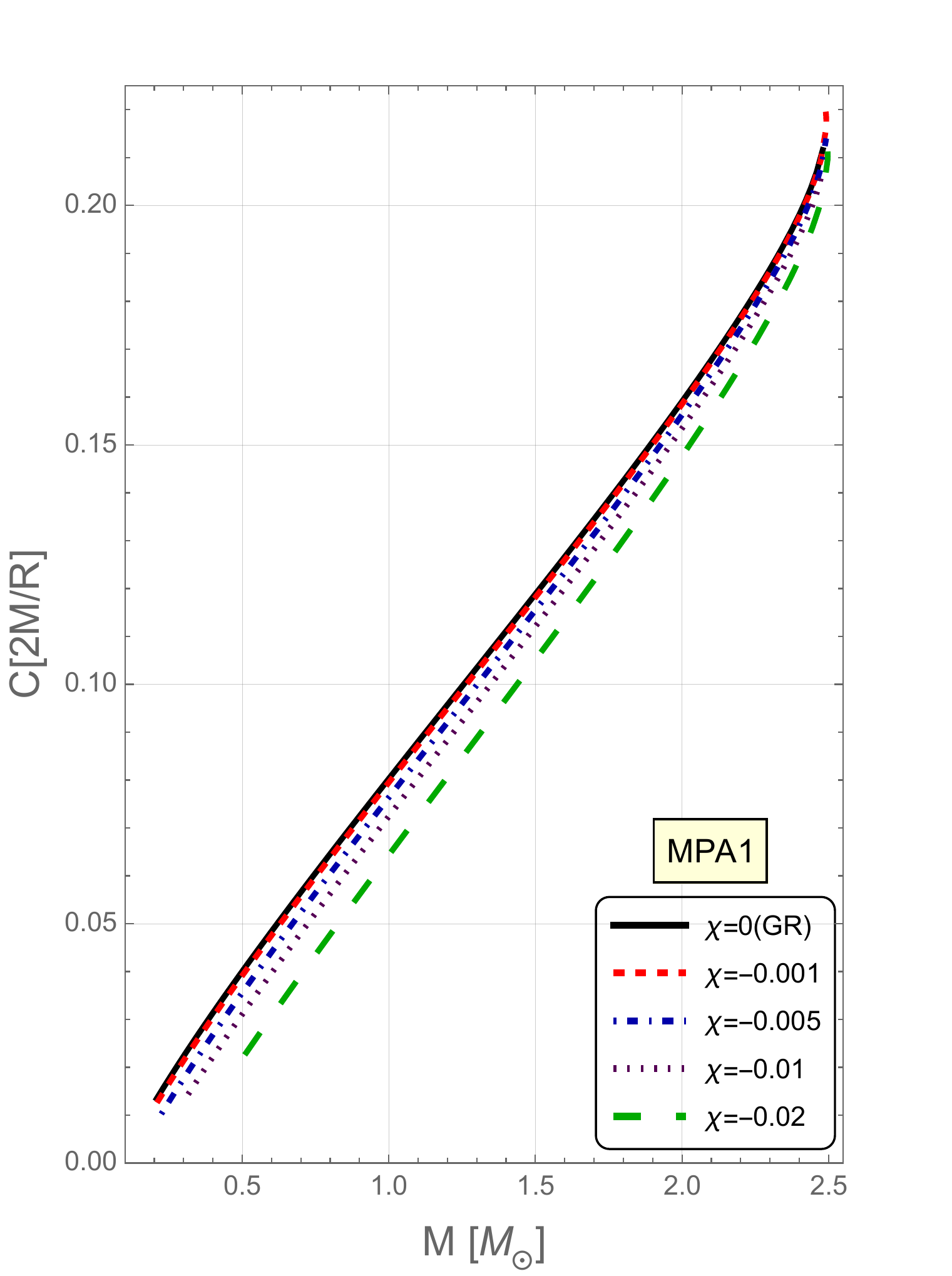}
 %   \caption{alf4}
    \label{fig:2e}
\end{subfigure}
\hfill
\begin{subfigure}{0.32\textwidth}
    \includegraphics[width=\textwidth]{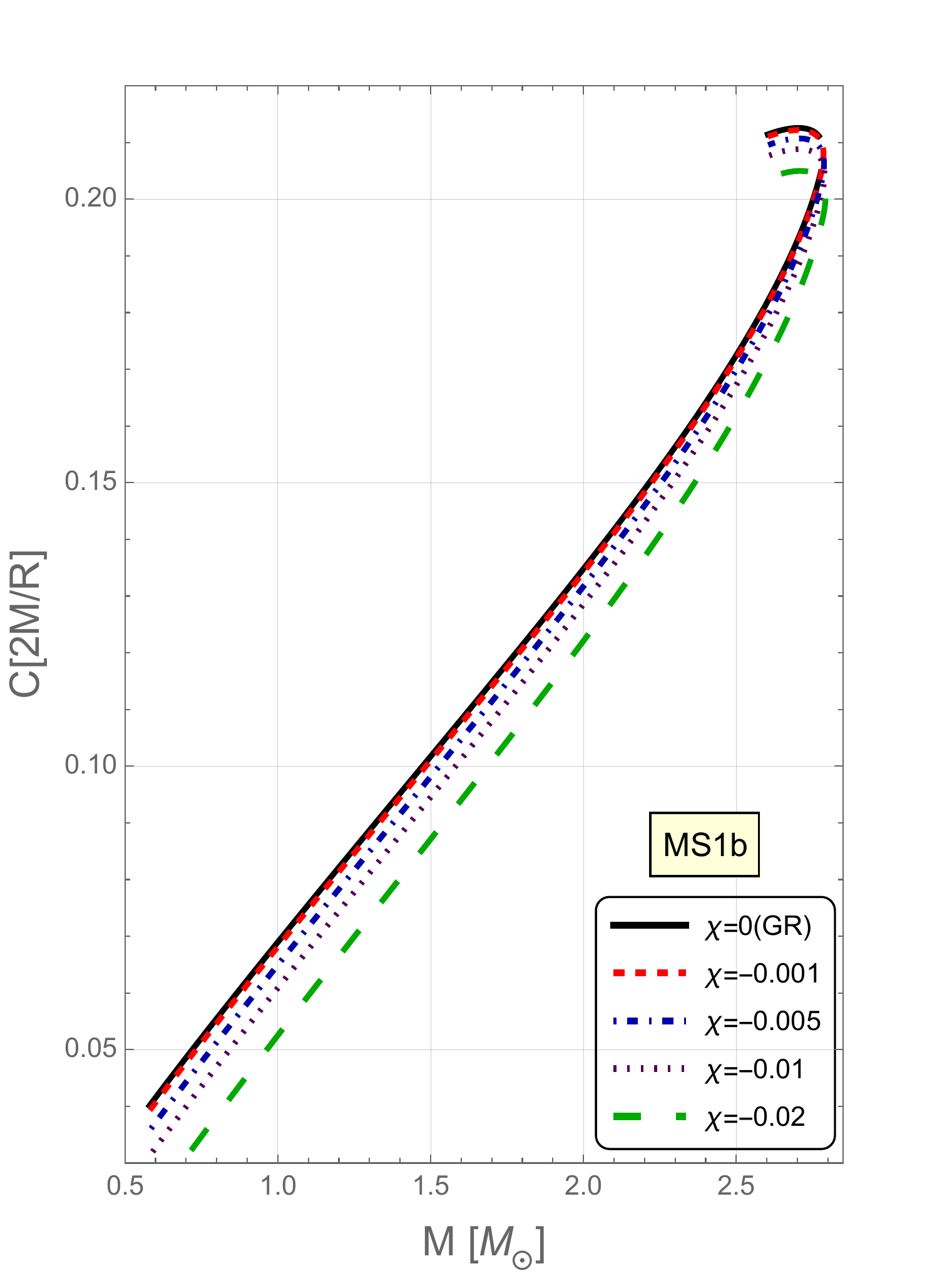}
%    \caption{pal6}
    \label{fig:2f}
\end{subfigure}
\caption{The relation between compactness and  mass in $f(R, T)=R+2\chi T$ modified gravity.}
\label{fig:2}
\end{figure*}

We can see that the Eqs.\ref{EQ:29} \& \ref{EQ:30} are inhomogeneous differential equations and, in general, the solutions of the inhomogenous differential equations are irregular at infinity. To find the regular solutions of the above two differential equations, we first construct two homogeneous differential equations from them as
  \begin{equation}\label{EQ:31}
  \dfrac{d v^{h}_{2}}{dr}+\frac{h^{h}_{2} \left(2 M+r^3 (P (\kappa +3 \chi )-\rho  \chi
  	)\right)}{r (r-2 M)}=0
  \end{equation}
  and
  
  \begin{widetext}
  \begin{align}\label{EQ:32}
 \dfrac{d h^{h}_{2}}{dr}  &+\frac{4 v^{h}_{2}}{2 M+r^3 (P (\kappa +3 \chi )-\rho 
   \chi )}  +\frac{h^{h}_{2} \left(-4 M^2+2 M r \left(r^2 (\kappa  \rho +P (3 \kappa +8 \chi
   ))+2\right)\right)}{r (r-2 M) \left(2
   M+r^3 (P (\kappa +3 \chi )-\rho  \chi )\right)}\nonumber \\
&+ \frac{r^4 h^{h}_{2}\left(P^2 r^2 (\kappa +3 \chi )^2-P \left(\kappa +2 \rho  r^2 \chi  (\kappa +3 \chi )+2
   \chi \right)+\rho  \left(-\kappa +\rho  r^2 \chi ^2-2 \chi \right)\right)}{r (r-2 M) \left(2
   M+r^3 (P (\kappa +3 \chi )-\rho  \chi )\right)}   =0
  \end{align}
  \end{widetext}
  
  Where $v^{h}_{2} $ and $h^{h}_{2}$ stand for homogeneous part of $v_{2}$ and $h_{2}$ respectively. The asymptotic expression's for Eqs.\ref{EQ:29}-\ref{EQ:32} at infinity are
  \begin{equation}\label{EQ:33}
  v_{2}=2 M^{2} A_{1}-2M A_{1}r +\frac{M~ A_{2}-3J^{2}}{2r^{4}}
  \end{equation}
  
  \begin{equation}\label{EQ:34}
  h_{2}=-2 M  A_{1} r+r^2 A_{1}+\frac{A_{2}}{r^3}
  \end{equation}
  \begin{equation}\label{EQ:35}
  v^{h}_{2}=2 M^{2} A^{h}_{1}-2M  A^{h}_{1} r+\frac{M~ A^{h}_{2}}{2r^{4}}
  \end{equation}
  
  \begin{equation}\label{EQ:36}
  h^{h}_{2}=-2 M r A^{h}_{1}+r^2 A^{h}_{1}+\frac{A^{h}_{2}}{r^3}
  \end{equation}
    
Where $A_{1}$, $A_{2}$, $A^{h}_{1}$ and $A^{h}_{2}$ are constant and values of them to be calculated from numerical integration. We find the regular solution of inhomogeneous differential equation at infinity by constructing a suitable linear combination equation using a particular inhomogeneous solution and the corresponding solution.Thus, the regular solutions of $v_{2}$ and $h_{2}$ have expressions as
  \begin{equation}\label{EQ:37}
  v^{r}=A^{r} v^{h}_{2}+v_{2}
  \end{equation}
   \begin{equation}\label{EQ:38}
  h^{r}=A^{r} h^{h}_{2}+h_{2}
  \end{equation}
  where $ v^{r}$ and $ h^{r}$ stand for regular solution, and $A^{r}$ has an expression as
  \begin{equation}\label{EQ:39}
  A^{r}=-\frac{A_{1}}{A^{h}_{1}}
  \end{equation}
  The quadrupole moment of the configuration is then defined as
  \begin{equation}\label{EQ:40}
  Q=\frac{A^{h}_{1} A_{2}-A_{1} A^{h}_{2} }{A^{h}_{1}}
  \end{equation}
 Besides two differential equations (Eqs.\ref{EQ:29} \& \ref{EQ:30}), we also get two algebraic relations for $m_{2}$ and $p_{2}$ as
 \begin{widetext}
  \begin{equation}\label{EQ:41}
 p_{2}=-\frac{\gamma  e^{-\nu } (\kappa +2 \chi ) \left(r^2
   \bar{\omega }^2+3 h_2 e^{\nu }\right)}{3 \left(\gamma  (\kappa
   +3 \chi )-\chi \right)}
  \end{equation}
  
  \begin{equation}\label{EQ:42}
m_{2}=\frac{1}{6} e^{-\nu } (r-2 M) \left[r^3 \left((r-2 M)
   \left(\bar{\omega }'\right)^2+2 r \bar{\omega }^2 (\kappa
   +2 \chi ) (P+\rho )\right)-6 h_2 e^{\nu }\right]
  \end{equation}
  \end{widetext}
  Rotation turns a non-rotating spherical configuration into a spheroid. The spheroid's surface at any radius $(r^{*})$ represents the surface of constant matter density.The radius $(r^{*})$ of the constant matter density surface of the rotating object is related to the radius $(r)$ of the non-rotating constant matter density surface by the expression
  \begin{equation}\label{EQ:43}
r^{*}=r+\epsilon^{2} \delta r=r+ \epsilon^{2}(\zeta_{0}+\zeta_{2} P_{2})  
  \end{equation}
  
  where $\zeta_{0}=-p_{0}(\rho+P)\left(\dfrac{dP}{dr}\right)^{-1}$ and $\zeta_{2}=-p_{2}(\rho+P)\left(\dfrac{dP}{dr}\right)^{-1}$
  
 and the eccentricity of that spheroid is then defined as
 \begin{equation}\label{EQ:44}
 e=\sqrt{-3\left(v_{2}-h_{2}+\frac{\zeta_{2}}{r}\right)}
\end{equation}

\begin{figure*}[!ht]
\centering
\begin{subfigure}{0.49\textwidth}
    \includegraphics[width=\textwidth]{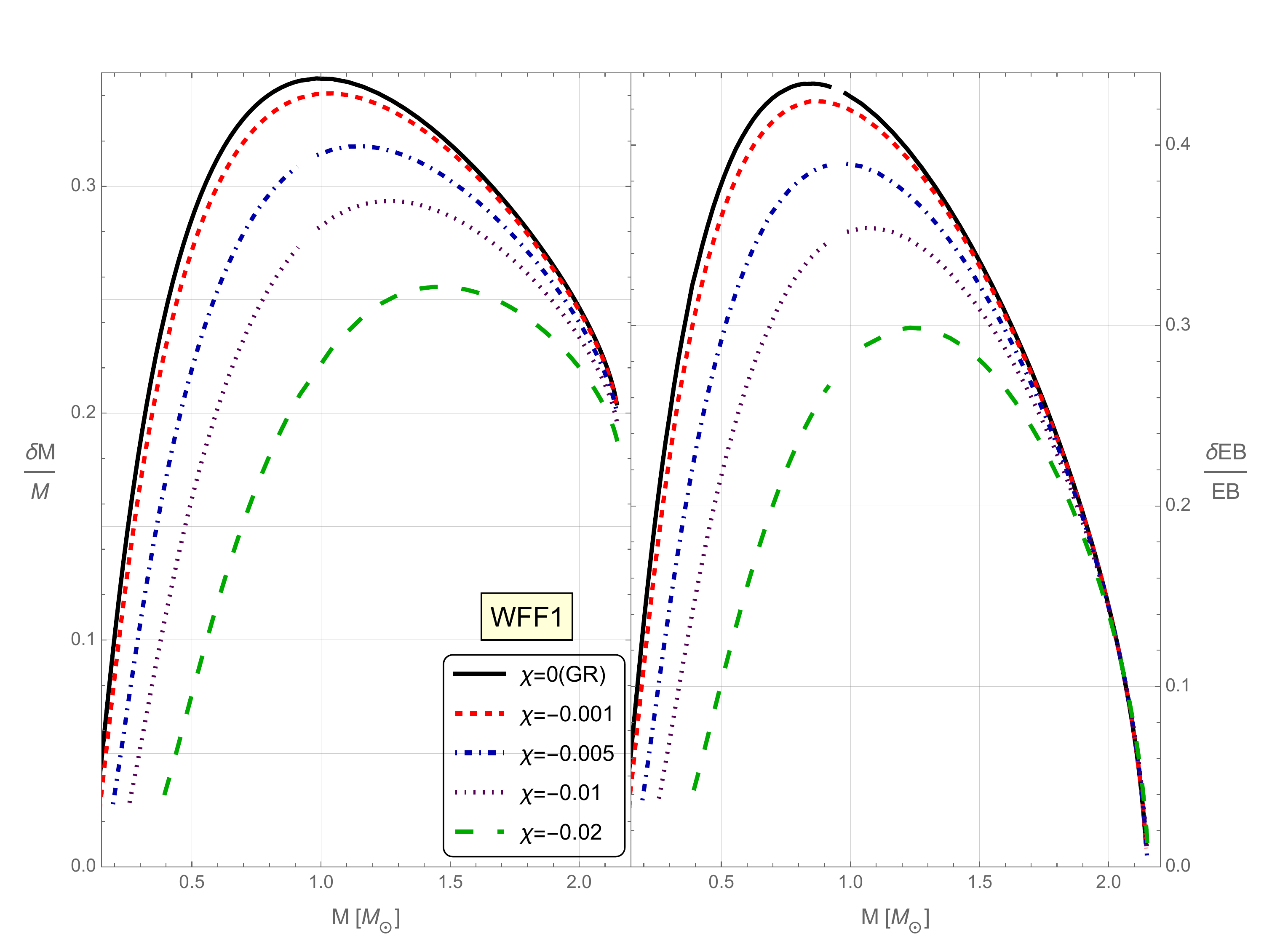}
%    \caption{sly}
    \label{fig:3a}
\end{subfigure}
\hfill
\begin{subfigure}{0.49\textwidth}
    \includegraphics[width=\textwidth]{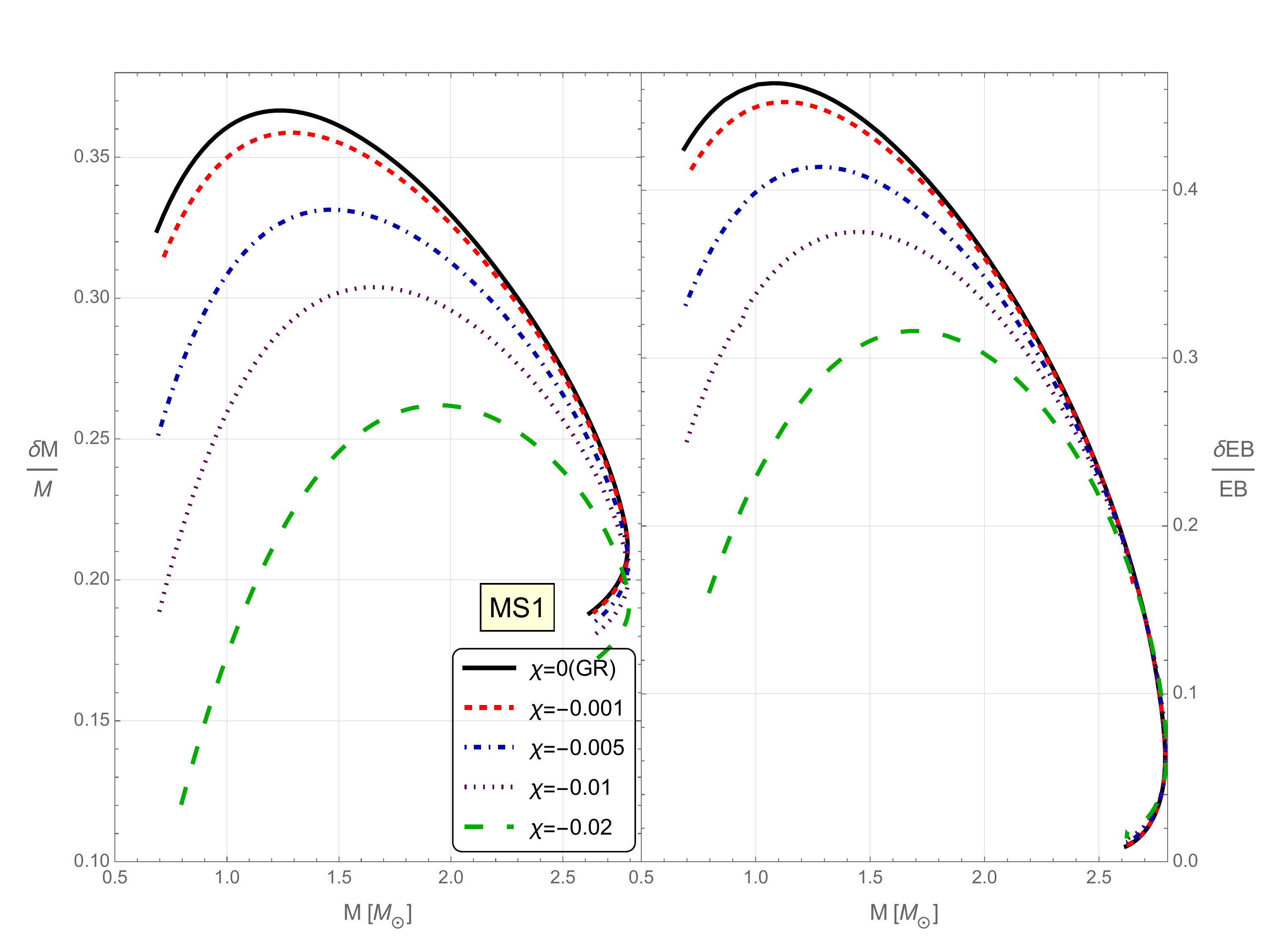}
 %   \caption{bbb2}
    \label{fig:3b}
\end{subfigure}
        \hfill
\begin{subfigure}{0.49\textwidth}
    \includegraphics[width=\textwidth]{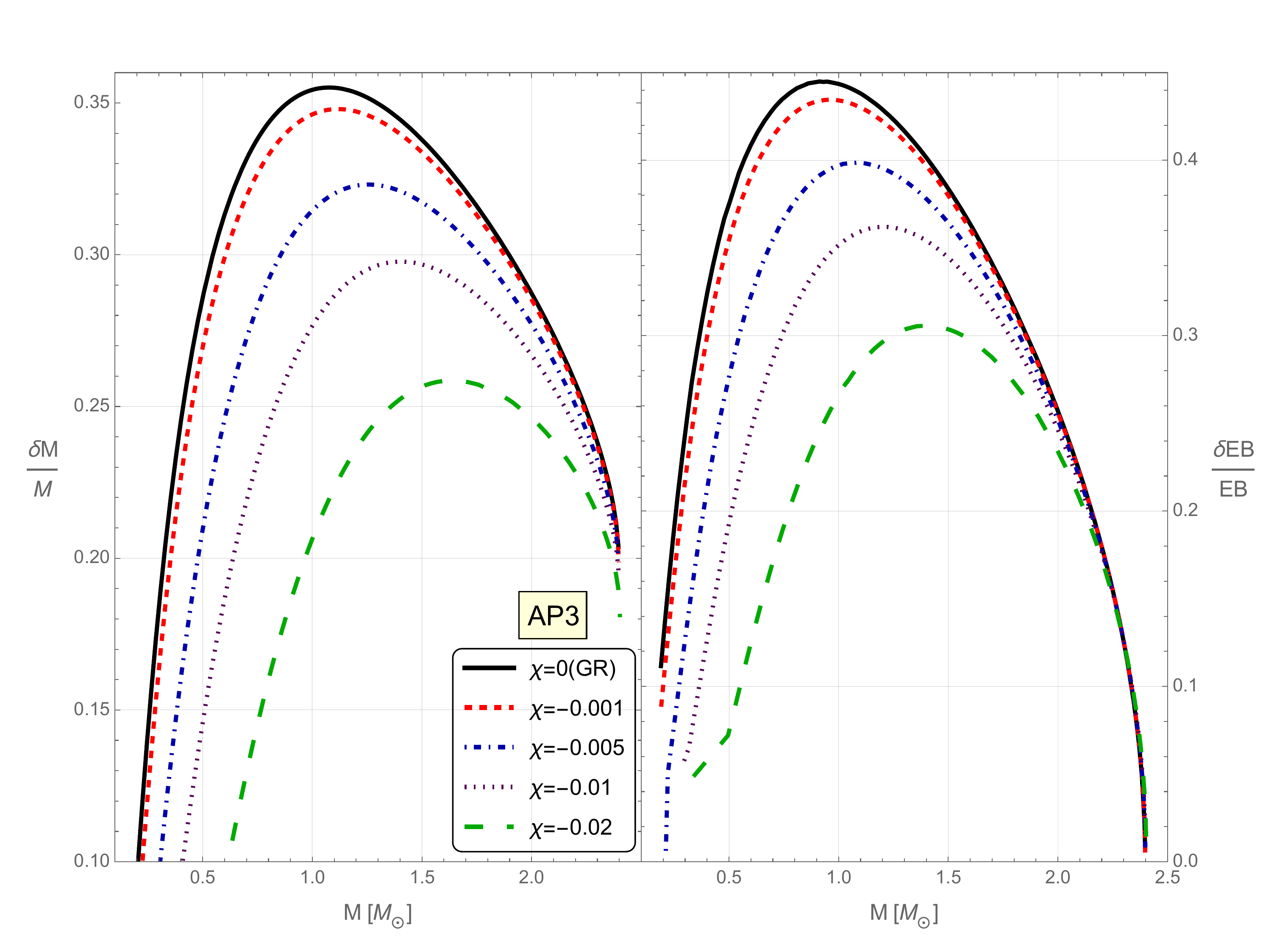}
 %   \caption{alf4}
    \label{fig:3c}
\end{subfigure}
\hfill
\begin{subfigure}{0.49\textwidth}
    \includegraphics[width=\textwidth]{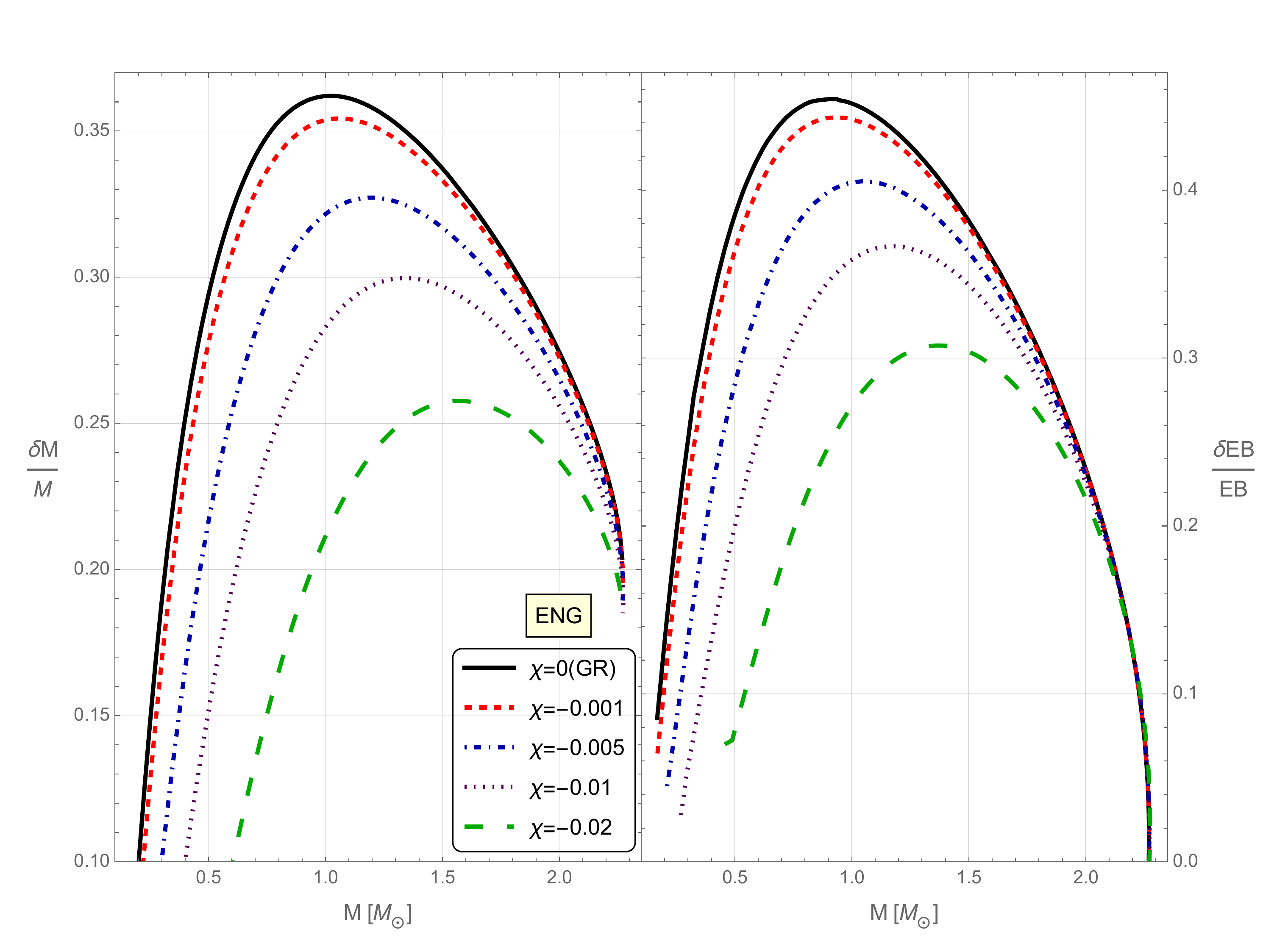}
%    \caption{pal6}
    \label{fig:3d}
\end{subfigure}
\hfill
\begin{subfigure}{0.49\textwidth}
    \includegraphics[width=\textwidth]{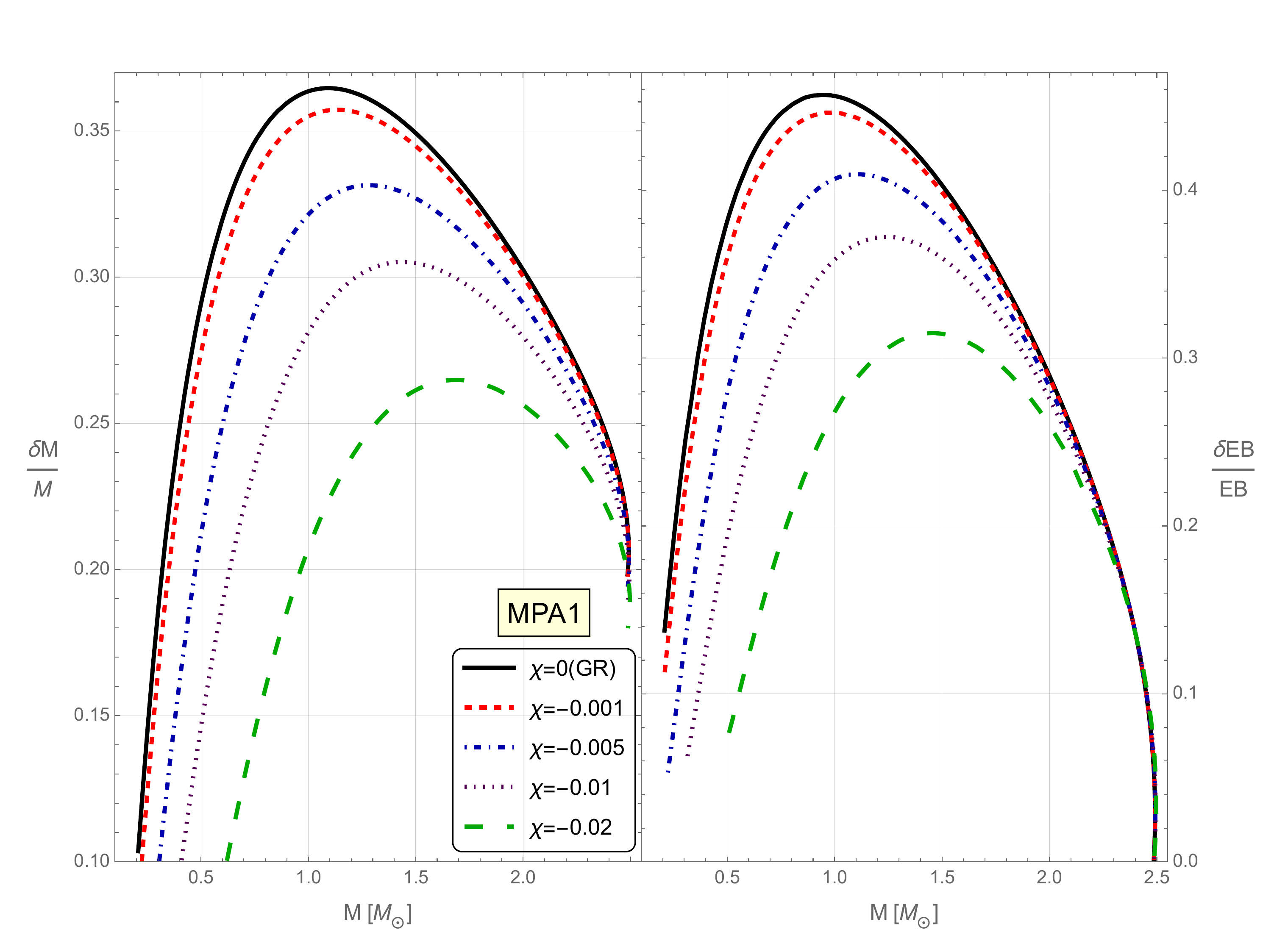}
 %   \caption{alf4}
    \label{fig:3e}
\end{subfigure}
\hfill
\begin{subfigure}{0.49\textwidth}
    \includegraphics[width=\textwidth]{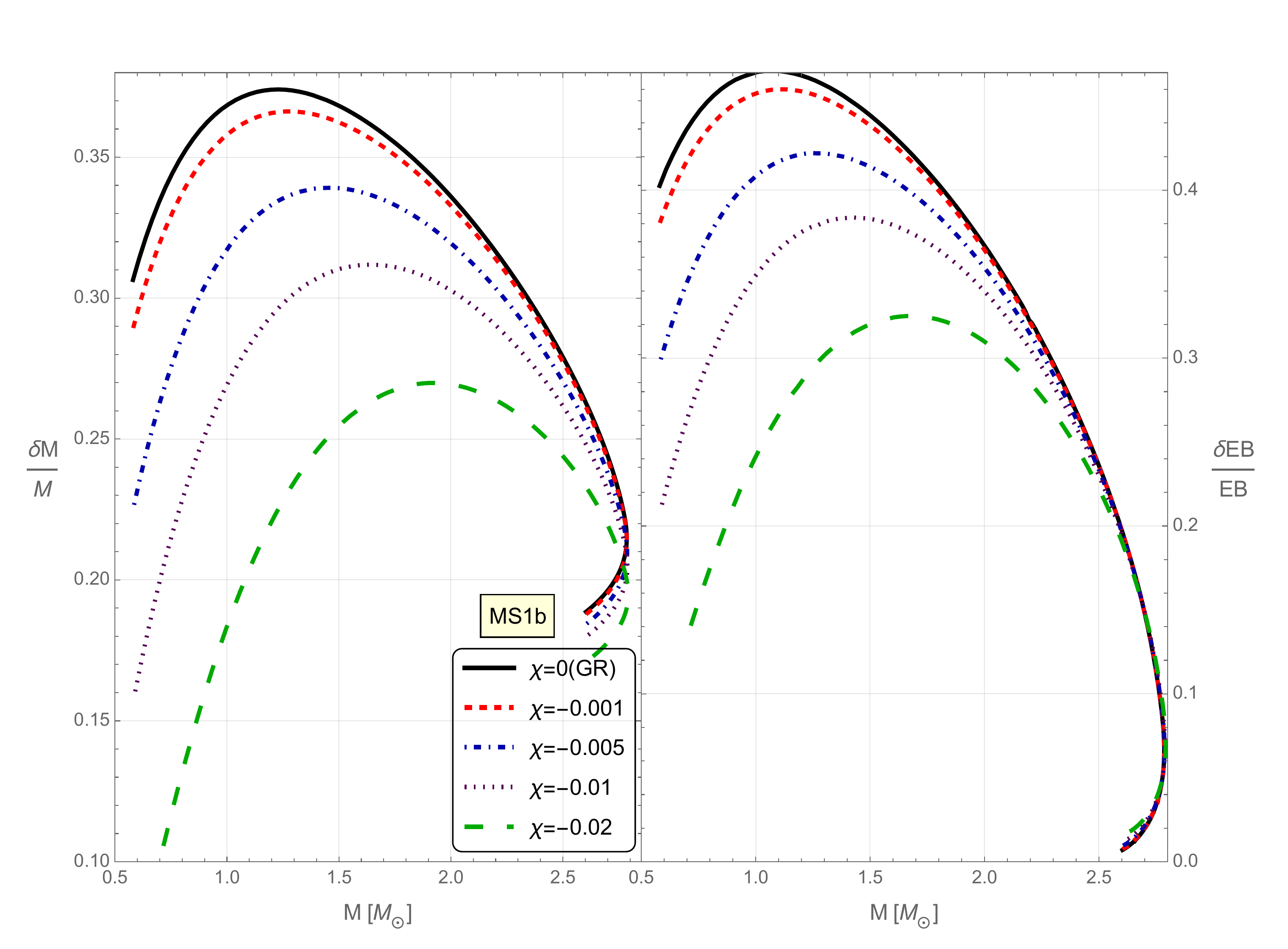}
%    \caption{pal6}
    \label{fig:3f}
\end{subfigure}
\caption{The rotational mass correction and the rotational binding energy correction are shown as the function of mass in $f(R, T)=R+2\chi T$ modified gravity.}
\label{fig:3}
\end{figure*}

\begin{figure*}[!ht]
\centering
\begin{subfigure}{0.49\textwidth}
    \includegraphics[width=\textwidth]{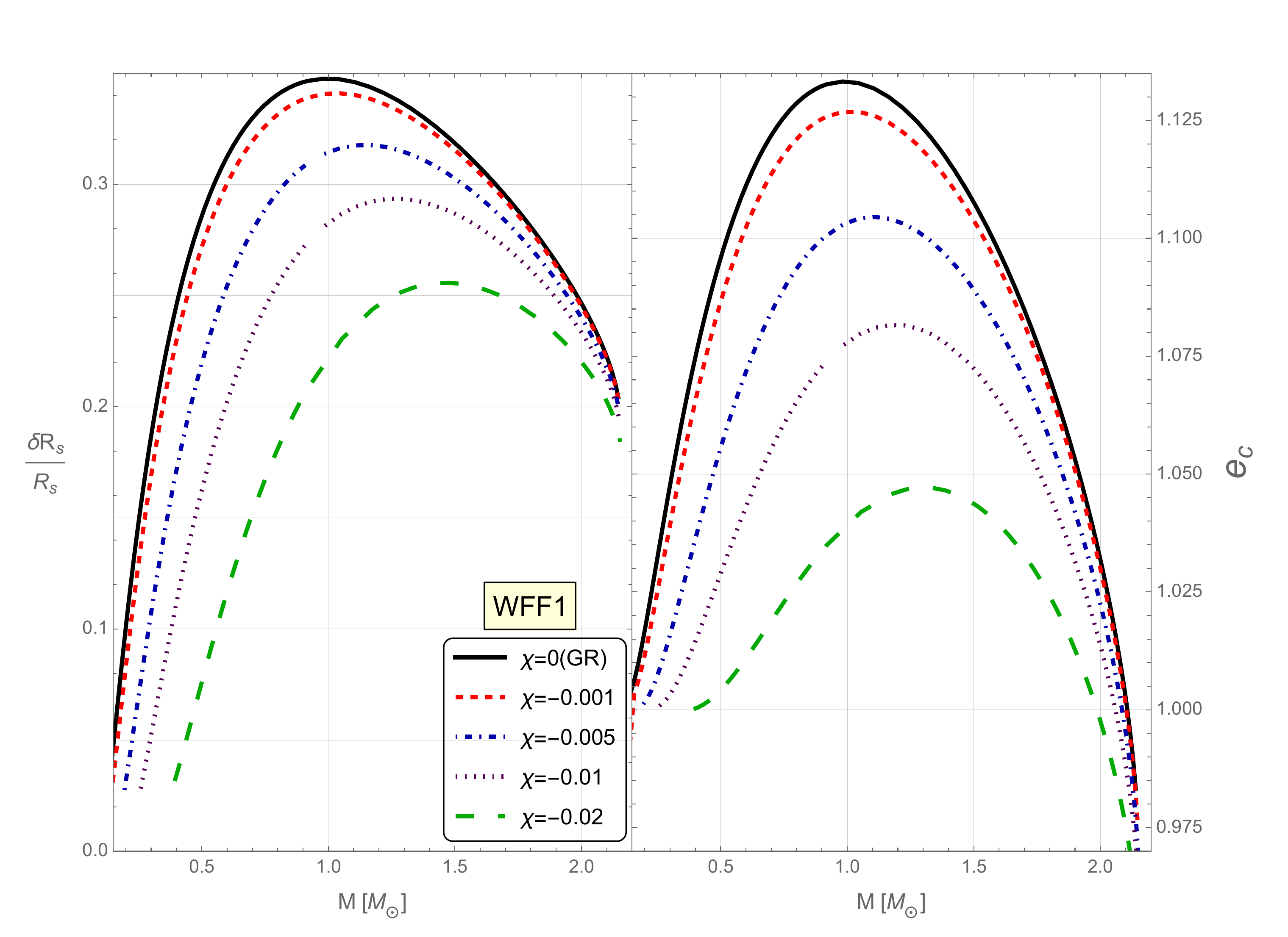}
%    \caption{sly}
    \label{fig:4a}
\end{subfigure}
\hfill
\begin{subfigure}{0.49\textwidth}
    \includegraphics[width=\textwidth]{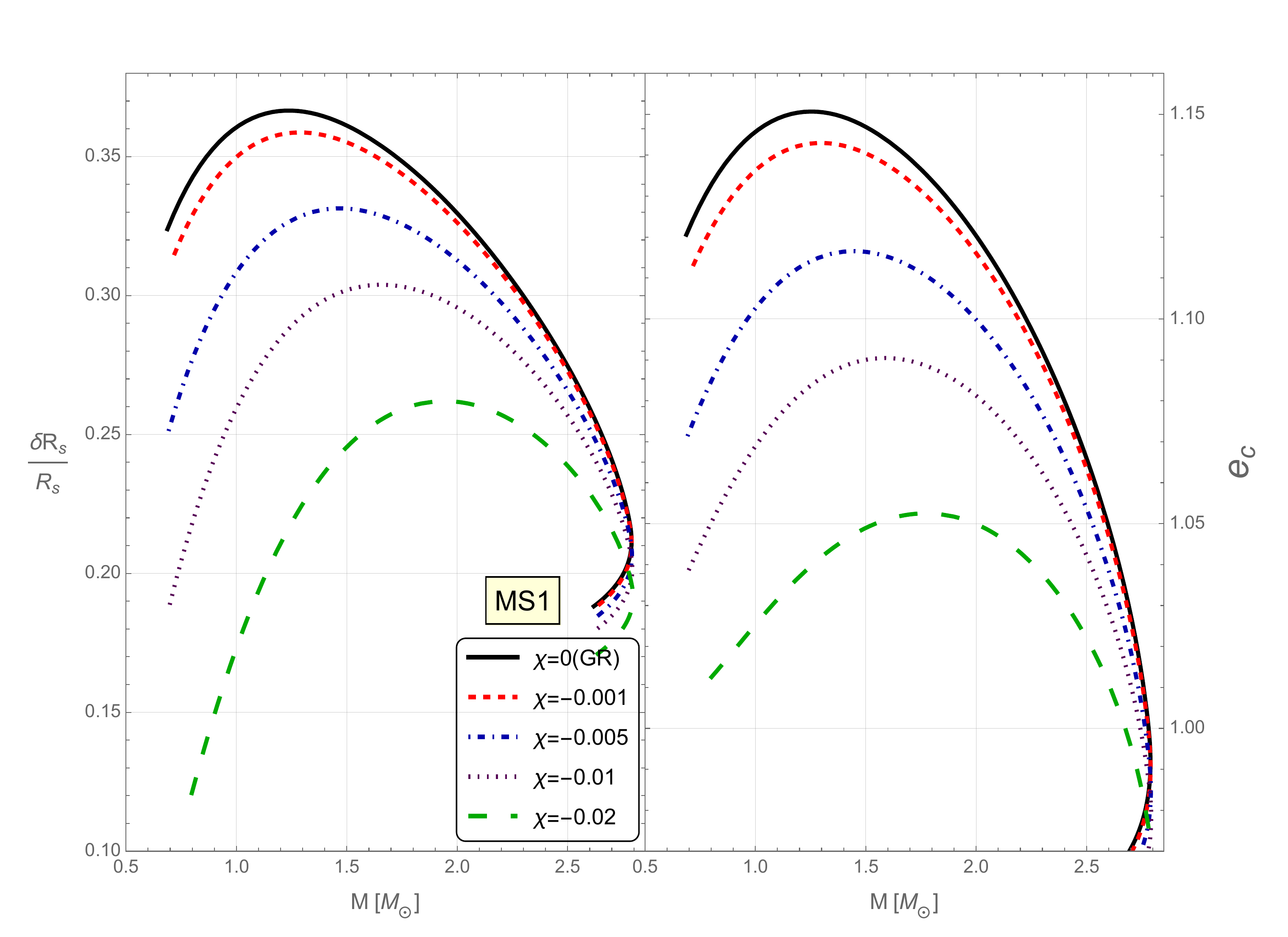}
 %   \caption{bbb2}
    \label{fig:4b}
\end{subfigure}
        \hfill
\begin{subfigure}{0.49\textwidth}
    \includegraphics[width=\textwidth]{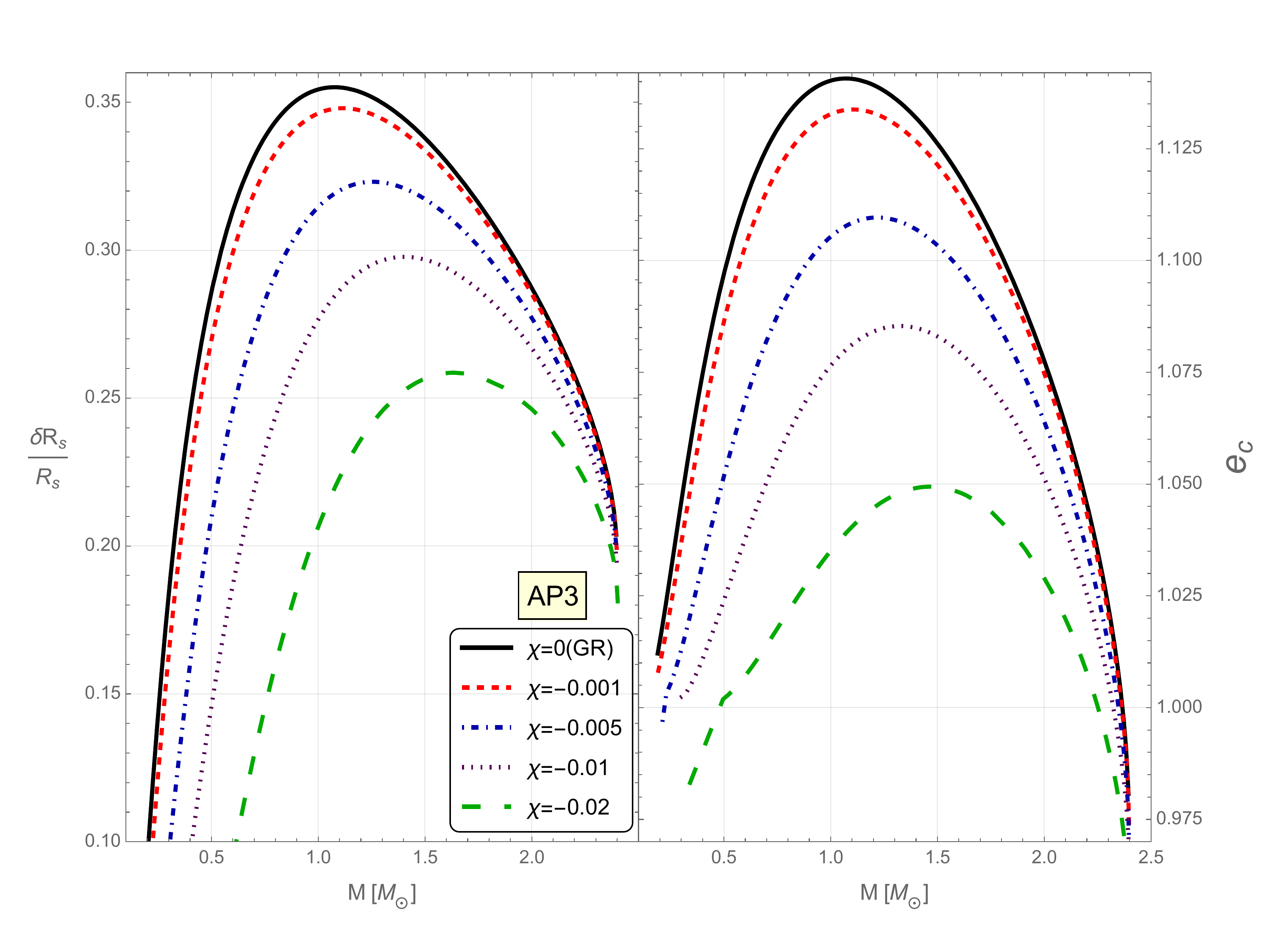}
 %   \caption{alf4}
    \label{fig:4c}
\end{subfigure}
\hfill
\begin{subfigure}{0.49\textwidth}
    \includegraphics[width=\textwidth]{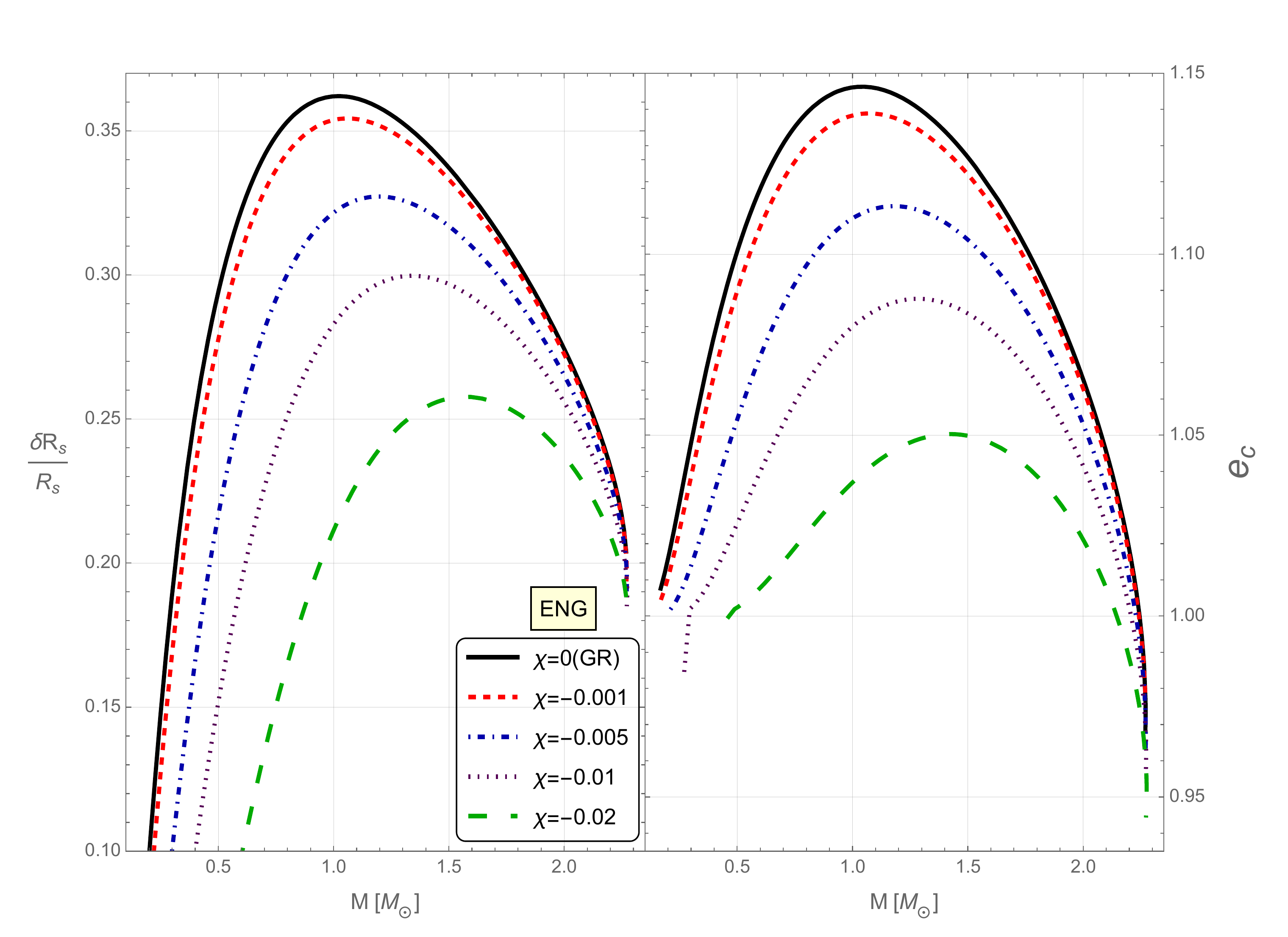}
%    \caption{pal6}
    \label{fig:4d}
\end{subfigure}
\hfill
\begin{subfigure}{0.49\textwidth}
    \includegraphics[width=\textwidth]{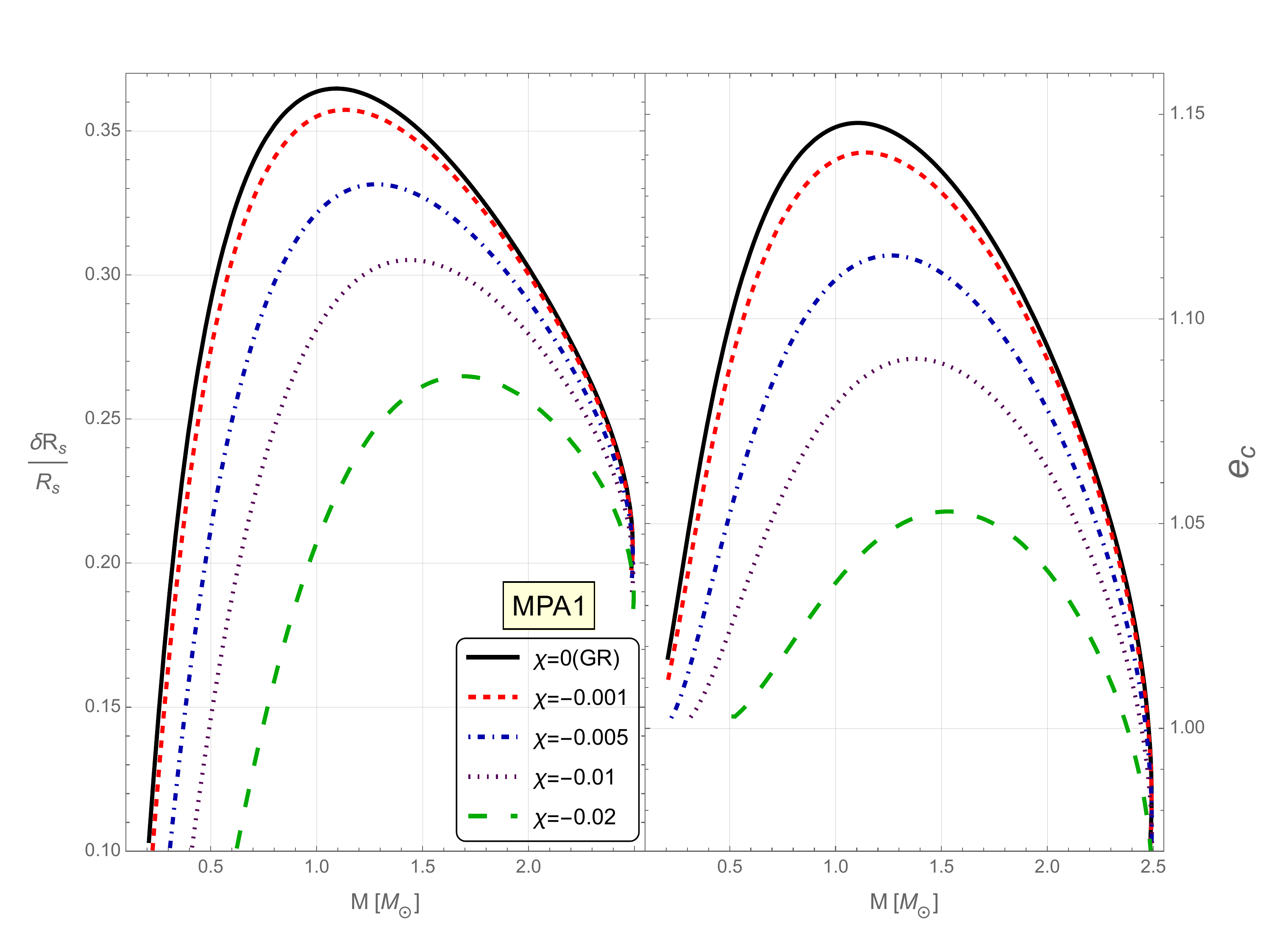}
 %   \caption{alf4}
    \label{fig:4e}
\end{subfigure}
\hfill
\begin{subfigure}{0.49\textwidth}
    \includegraphics[width=\textwidth]{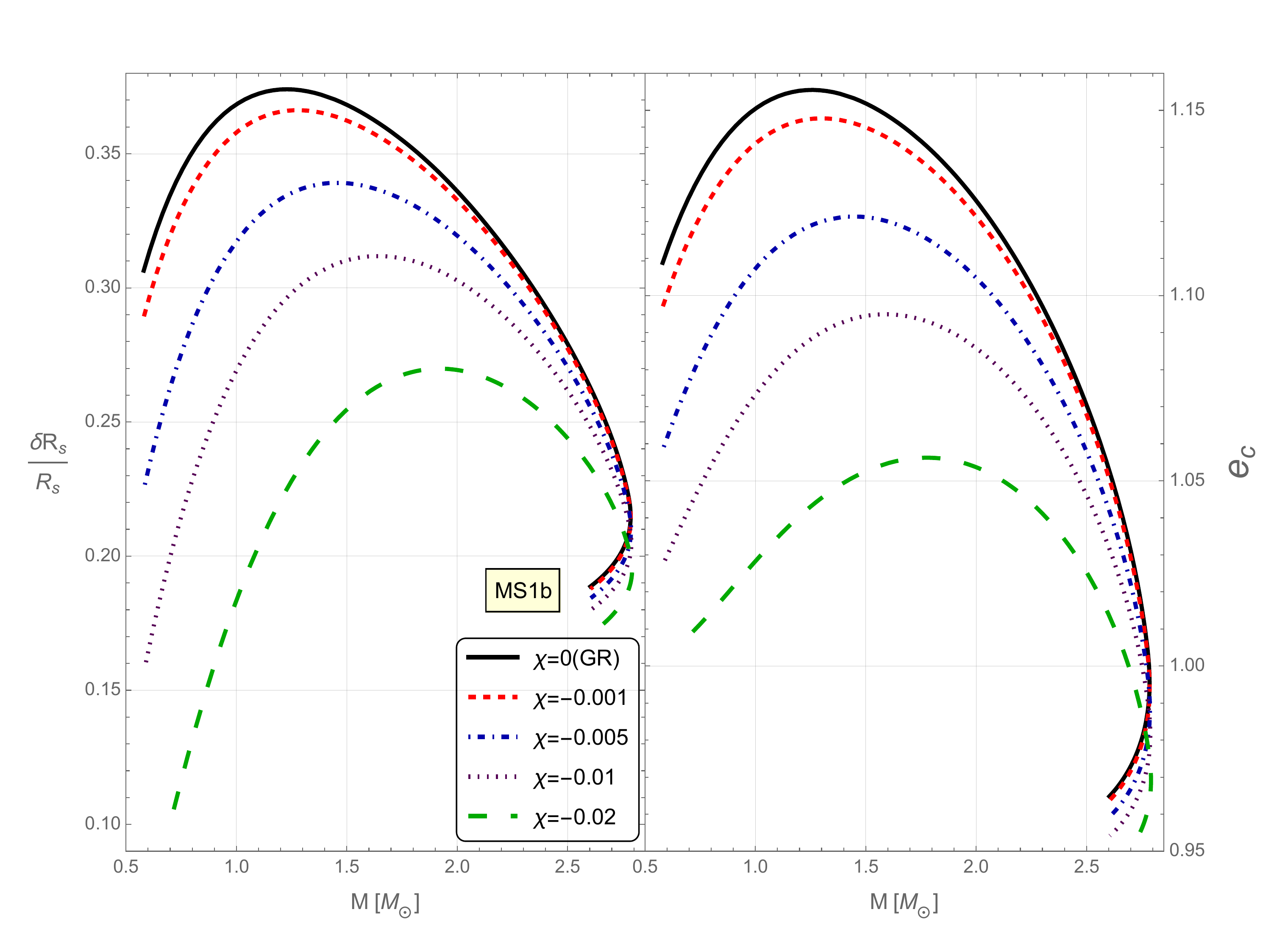}
%    \caption{pal6}
    \label{fig:4f}
\end{subfigure}
\caption{The change in radius due to rotation and the eccentricity are shown as the function of mass in $f(R, T)=R+2\chi T$ modified gravity.}
\label{fig:4}
\end{figure*}

\begin{figure*}[!ht]
\centering
\begin{subfigure}{0.32\textwidth}
    \includegraphics[width=\textwidth]{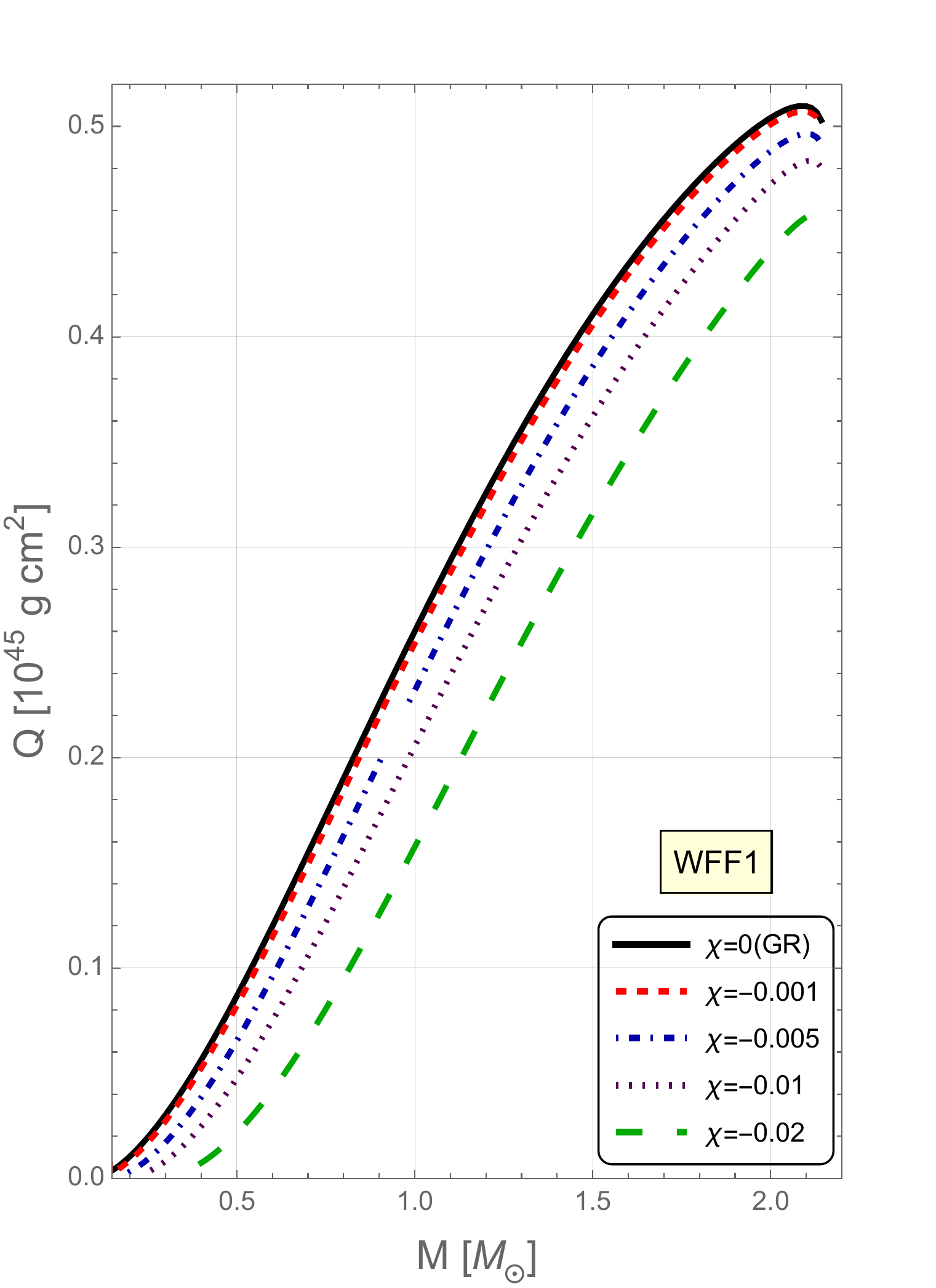}
%    \caption{sly}
    \label{fig:5a}
\end{subfigure}
\hfill
\begin{subfigure}{0.32\textwidth}
    \includegraphics[width=\textwidth]{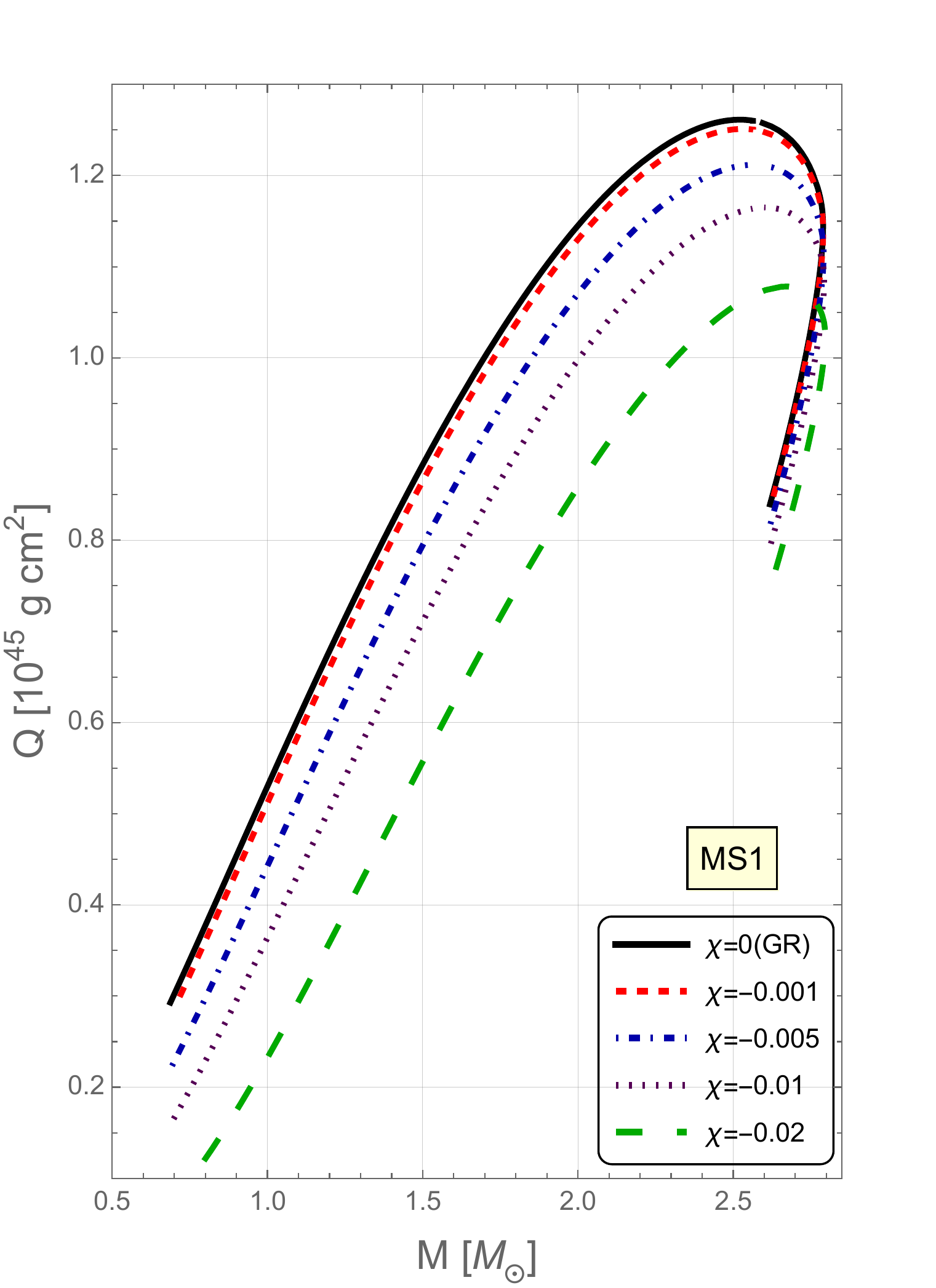}
 %   \caption{bbb2}
    \label{fig:5b}
\end{subfigure}
        \hfill
\begin{subfigure}{0.32\textwidth}
    \includegraphics[width=\textwidth]{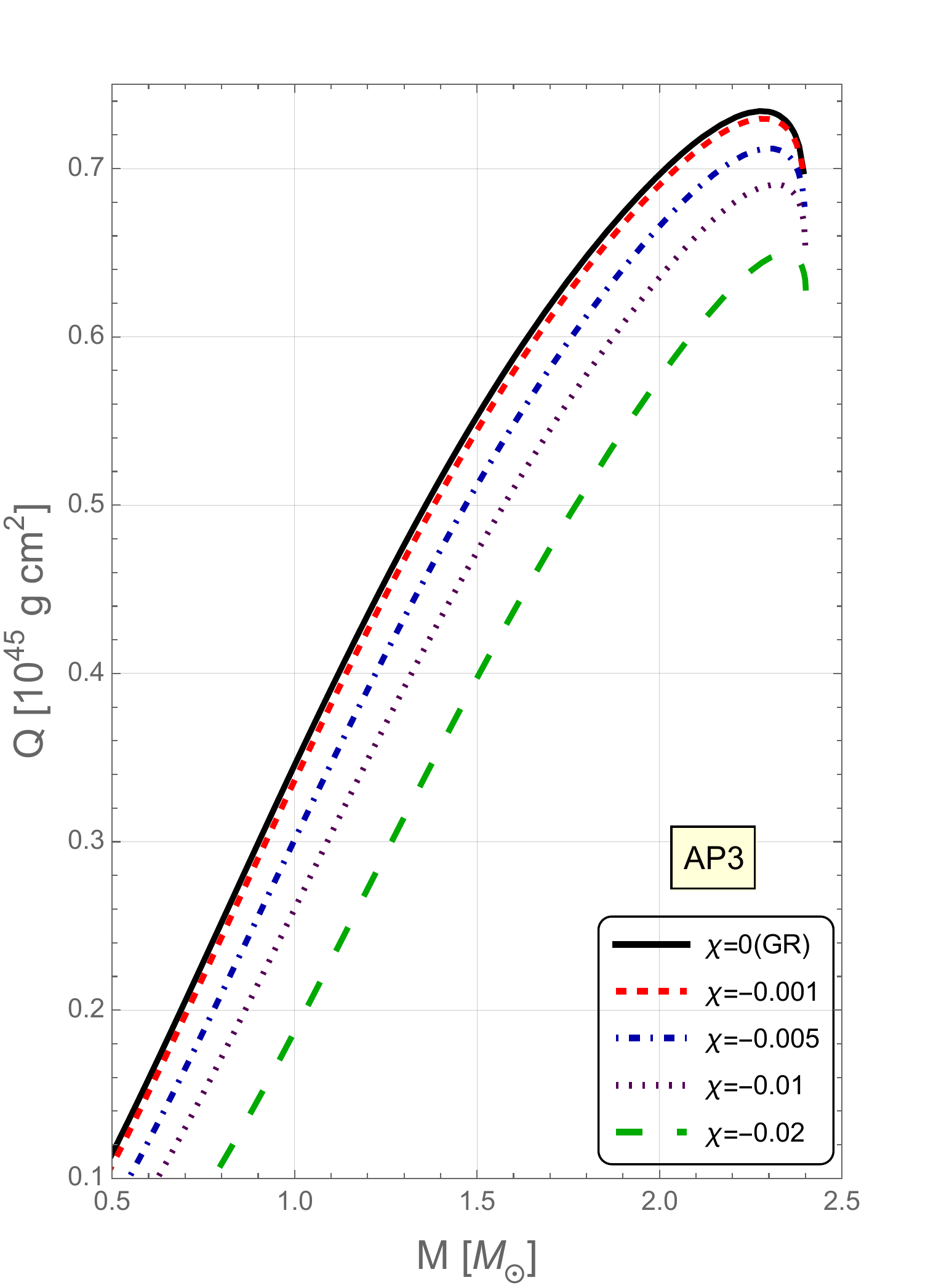}
 %   \caption{alf4}
    \label{fig:5c}
\end{subfigure}
\hfill
\begin{subfigure}{0.32\textwidth}
    \includegraphics[width=\textwidth]{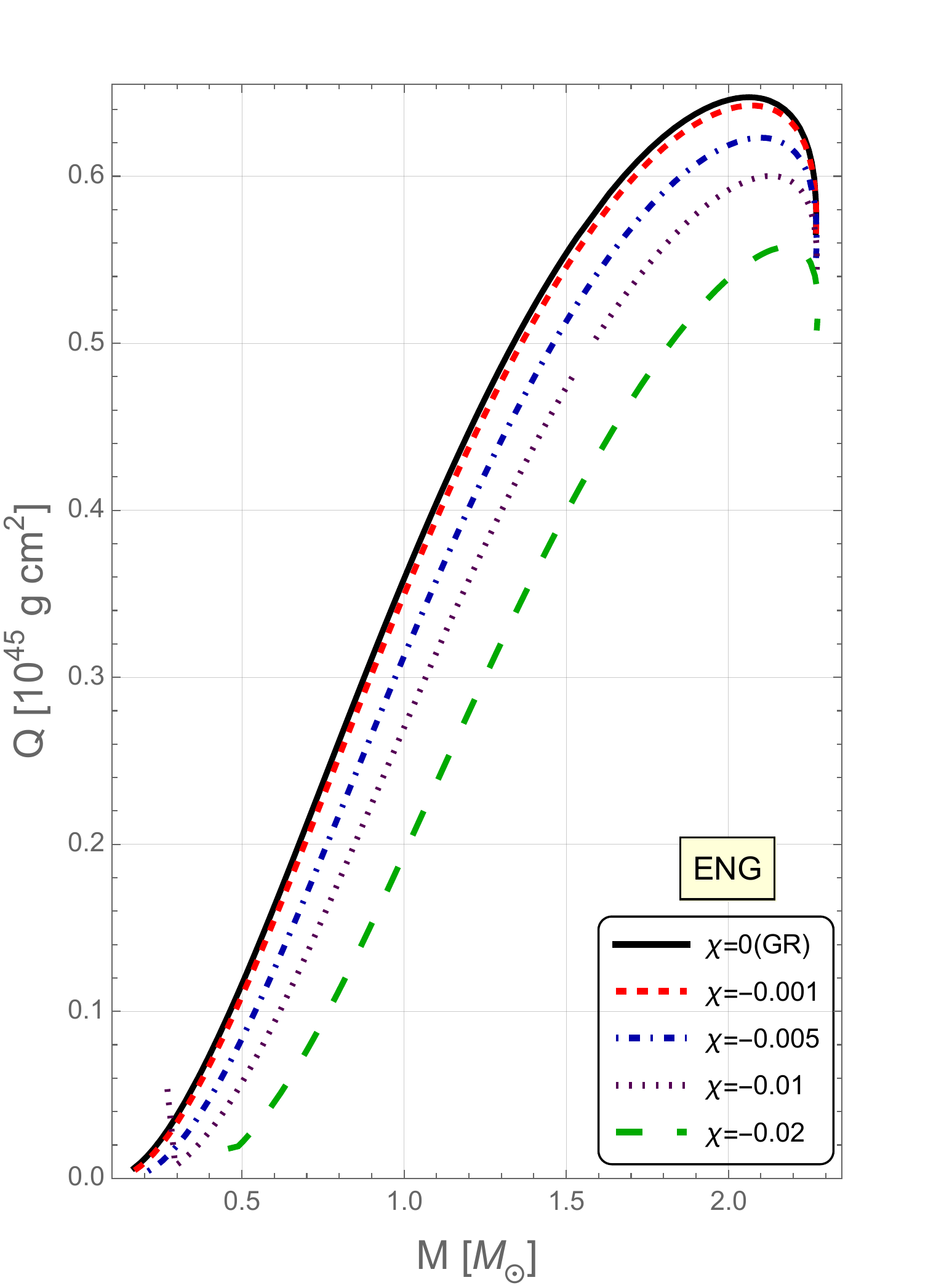}
%    \caption{pal6}
    \label{fig:5d}
\end{subfigure}
\hfill
\begin{subfigure}{0.32\textwidth}
    \includegraphics[width=\textwidth]{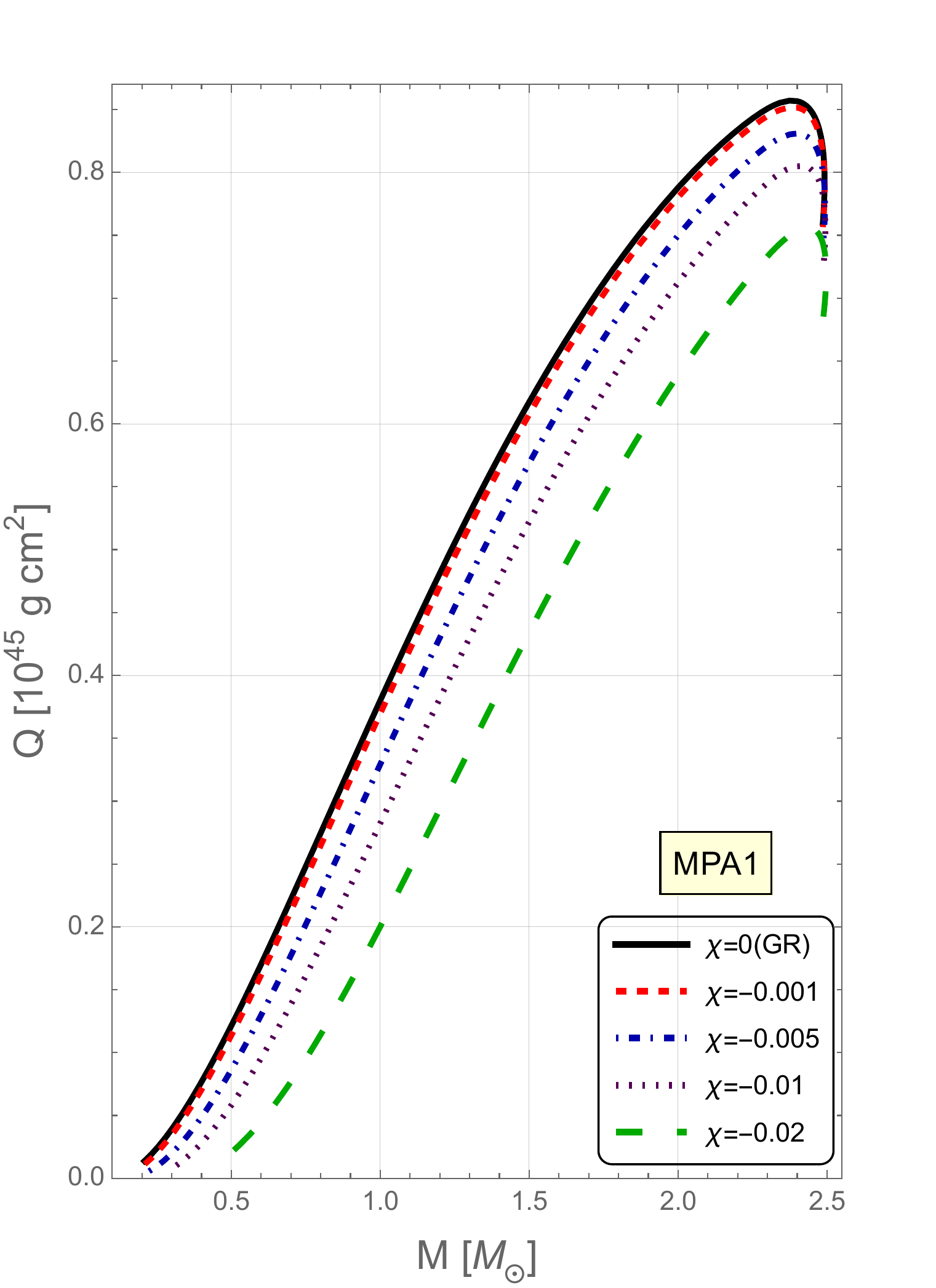}
 %   \caption{alf4}
    \label{fig:5e}
\end{subfigure}
\hfill
\begin{subfigure}{0.32\textwidth}
    \includegraphics[width=\textwidth]{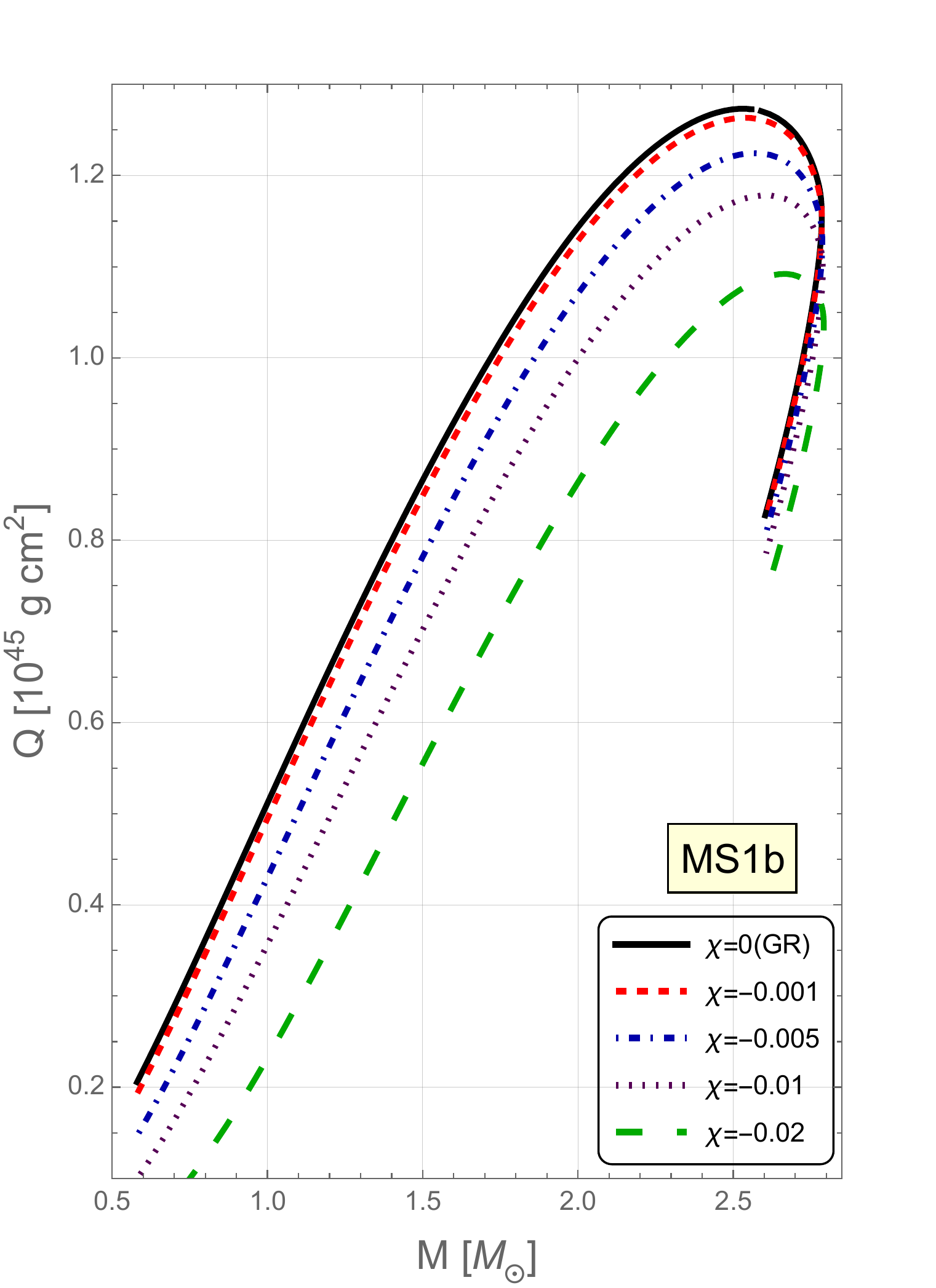}
%    \caption{pal6}
    \label{fig:5f}
\end{subfigure}
\caption{The relation between quadrupole moment and mass in $f(R, T)=R+2\chi T$ modified gravity.}
\label{fig:5}
\end{figure*}

\begin{figure*}[!ht]
\centering
\begin{subfigure}{0.32\textwidth}
    \includegraphics[width=\textwidth]{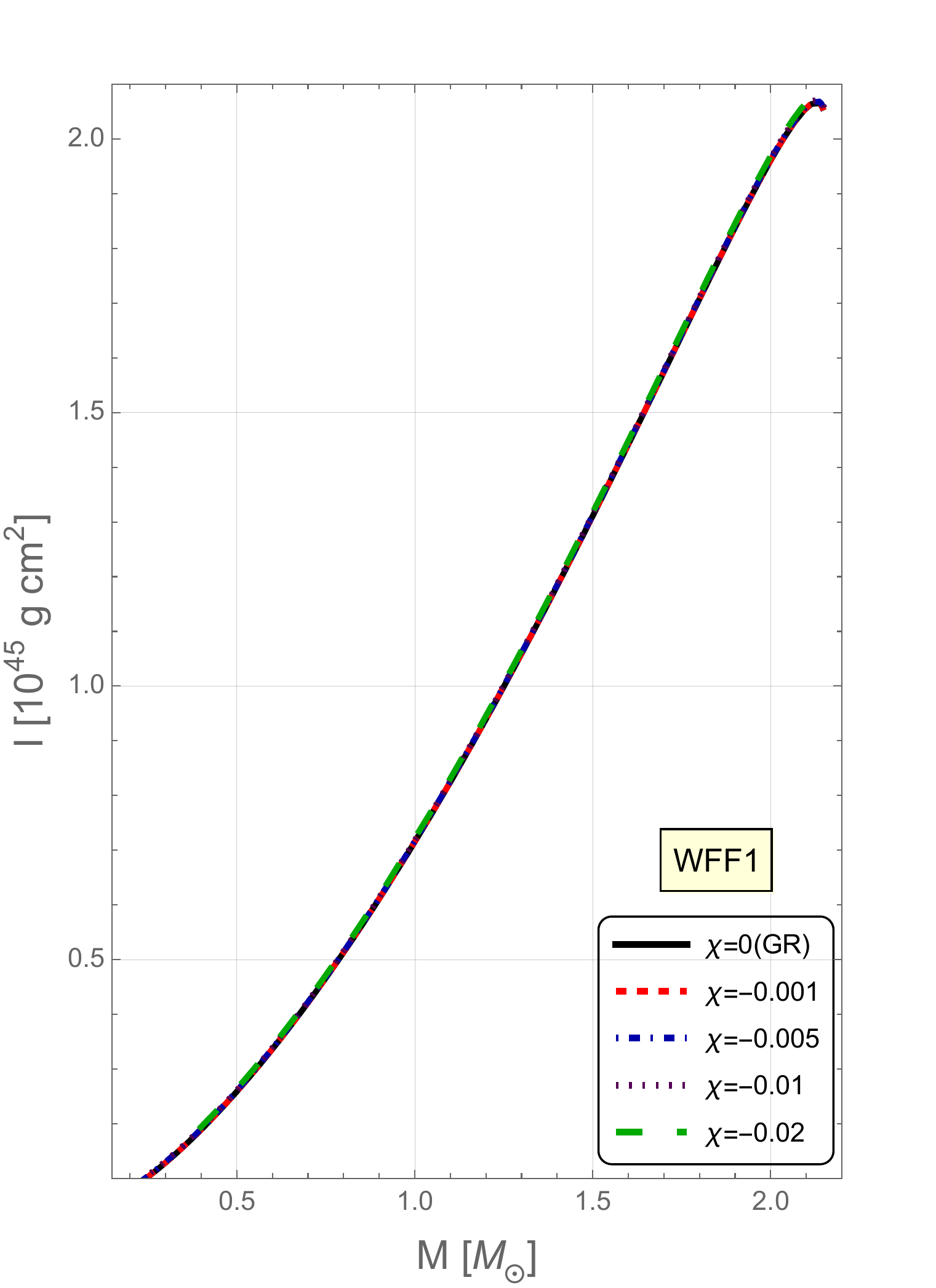}
%    \caption{sly}
    \label{fig:6a}
\end{subfigure}
\hfill
\begin{subfigure}{0.305\textwidth}
    \includegraphics[width=\textwidth]{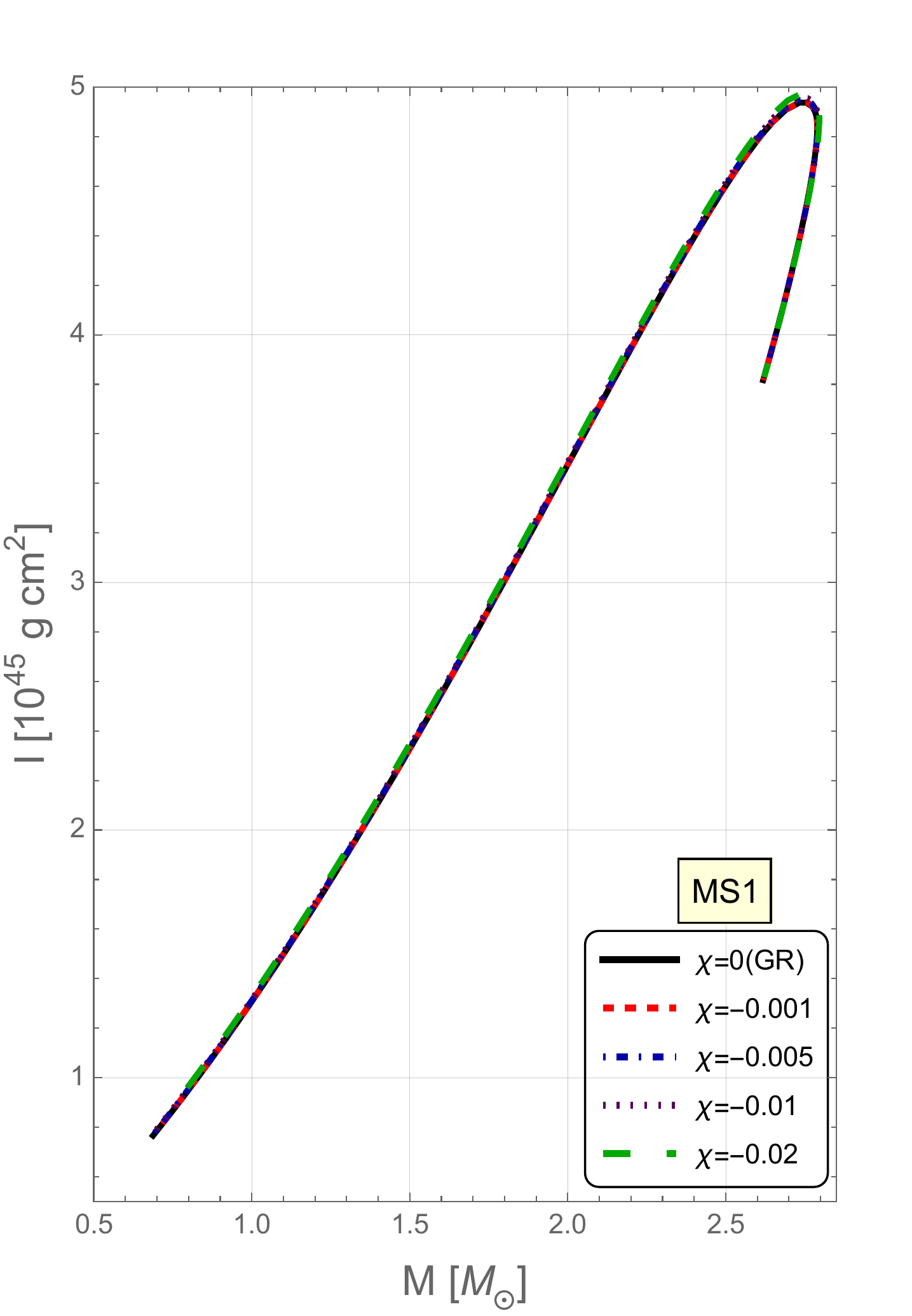}
 %   \caption{bbb2}
    \label{fig:6b}
\end{subfigure}
        \hfill
\begin{subfigure}{0.32\textwidth}
    \includegraphics[width=\textwidth]{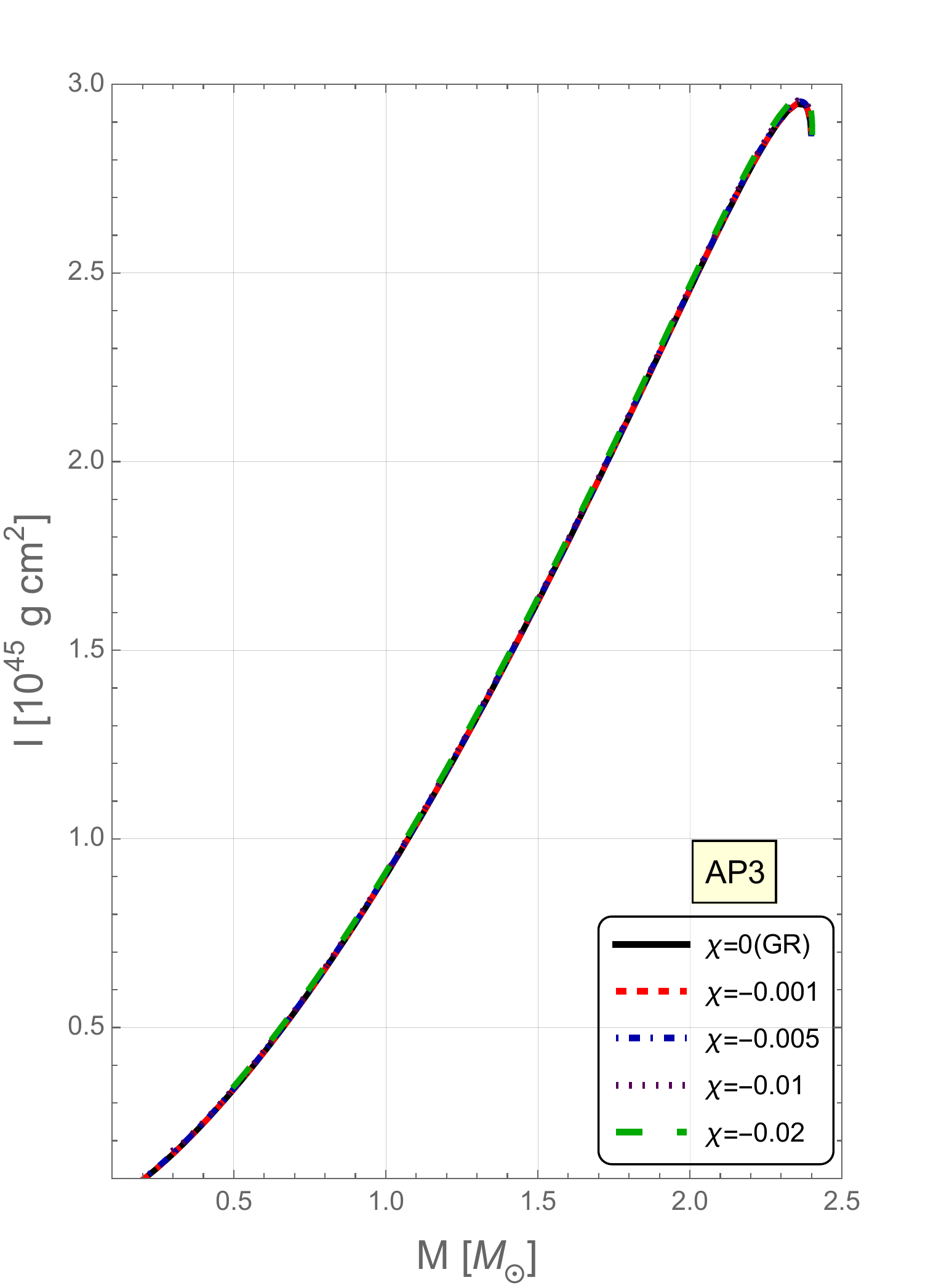}
 %   \caption{alf4}
    \label{fig:6c}
\end{subfigure}
\hfill
\begin{subfigure}{0.32\textwidth}
    \includegraphics[width=\textwidth]{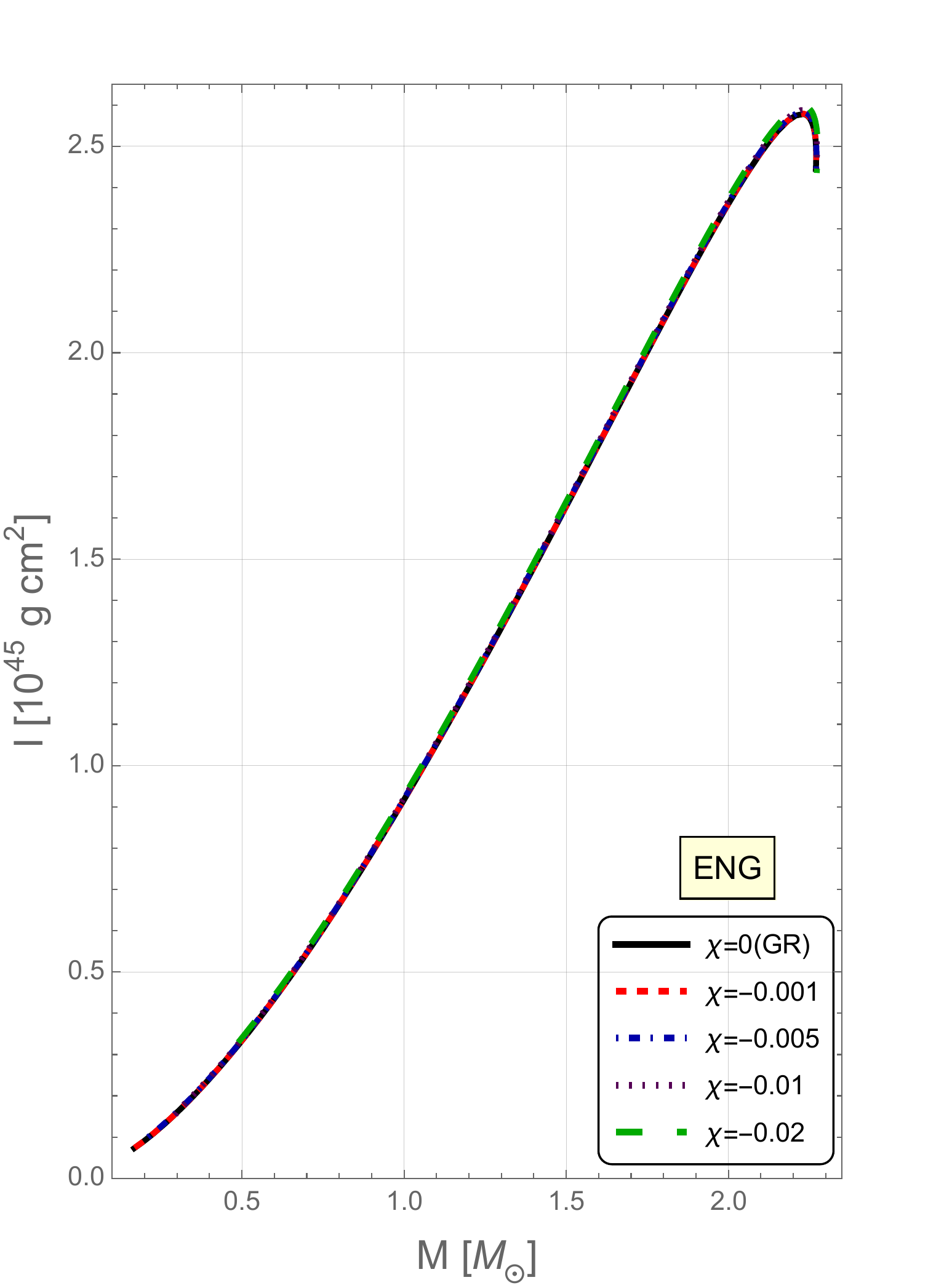}
%    \caption{pal6}
    \label{fig:6d}
\end{subfigure}
\hfill
\begin{subfigure}{0.32\textwidth}
    \includegraphics[width=\textwidth]{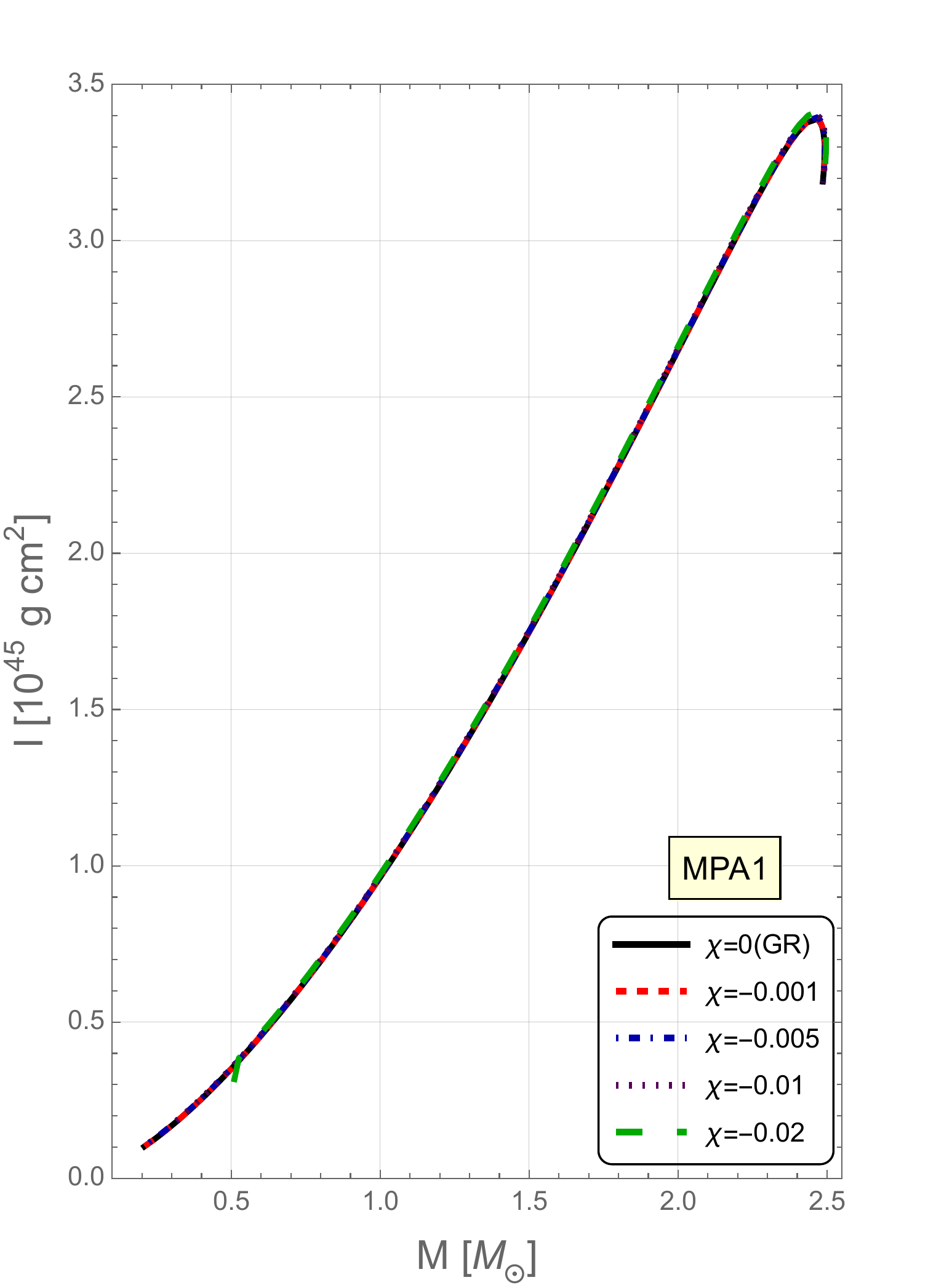}
 %   \caption{alf4}
    \label{fig:6e}
\end{subfigure}
\hfill
\begin{subfigure}{0.305\textwidth}
    \includegraphics[width=\textwidth]{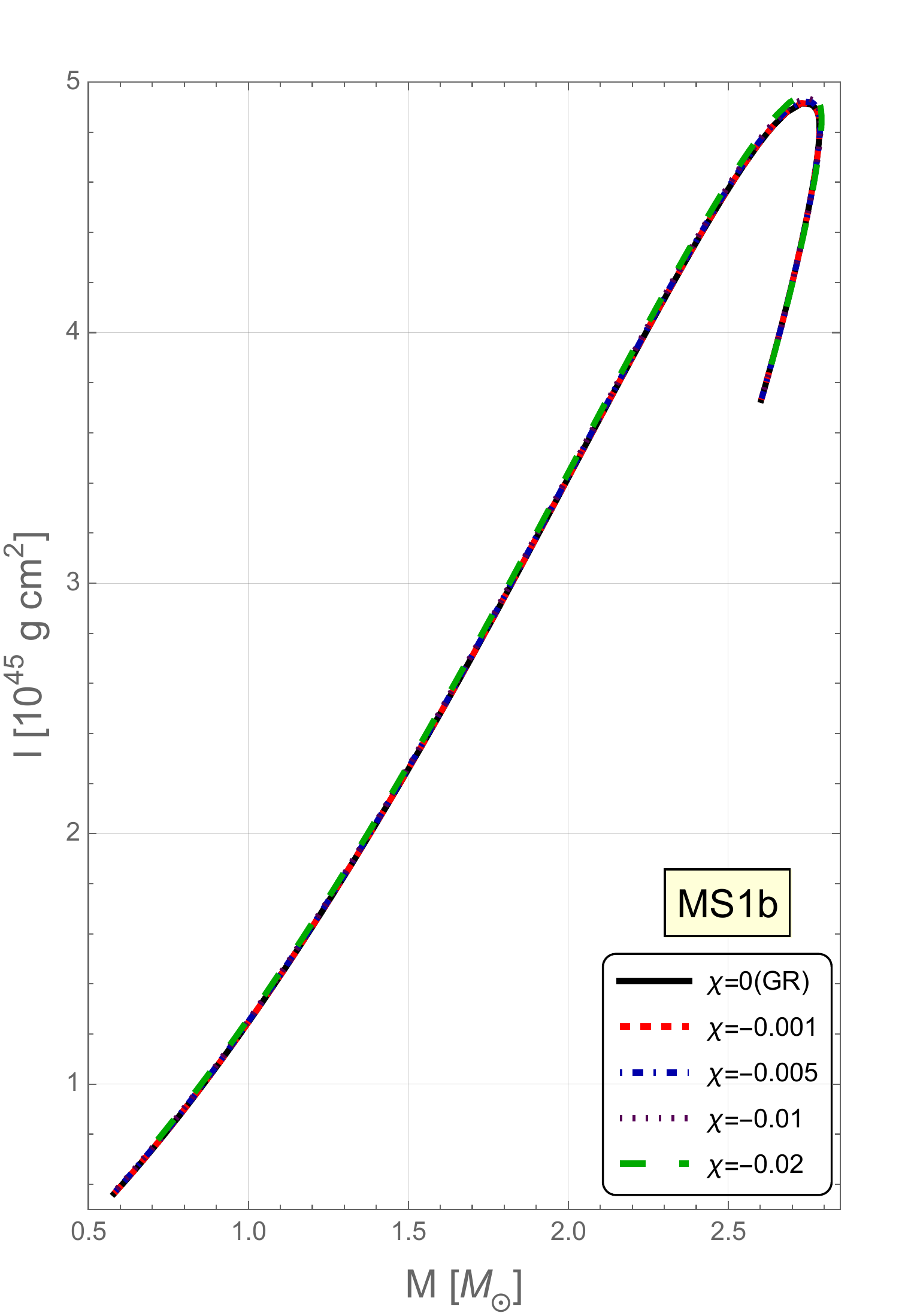}
%    \caption{pal6}
    \label{fig:6f}
\end{subfigure}
\caption{The moment of inertia is shown as a function of the mass in $f(R, T)=R+2\chi T$ modified gravity.}
\label{fig:6}
\end{figure*}

\begin{figure*}[!ht]
\centering
\begin{subfigure}{0.32\textwidth}
    \includegraphics[width=\textwidth]{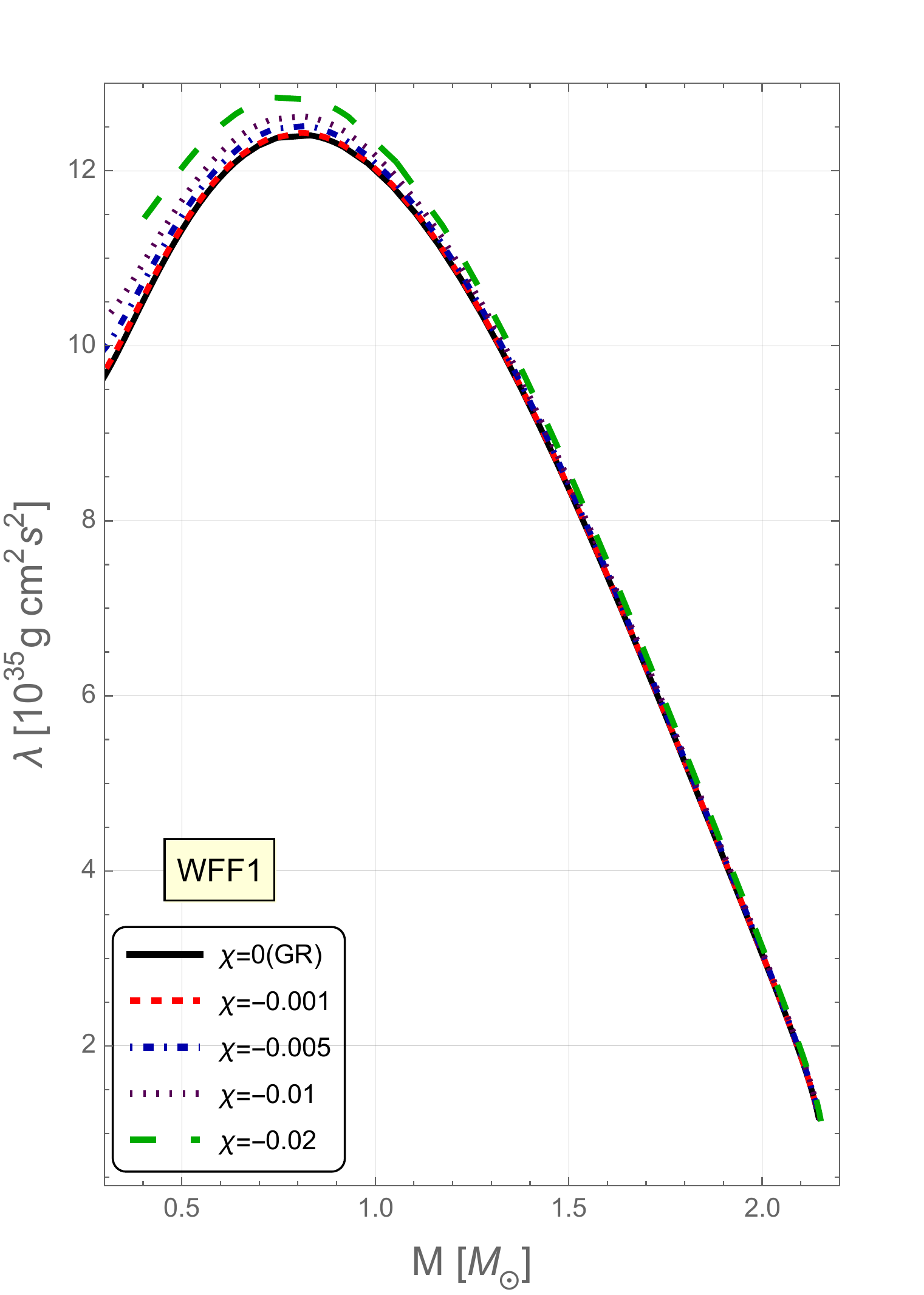}
%    \caption{sly}
    \label{fig:7a}
\end{subfigure}
\hfill
\begin{subfigure}{0.32\textwidth}
    \includegraphics[width=\textwidth]{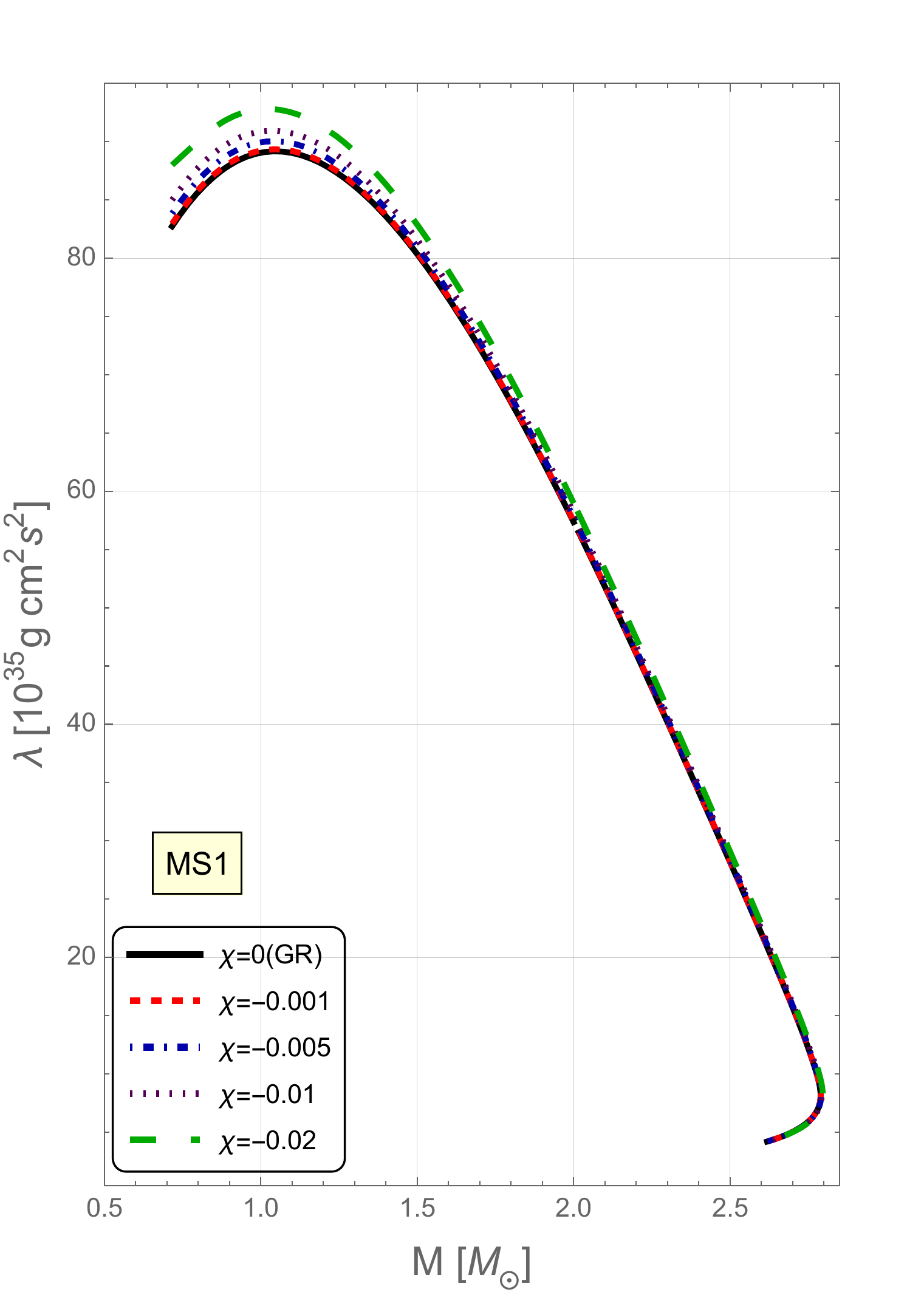}
 %   \caption{bbb2}
    \label{fig:7b}
\end{subfigure}
        \hfill
\begin{subfigure}{0.32\textwidth}
    \includegraphics[width=\textwidth]{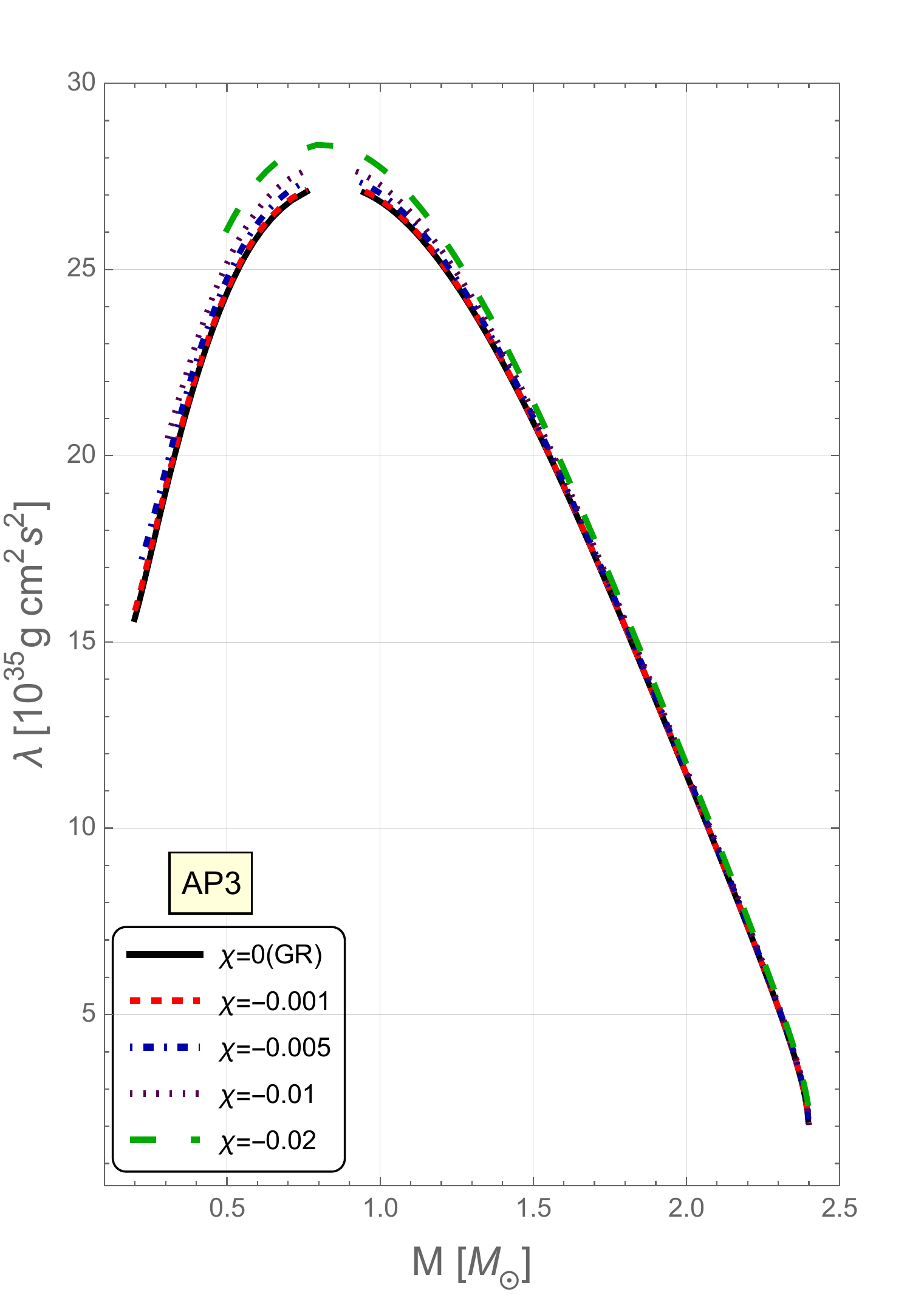}
 %   \caption{alf4}
    \label{fig:7c}
\end{subfigure}
\hfill
\begin{subfigure}{0.32\textwidth}
    \includegraphics[width=\textwidth]{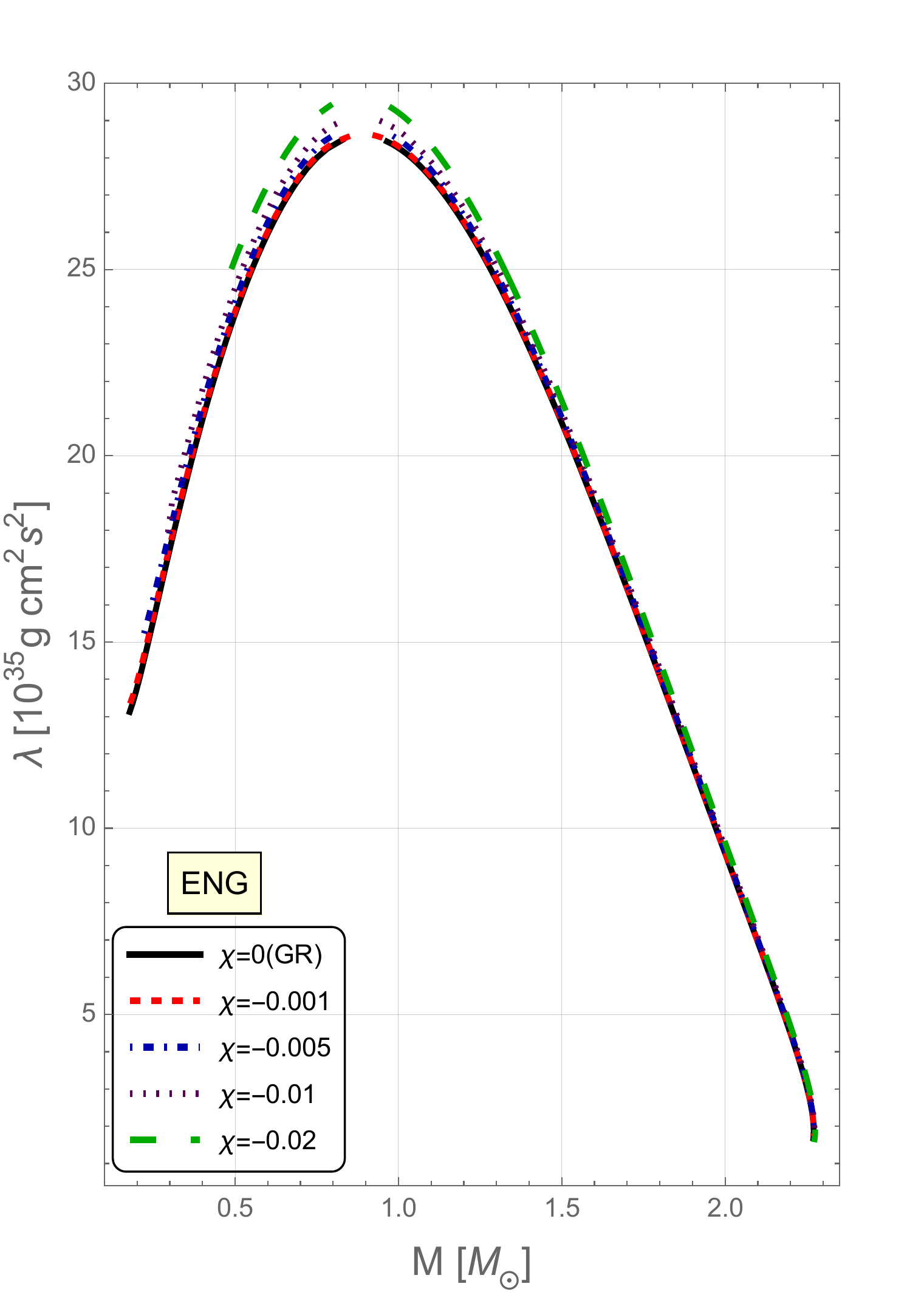}
%    \caption{pal6}
    \label{fig:7d}
\end{subfigure}
\hfill
\begin{subfigure}{0.32\textwidth}
    \includegraphics[width=\textwidth]{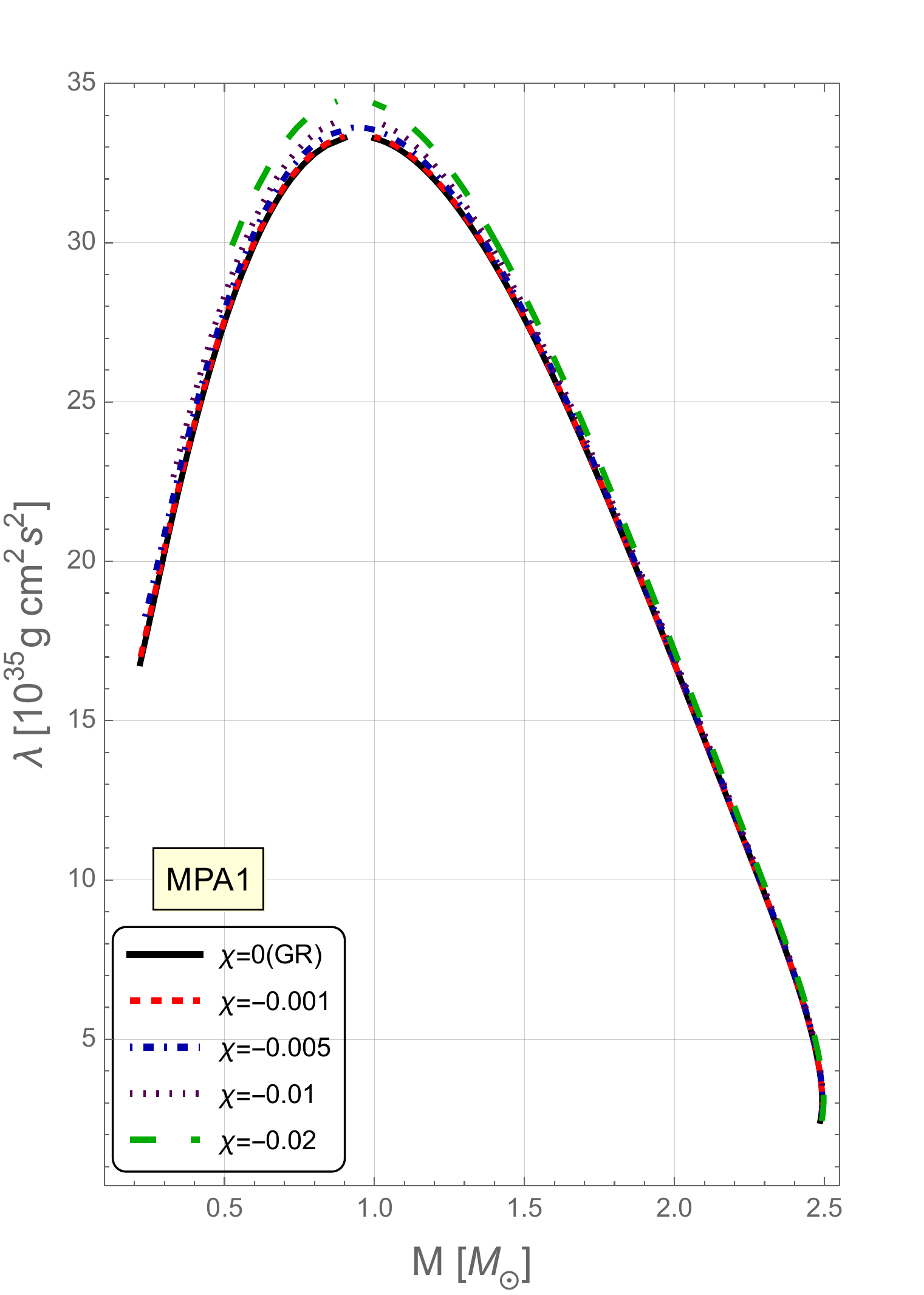}
 %   \caption{alf4}
    \label{fig:7e}
\end{subfigure}
\hfill
\begin{subfigure}{0.32\textwidth}
    \includegraphics[width=\textwidth]{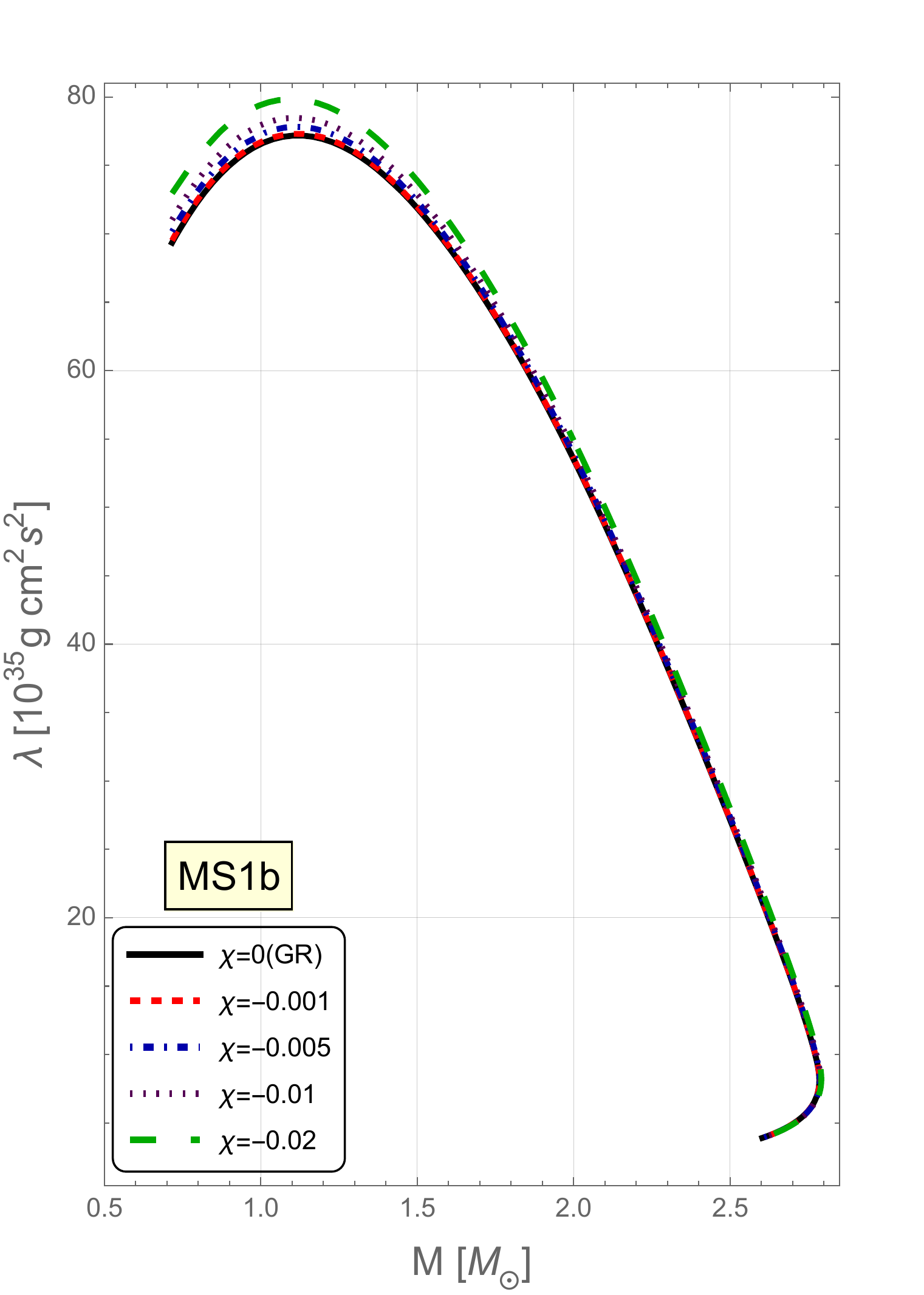}
%    \caption{pal6}
    \label{fig:7f}
\end{subfigure}
\caption{The tidal love number is shown as a function of the mass in $f(R, T)=R+2\chi T$ modified gravity.}
\label{fig:7}
\end{figure*}

\begin{figure*}[!ht]
\centering
\includegraphics[scale=0.32]{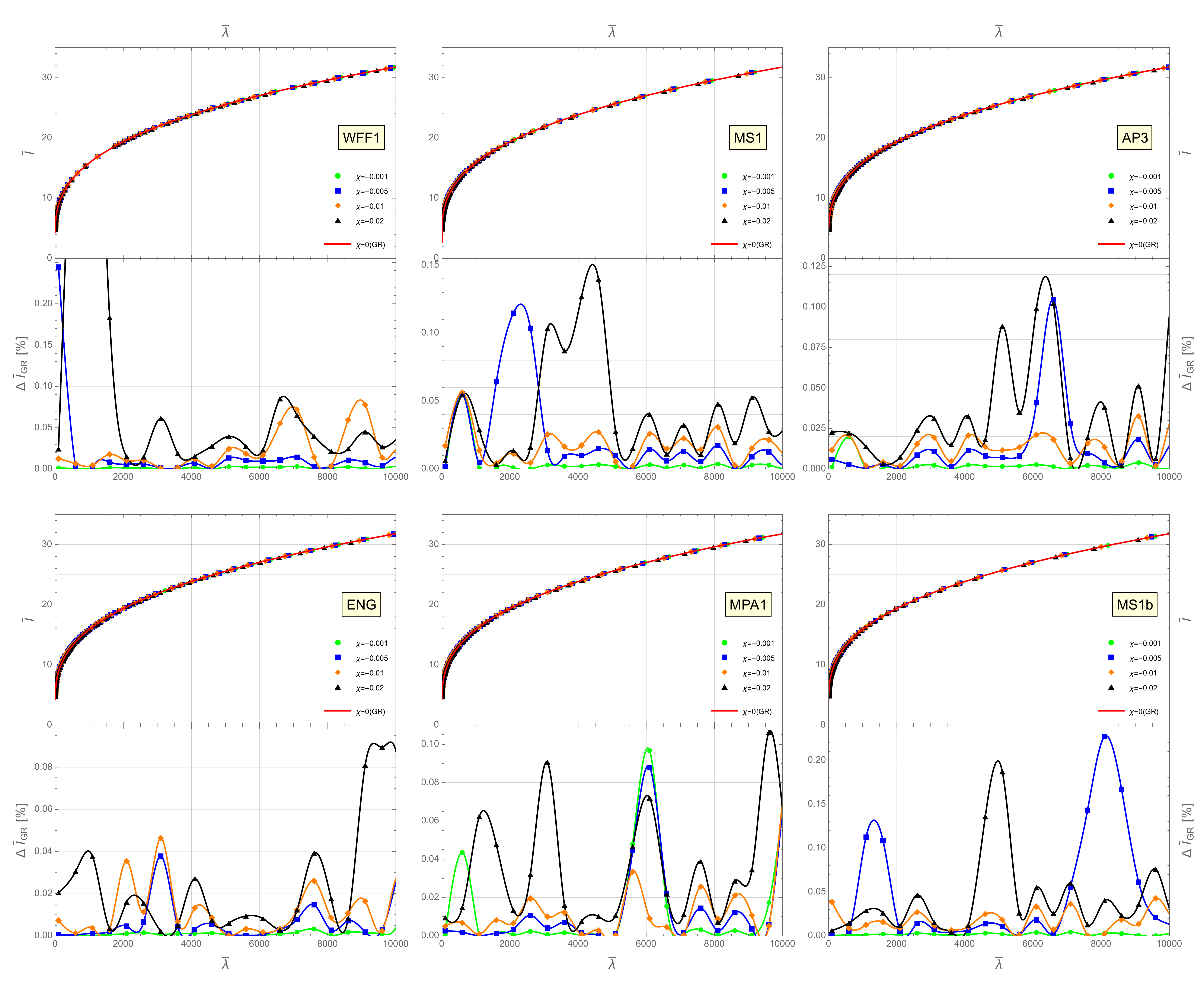}
\caption{The relationship between the dimensionless moment of inertia and the dimensionless tidal love number. The lower inset of every panel shows how the $\bar{I}-\bar{\lambda}$ relationship for different values of $\chi$ in $f(R, T)=R+2\chi T$  deviates from the general relativistic one.}
\label{fig:8}
\end{figure*}
\begin{figure*}[!t]
\centering
\includegraphics[scale=0.32]{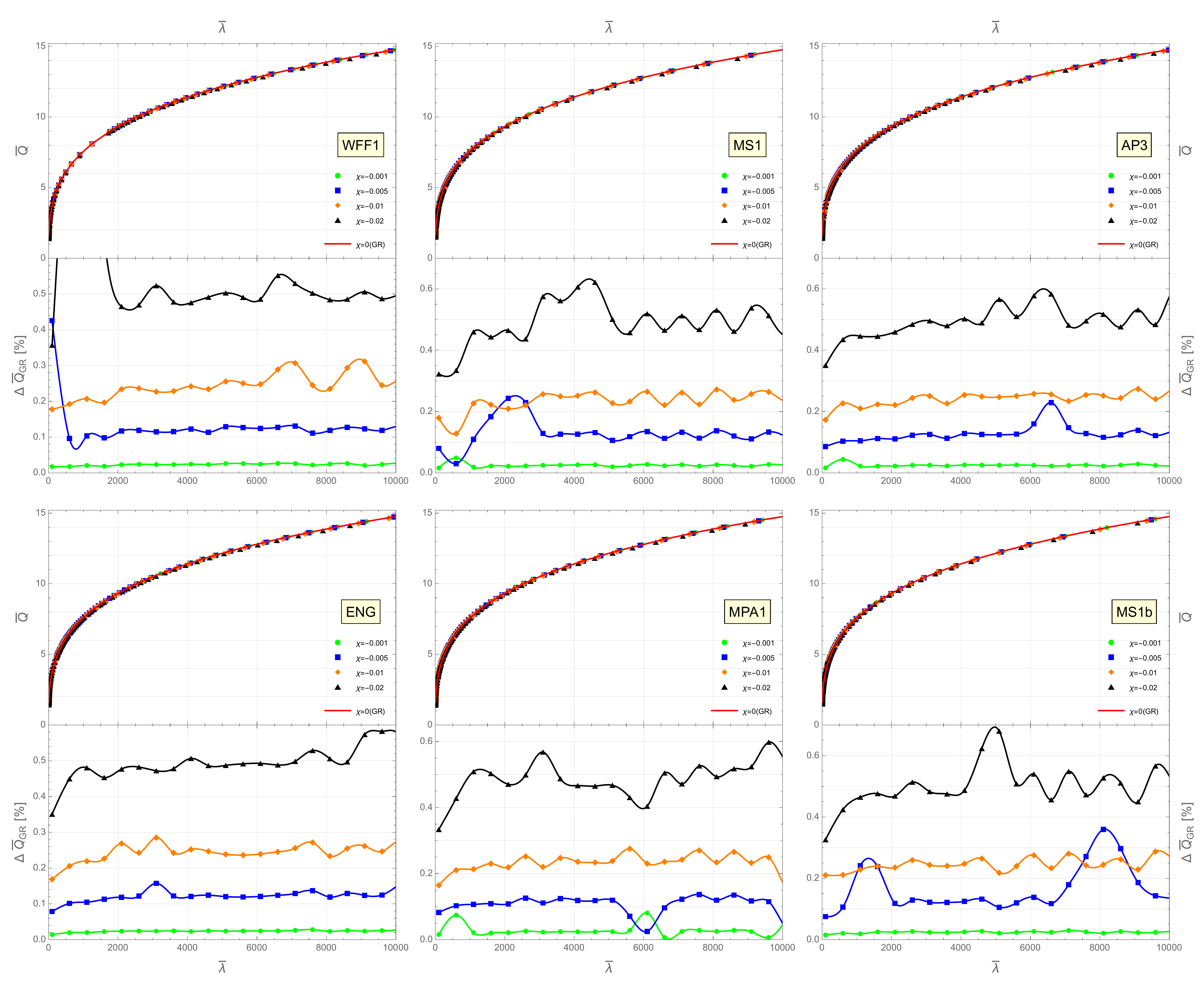}
\caption{The relationship between the dimensionless quadrupole and the dimensionless tidal love number. The lower inset of every panel shows how the $\bar{Q}-\bar{\lambda}$ relationship for different values of $\chi$ in $f(R, T)=R+2\chi T$  deviates from the general relativistic one.}
\label{fig:9}
\end{figure*}

\begin{figure*}[!hb]
\centering
\begin{subfigure}{0.49\textwidth}
    \includegraphics[width=\textwidth]{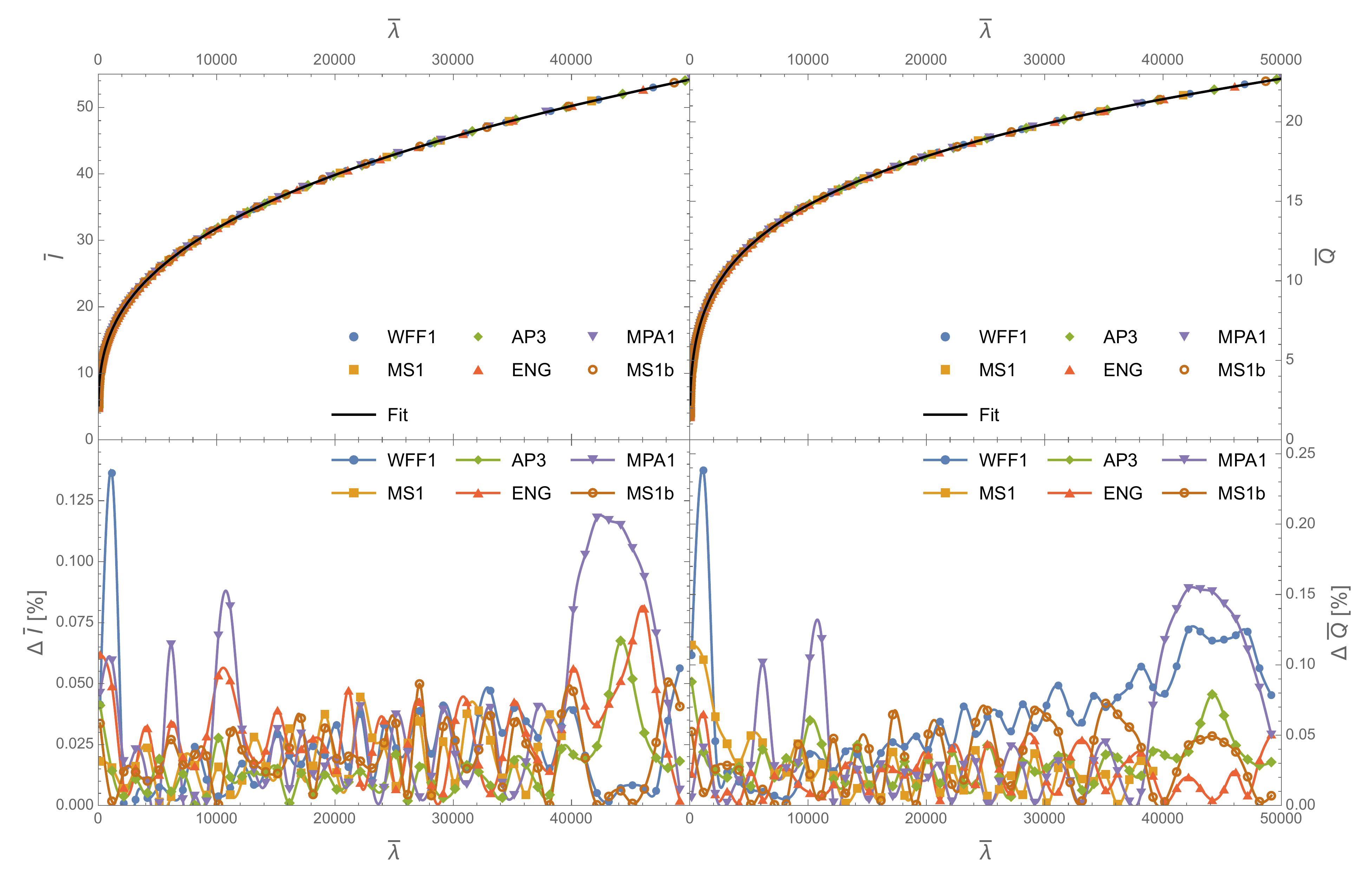}
    \caption{GR}
    \label{fig:10a}
\end{subfigure}
\hfill
\begin{subfigure}{0.49\textwidth}
    \includegraphics[width=\textwidth]{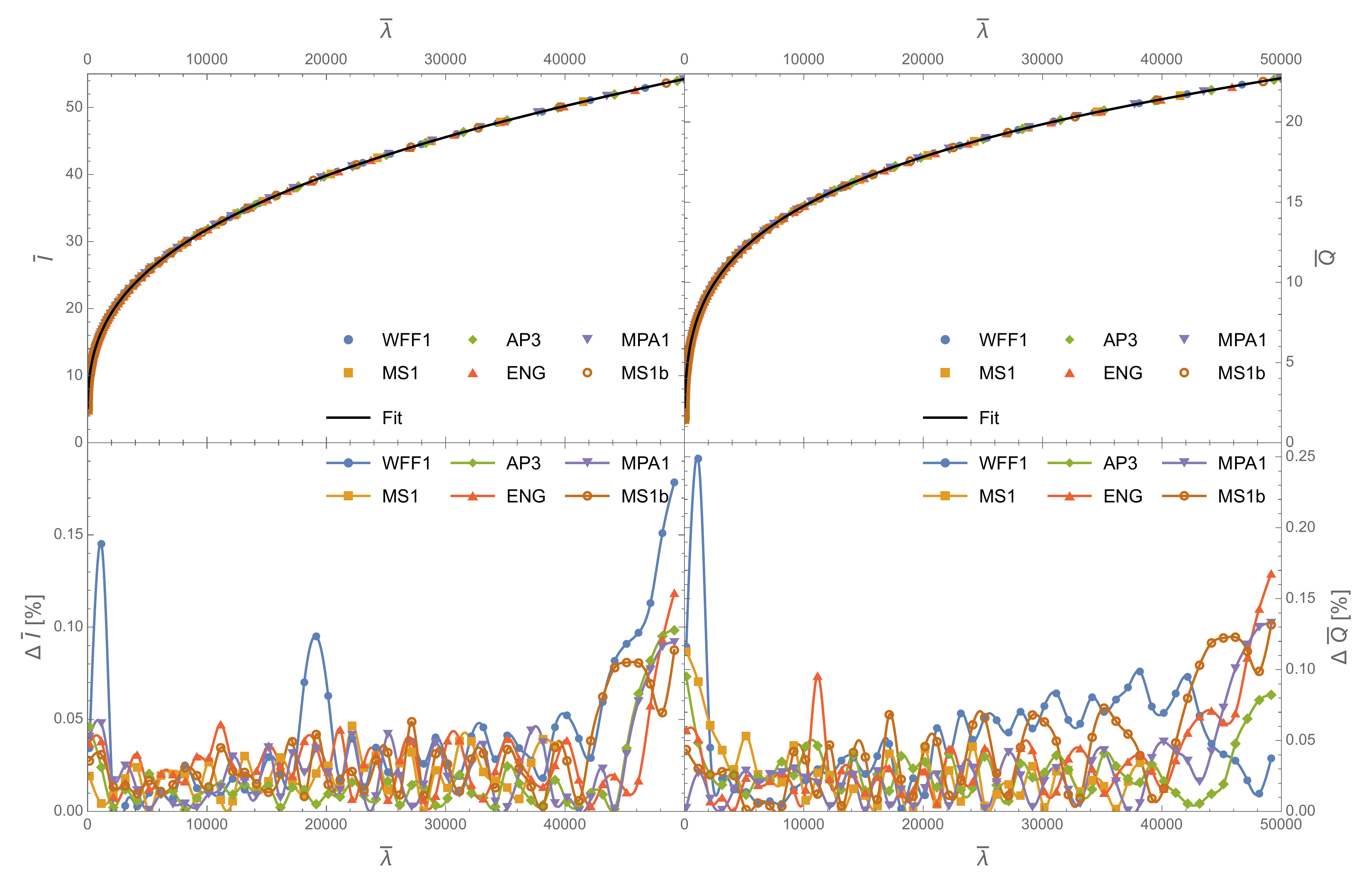}
    \caption{$\chi=-0.001$}
    \label{fig:10b}
\end{subfigure}
\begin{subfigure}{0.49\textwidth}
    \includegraphics[width=\textwidth]{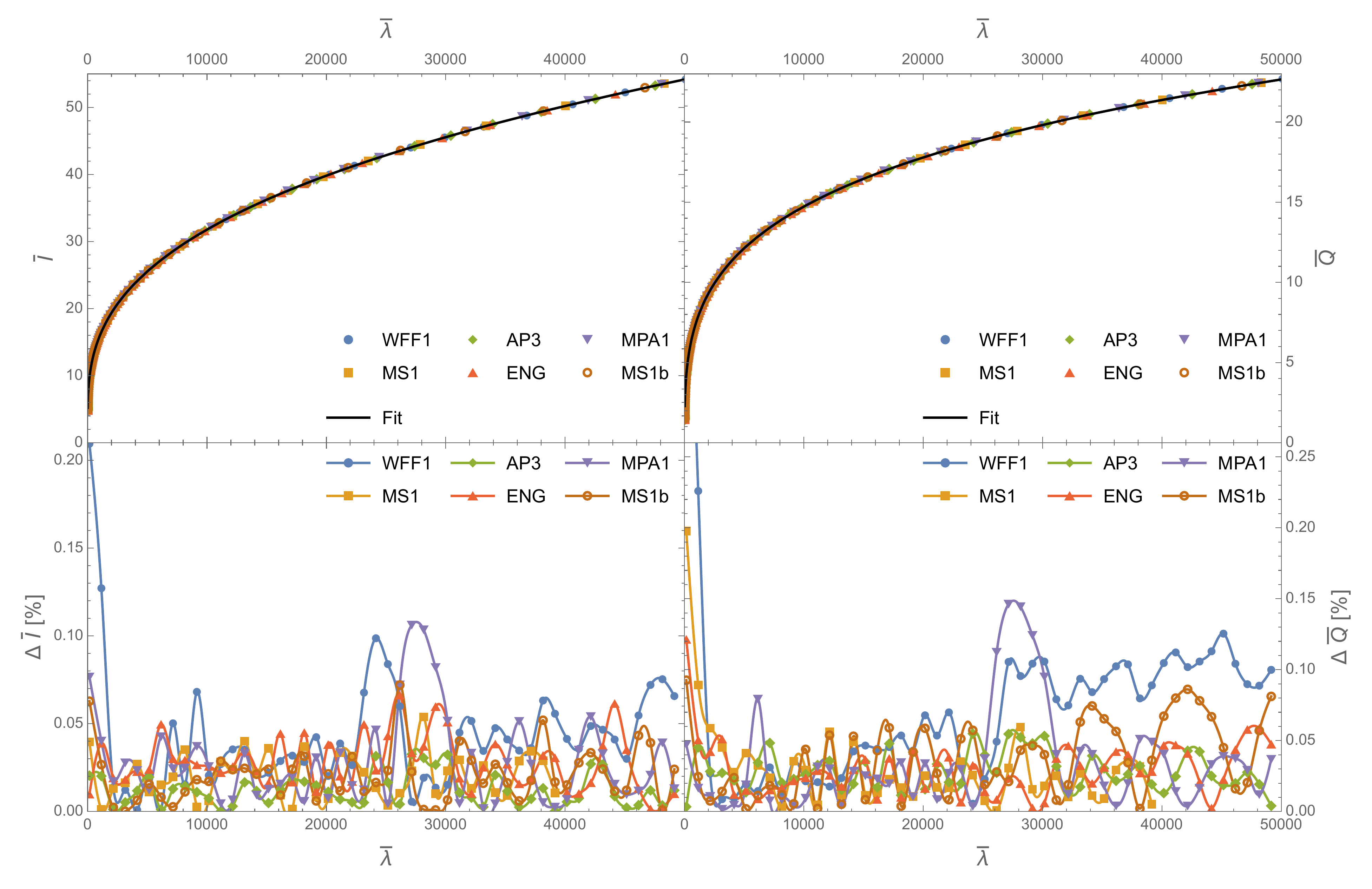}
    \caption{$\chi=-0.01$}
    \label{fig:10c}
\end{subfigure}
\hfill
\begin{subfigure}{0.49\textwidth}
    \includegraphics[width=\textwidth]{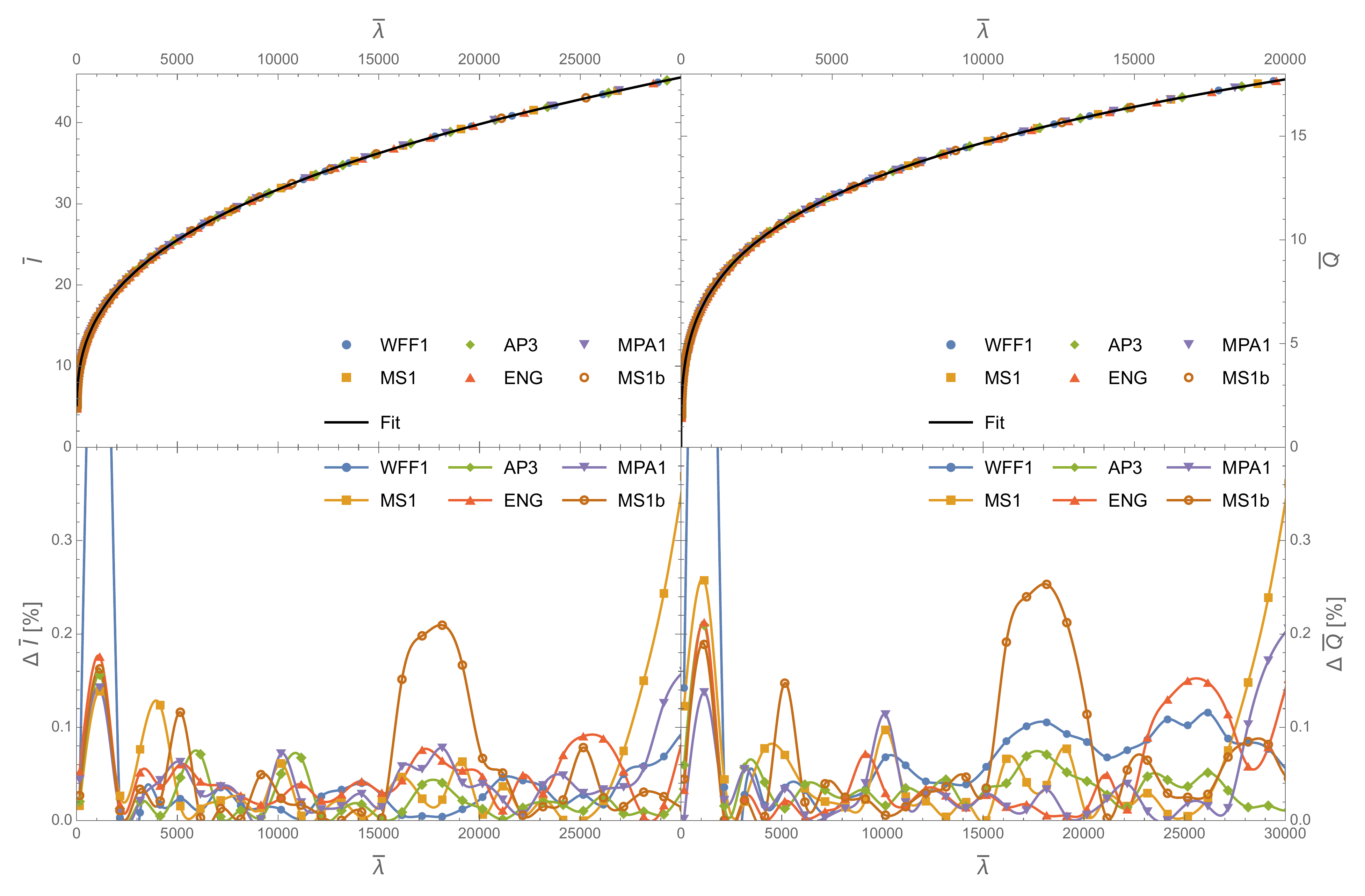}
    \caption{$\chi=-0.02$}
    \label{fig:10d}
\end{subfigure}
\caption{The EoS independent $\bar{I}-\bar{\lambda}$ (Upper left inset) and $\bar{Q}-\bar{\lambda}$ (Upper right inset) relationships in $f(R, T)=R+2\chi T$ modified gravity are shown. The lower left  and right insets are respectively showing the deviation of $\bar{I}-\bar{\lambda}$ and $\bar{Q}-\bar{\lambda}$ found for different EoS from the fitting curve.}
\label{fig:10}
\end{figure*}

\section{$f(R, T)=R+\alpha R^{2}+2\chi T$ modified gravity }
\textbf{In this section, we consider $f(R, T)=R+\alpha R^{2}+2\chi T$, where $\alpha$ is the positively valued parameter usually given in the units of $r_{g}$, where  $r_{g}=\frac{G M_{\odot}}{c^2}=1.47664km$ is the solar mass in geometric units and $\chi $ is the matter-gravity coupling parameter. For the Lagrangian $\mathcal{L}_{m}=P$, the Eq.-\ref{EQ:2} becomes}
\begin{equation}\label{EQ:1A}
(1+\alpha R)G_{\mu\nu}+\frac{\alpha}{2}R^{2}g_{\mu\nu}=\kappa T_{\mu\nu}+\chi T g_{\mu\nu}+2\chi \left(T_{\mu\nu}-P g_{\mu\nu}\right)
\end{equation}
\textbf{and the covariant divergence equation (Eq.\ref{EQ:4}) remains same as the Eq.\ref{EQ:7} in this gravity. In this $f(R, T)=R+\alpha R^{2}+2\chi T$ gravity, the Ricci scalar is the dynamic quantity, and its evolution is obtained by taking the trace of the field equation. Therefore, the trace of Eq.-\ref{EQ:2} yields}
\begin{equation}\label{EQ:2A}
6\alpha\Box R-R=\kappa T+2\chi(T-\Theta)
\end{equation}

\textbf{Here, we also choose Boyer-Lindquist type coordinates $(t,r,\theta,\phi)$ to represent the most general stationary axisymmetric spacetime for rotation up to first order, which can be written as}
\begin{align} \label{EQ:3A}
ds^{2}&= -e^{\nu}dt^{2} +\frac{1}{1-\frac{2m}{r}}dr^2  \nonumber\\  &+r^2 [d\theta^2+\sin^{2}\theta (d\phi-\epsilon \omega dt)^2]
\end{align}
where $\epsilon$ is a bookkeeping parameter for slow rotation.The pressure and energy density are not affected in the first order of rotation. Thus, $P$ and $\rho$  remain the same as the non-rotating ones.
\subsection{Zeroth Order Effect}
\textbf{The (t,t) and (r,r) components of the Eq.\ref{EQ:1A} for nonrotating or zeroth order in rotation becomes}
\begin{widetext}
\begin{align}\label{EQ:4A}
\dfrac{dM}{dr}-&\frac{2 \alpha  R' \left[r^3 \left(-2 P \chi +6 \rho  \chi +16 \pi  \rho +3 \alpha  R^2+R\right)-6 M
   (2 \alpha  R+1)\right]}{12 (2 \alpha  R+1) \left(\alpha  r R'+2 \alpha 
   R+1\right)}\nonumber \\
&   -\frac{24 \alpha ^2 r (r-2 M) \left(R'\right)^2+r^2 (2 \alpha  R+1) \left[14 P \chi +48 \pi
    P+6 \rho  \chi +32 \pi  \rho +R (3 \alpha  R+2)\right]}{12 (2 \alpha  R+1) \left(\alpha  r R'+2 \alpha 
   R+1\right)}=0
\end{align}

\begin{align}\label{EQ:5A}
\dfrac{d\nu}{dr} -\frac{4 M \left(4 \alpha  r R'+2 \alpha  R+1\right)+r^2 \left[2 P r (3 \chi +8 \pi )-2 \rho  r \chi
   -\alpha  r R^2-8 \alpha  R'\right]}{2 r (r-2 M) \left(\alpha  r R'+2 \alpha  R+1\right)}=0
\end{align}

The covariant divergence Eq.\ref{EQ:7} becomes

\begin{align}\label{EQ:6A}
\dfrac{dP}{dr}+\frac{\gamma  (\chi +4 \pi ) (P+\rho ) \left[4 M \left(4 \alpha  r R'+2 \alpha  R+1\right)+r^2 \left\{2 P
   r (3 \chi +8 \pi )-2 \rho  r \chi -\alpha  r R^2-8 \alpha  R'\right\}\right]}{2 r (\gamma  (3 \chi
   +8 \pi )-\chi ) (r-2 M) \left(\alpha  r R'+2 \alpha  R+1\right)}=0
\end{align}
and the Eq.\ref{EQ:2A} yields

\begin{align}\label{EQ:7A}
\dfrac{d^2R}{dr^2}-&\frac{\left[12 M (2 \alpha  R+1)+r \left\{-2 P r^2 \chi +2 \rho  r^2 (3 \chi +8 \pi )+R
   \left(-24 \alpha +3 \alpha  r^2 R+r^2\right)-12\right\}\right]}{6   r(r-2 M) (2 \alpha  R+1)}\dfrac{dR}{dr} \nonumber \\
&-\frac{2 \alpha }{(2 \alpha  R+1)}\left(\dfrac{dR}{dr}\right)^2   -\frac{r^2 (2 \alpha  R+1) \left[2 P (5 \chi +12 \pi )-2 \rho  (3 \chi +4 \pi )+R\right]}{6 \alpha  r
   (r-2 M) (2 \alpha  R+1)}=0
\end{align}
\end{widetext}

where $R'=\dfrac{dR}{dr}$. The boundary conditions needed to solve the above four equations are the following:
\begin{align}\label{EQ:8A}
&M(0)=0, \hspace{5mm} \nu(0)=0, \nonumber \\
&R(0)=R_{c},\hspace{5mm} \text{and} \hspace{5mm} \dfrac{dR}{dr}\bigg|_{r=0}=0
\end{align}
\textbf{The asymptotic flatness of the spacetime is achieved by setting $R(r  \to \infty)=0$, and this requirement of asymptotic flatness comes at the cost of the unique value of $R(0)=R_{c}$ for a given central density ($\rho_{c}$). In order to determine the unique value of $R(0)=R_c$ for a given central density, we need to integrate four equations from the centre of the star to its surface and then from the surface to infinity by assuming a certain value for $R(0)$, and then we can find the specific value of $R(0)=R_c$ by setting $R(r \to \infty)=0$. We again integrate the four equations for that given central density using the corresponding unique value of $R(0)=R_c$ to determine the radius ($R_{s}$) and mass (M) of the star.}
\subsection{First Order Effect}
In the first order in rotation, the $(t,\phi)$ component becomes non-zero and is given by
\begin{widetext}
\begin{align}\label{EQ:9A}
&\dfrac{d^2\bar{\omega }}{dr^2}+\frac{\dfrac{d\bar{\omega }}{dr}}{6 r (2 M-r) (2 \alpha  R+1) \left(\alpha  r R'+2
   \alpha  R+1\right)}\left[\alpha  r R' \left(84 M (2 \alpha  R+1)+r^3 \left(R+3 \alpha 
   R^2-2 P \chi +6 \rho  \chi +16 \pi  \rho \right)-48 r (2 \alpha  R+1)\right)\right. \nonumber \\
& \left.  +(2 \alpha  R+1) \left(48 M (2 \alpha  R+1)+r^3 (16 P (\chi +3
   \pi )+16 \pi  \rho +R)-24 r (2 \alpha  R+1)\right)\right]+\frac{4 r (\chi +4 \pi ) (P+\rho )}{(2 M-r) (2 \alpha  R+1)}\bar{\omega }=0
\end{align}
\end{widetext}
The boundary conditions for this equation is given by
\begin{equation}
\bar{\omega }(r=0)=1 \hspace{5mm} and \hspace{5mm} \dfrac{d\bar{\omega }}{dr}\bigg|_{r=0}=0
\end{equation}
This equation is integrated for given $M(r)$, $\nu(r)$, $P(r)$ and $R(r)$ functions from the star's centre to surface and then from surface to infinity to determine moment of inertia $(I)$ of the star using Eqs\ref{EQ:18}\&\ref{EQ:19}.
\section{Tidal deformability}
  Up untill now, we only consider isolated slowly rotating star and the effect of rotation on it. In this section, we will concentrate on the binary NSs system. When a NS is in binary system, it gets tidally deformed by the tidal field of its binary companion. The tidal deformability of the neutron star is, then, in the linear order of its companion tidal field $\epsilon_{ij}$, defined as
  \begin{equation}\label{EQ:45}
\lambda=-\frac{Q_{ij}}{\epsilon_{ij}}  
  \end{equation}
  where $Q_{ij}$ is the quadrupole moment developed in it due to its companion tidal field.
  The dimensionless second Love number $k_{2}$ is related to the tidal deformability $\lambda$ as
   
  \begin{equation}\label{EQ:46}
k_{2}=\frac{3}{2}~\frac{\lambda}{R^{5}}  
  \end{equation}
  and the dimensionless tidal deformability is defined as
  \begin{equation}\label{EQ:47}
  \bar{\lambda}=\frac{\lambda}{M^{5}}
  \end{equation}
  For simplicity, during the calculation of tidal deformability of the star due to its companion tidal field, we will consider that the star is not spinning. Therefore, here, we can set $\bar{\omega }=0$. Now, if we set $h_{0}=m_{0}=0$, and redefine $h_{2}$ and $m_{2}$ as
  \begin{equation}\label{EQ:48}
  h_{2}=-\frac{H_0}{2}
  \end{equation}
  \begin{equation}\label{EQ:49}
  m_{2}=\left(1-\frac{2M}{r}\right) \frac{r~H_{2}}{2}
  \end{equation}
  then Eq.\ref{EQ:8} is reduced as
  \begin{equation}\label{EQ:50}
  g_{ab}=g^{0}_{ab}+h_{ab}
  \end{equation}
  where $g^{0}_{ab}$ is unperturbed metric and $h_{ab}$ is even-parity perturbation in Regge-Wheeler gauge \citep{Regge1957}.
  By setting $p_{0}=0$, and following T. Hinderer formalism \citep{Hinderer2008}, we get $H_{2}=H_{0}$ and a master equation for $H_{0}$ as
  \begin{equation}\label{EQ:51}
\dfrac{d^{2}H_{0}}{dr^{2}} +C_{1}  \dfrac{dH_{0}}{dr}+C_{2} H_{0}=0
  \end{equation}

\begin{figure*}[!ht]
\centering
\begin{subfigure}{0.49\textwidth}
    \includegraphics[width=\textwidth]{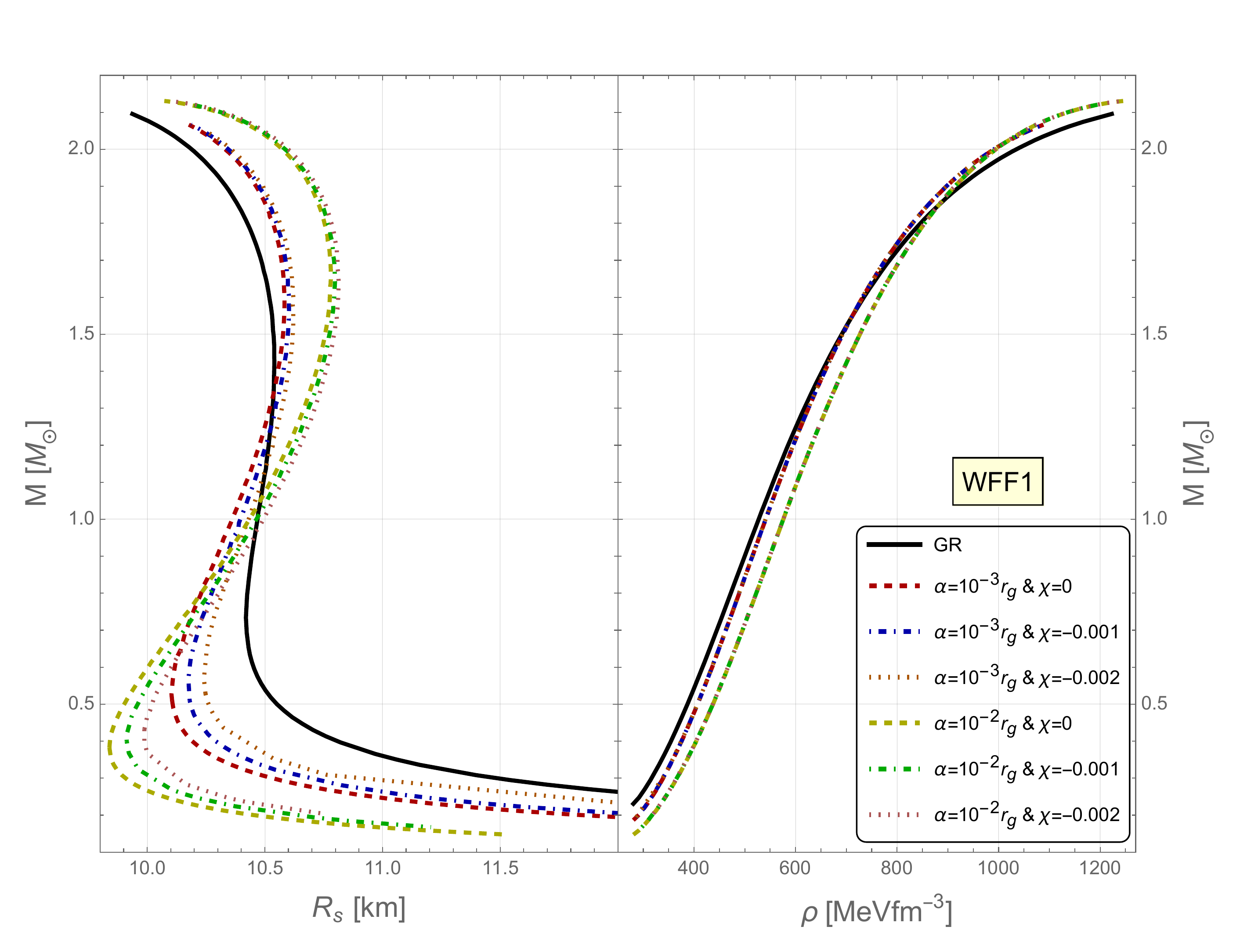}
%    \caption{sly}
    \label{fig:11a}
\end{subfigure}
\hfill
\begin{subfigure}{0.49\textwidth}
    \includegraphics[width=\textwidth]{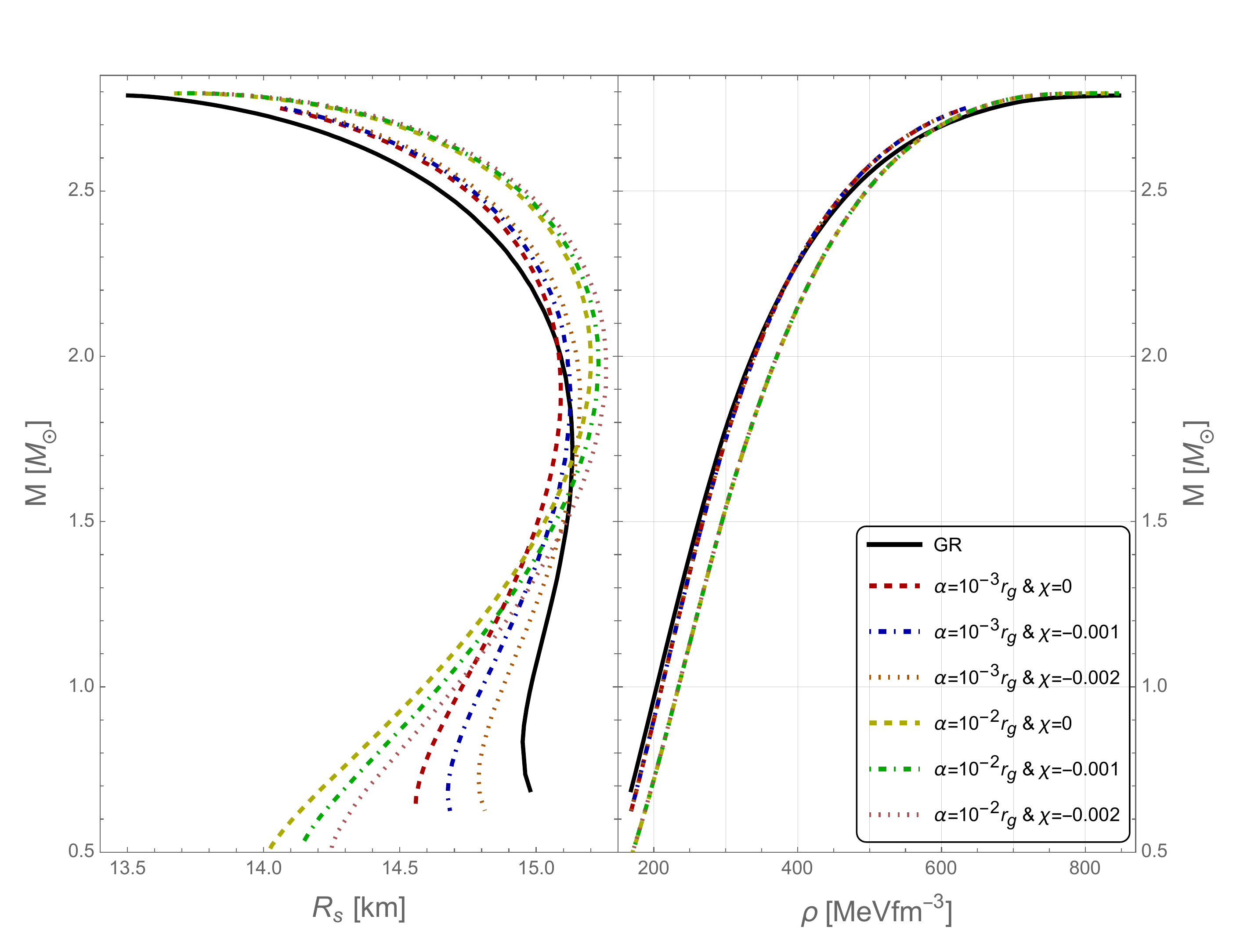}
 %   \caption{bbb2}
    \label{fig:11b}
\end{subfigure}
        \hfill
\begin{subfigure}{0.49\textwidth}
    \includegraphics[width=\textwidth]{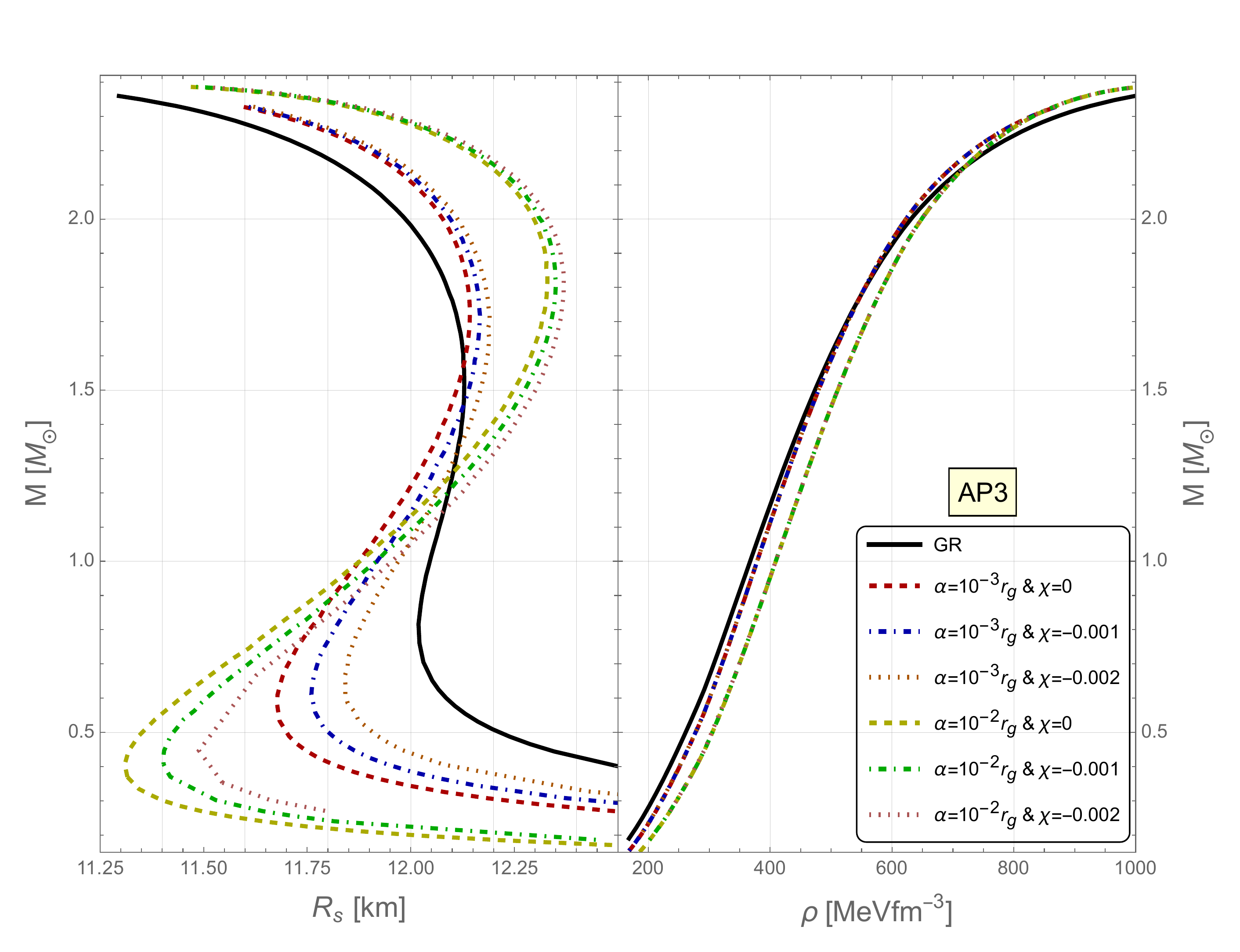}
 %   \caption{alf4}
    \label{fig:11c}
\end{subfigure}
\hfill
\begin{subfigure}{0.49\textwidth}
    \includegraphics[width=\textwidth]{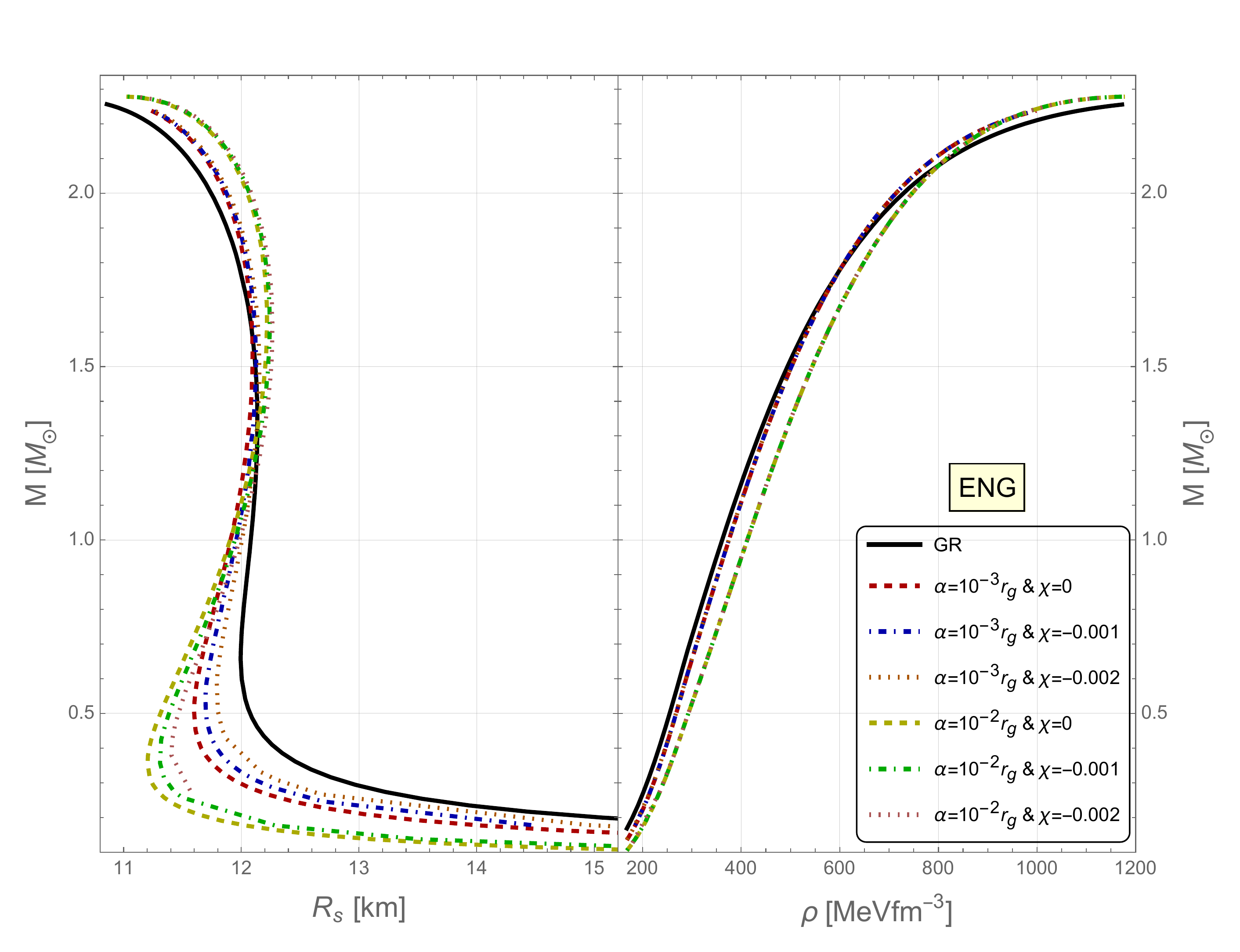}
%    \caption{pal6}
    \label{fig:11d}
\end{subfigure}
\hfill
\begin{subfigure}{0.49\textwidth}
    \includegraphics[width=\textwidth]{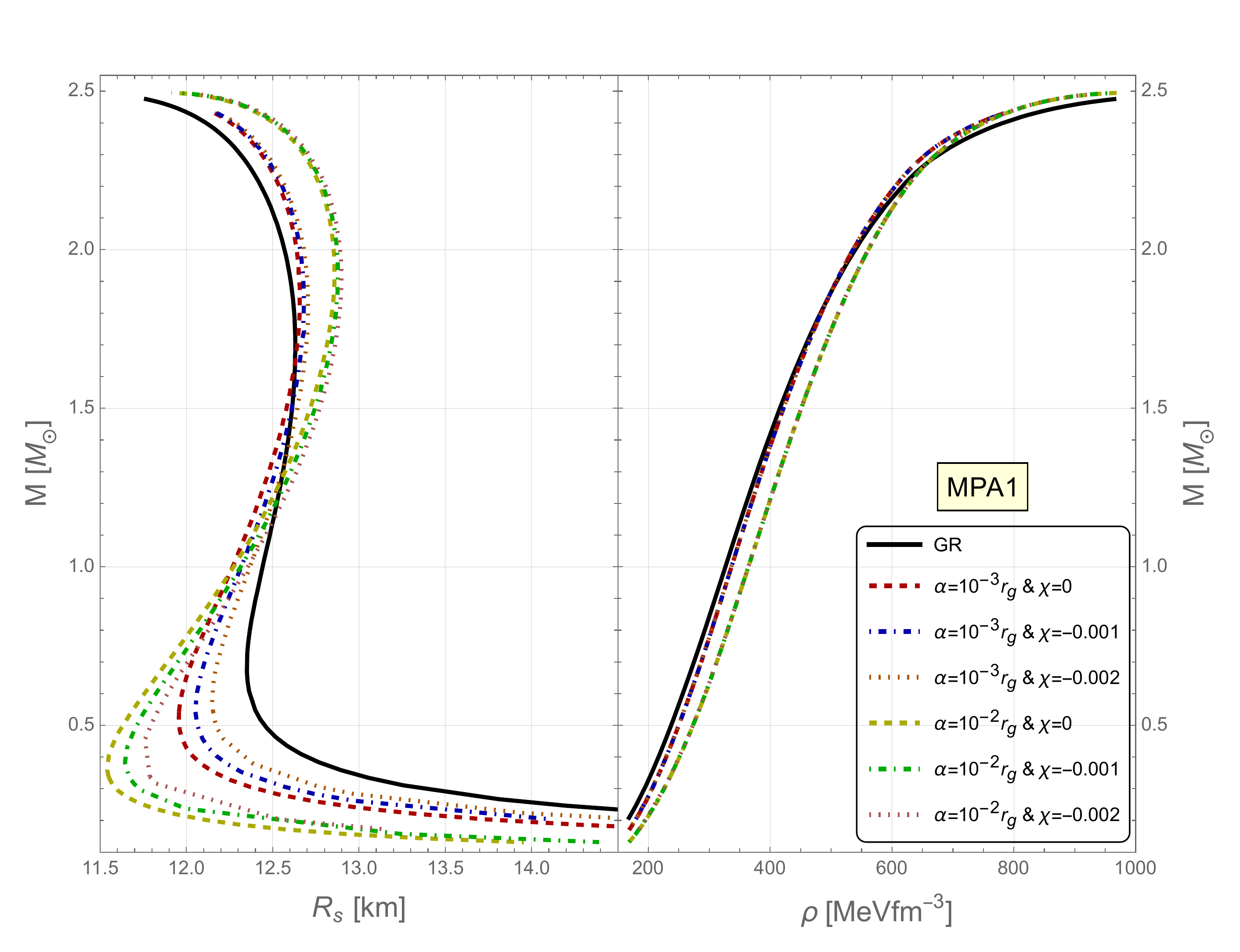}
 %   \caption{alf4}
    \label{fig:11e}
\end{subfigure}
\hfill
\begin{subfigure}{0.49\textwidth}
    \includegraphics[width=\textwidth]{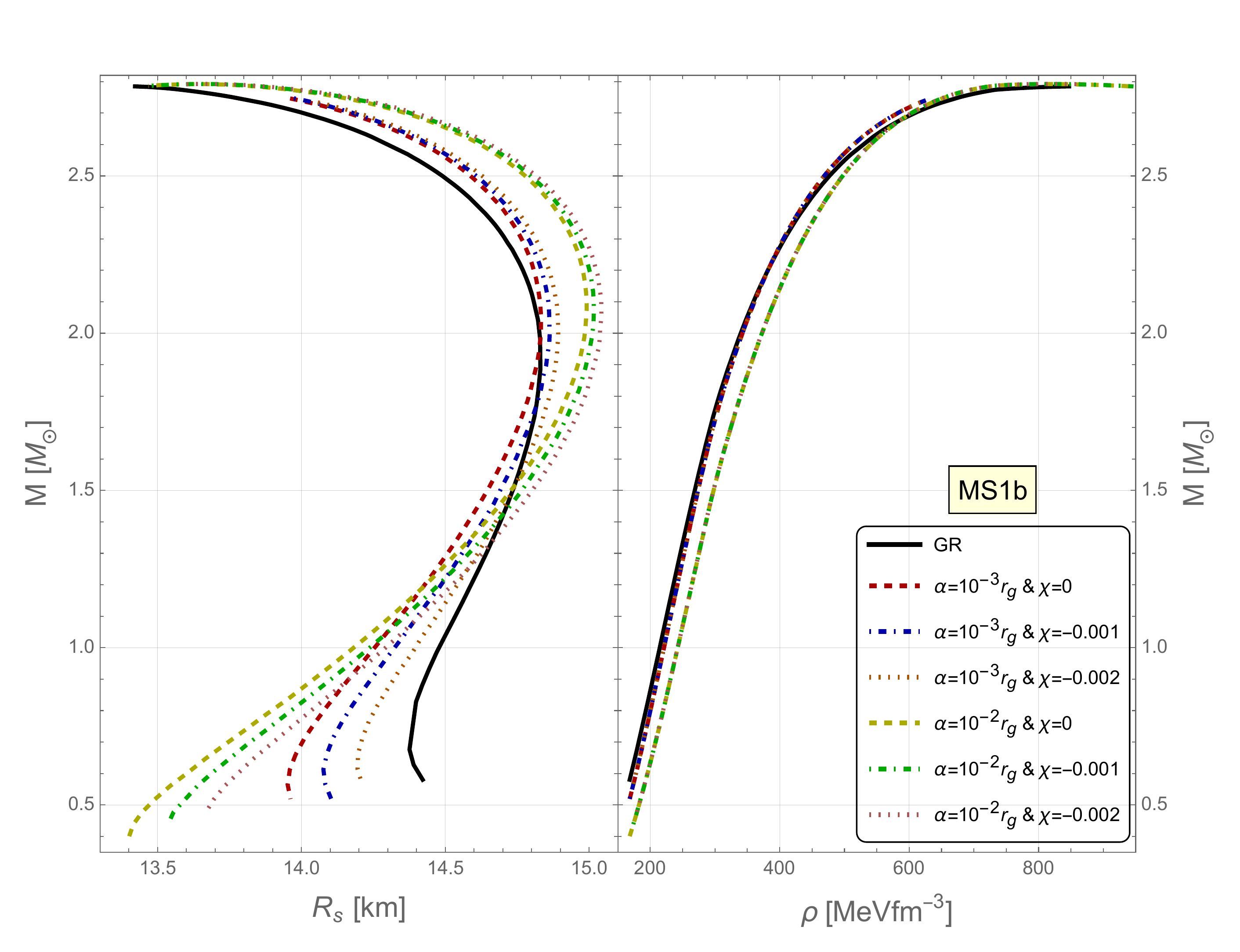}
%    \caption{pal6}
    \label{fig:11f}
\end{subfigure}
\caption{Mass-Radius relation  and Mass-Density relation for $f(R,T)=R+\alpha R^{2}+2\chi T$ gravity. }
\label{fig:11}
\end{figure*}

\begin{figure*}[!ht]
\centering
\begin{subfigure}{0.32\textwidth}
    \includegraphics[width=\textwidth]{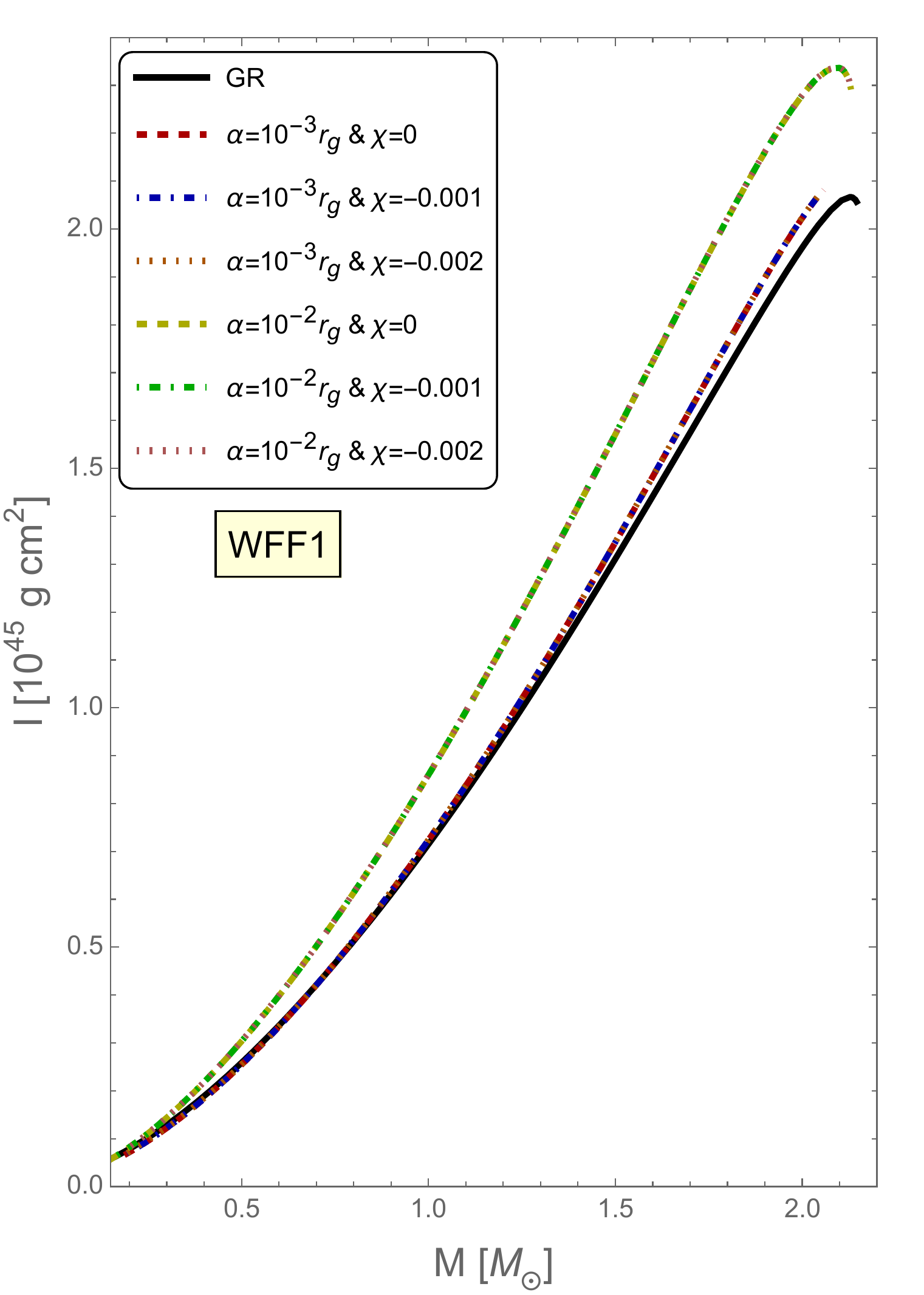}
%    \caption{sly}
    \label{fig:12a}
\end{subfigure}
\hfill
\begin{subfigure}{0.305\textwidth}
    \includegraphics[width=\textwidth]{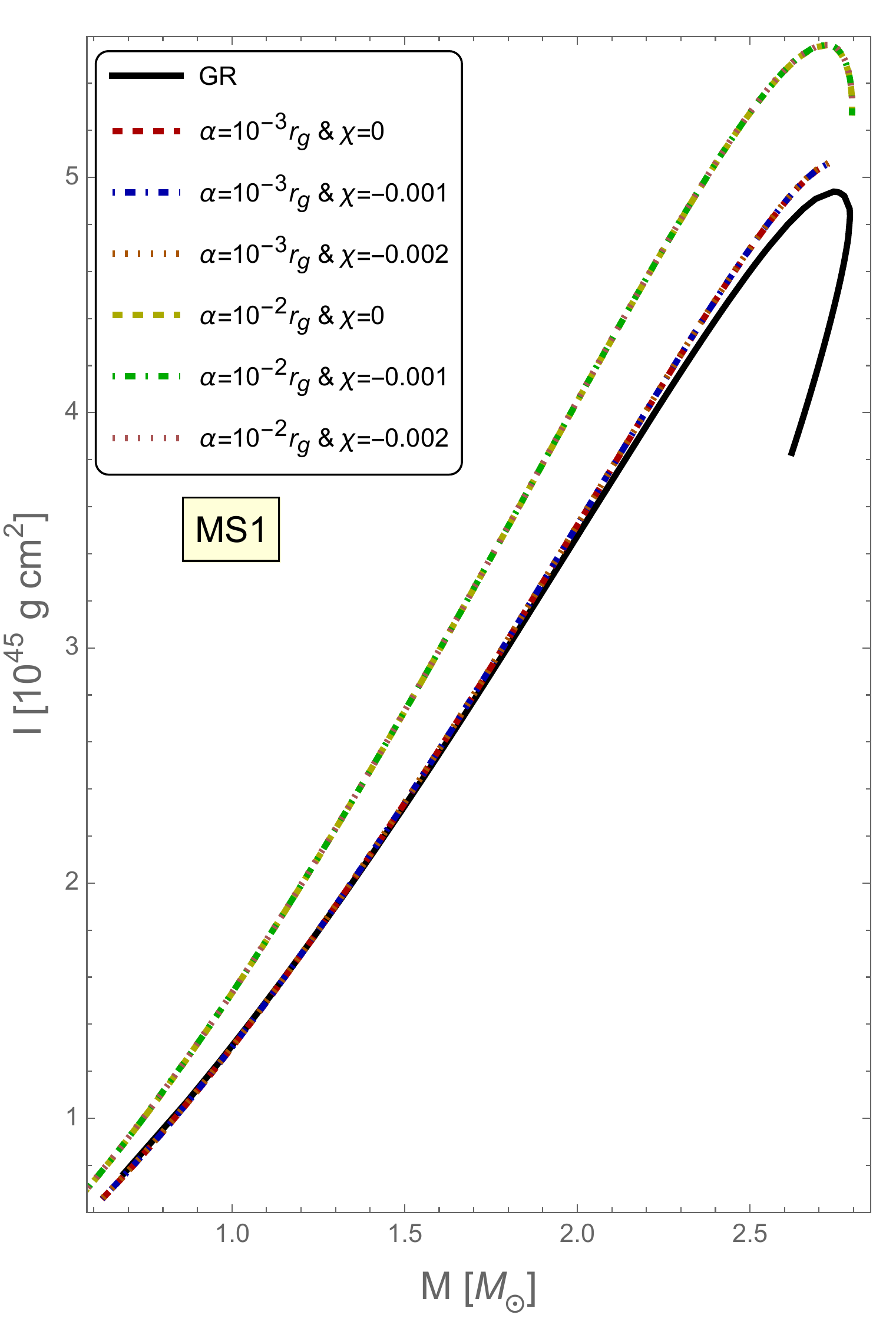}
 %   \caption{bbb2}
    \label{fig:12b}
\end{subfigure}
        \hfill
\begin{subfigure}{0.32\textwidth}
    \includegraphics[width=\textwidth]{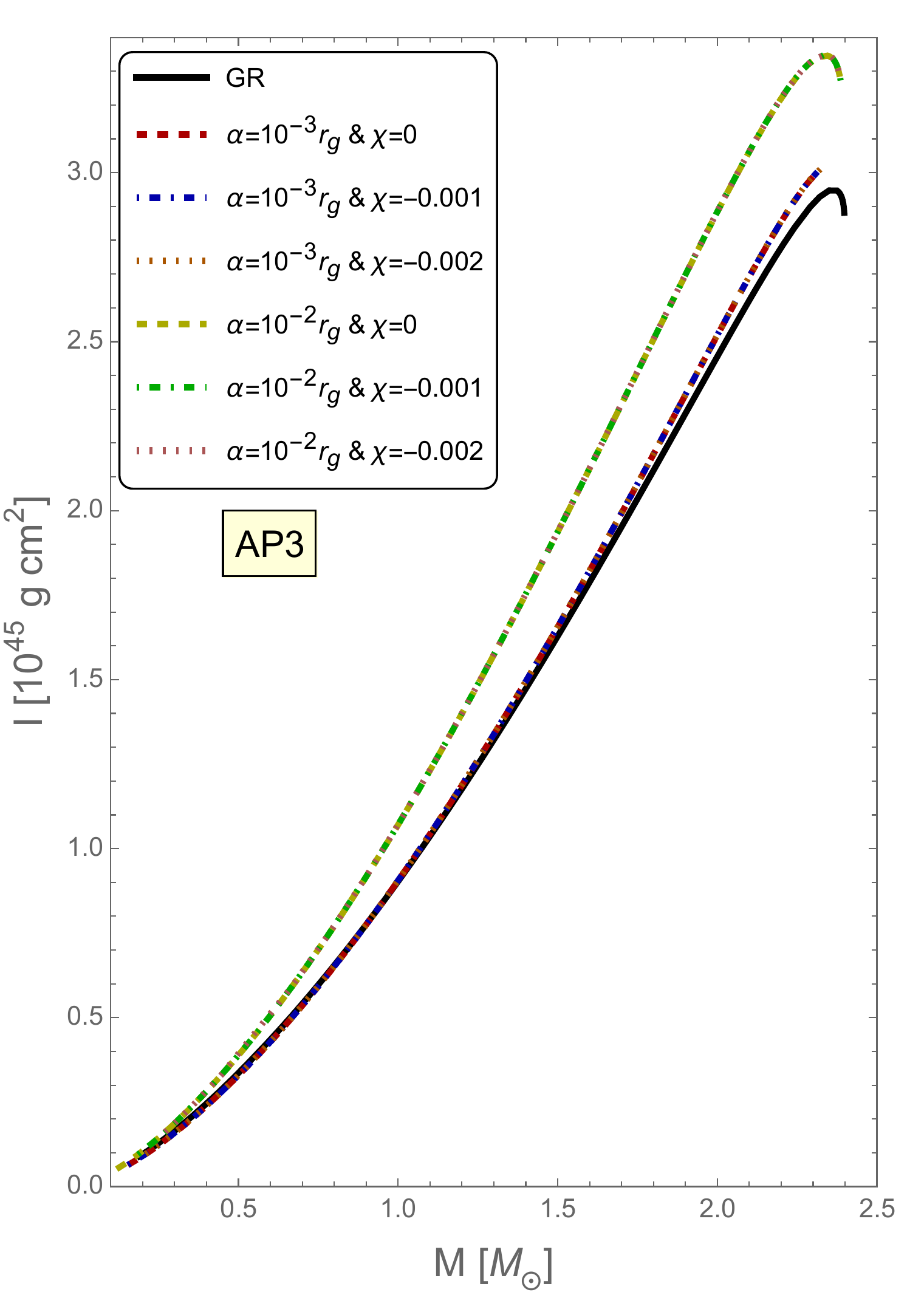}
 %   \caption{alf4}
    \label{fig:12c}
\end{subfigure}
\hfill
\begin{subfigure}{0.32\textwidth}
    \includegraphics[width=\textwidth]{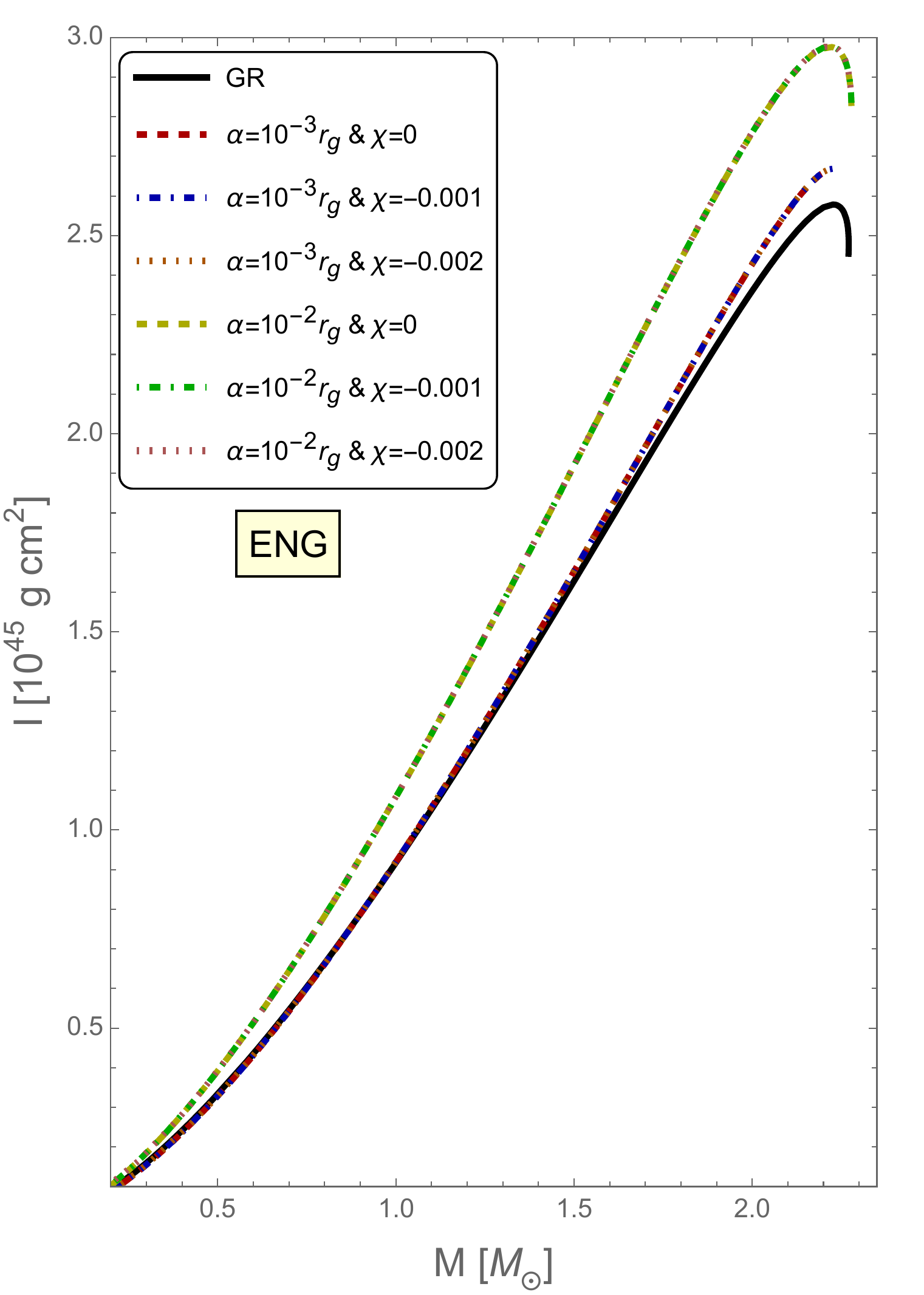}
%    \caption{pal6}
    \label{fig:12d}
\end{subfigure}
\hfill
\begin{subfigure}{0.32\textwidth}
    \includegraphics[width=\textwidth]{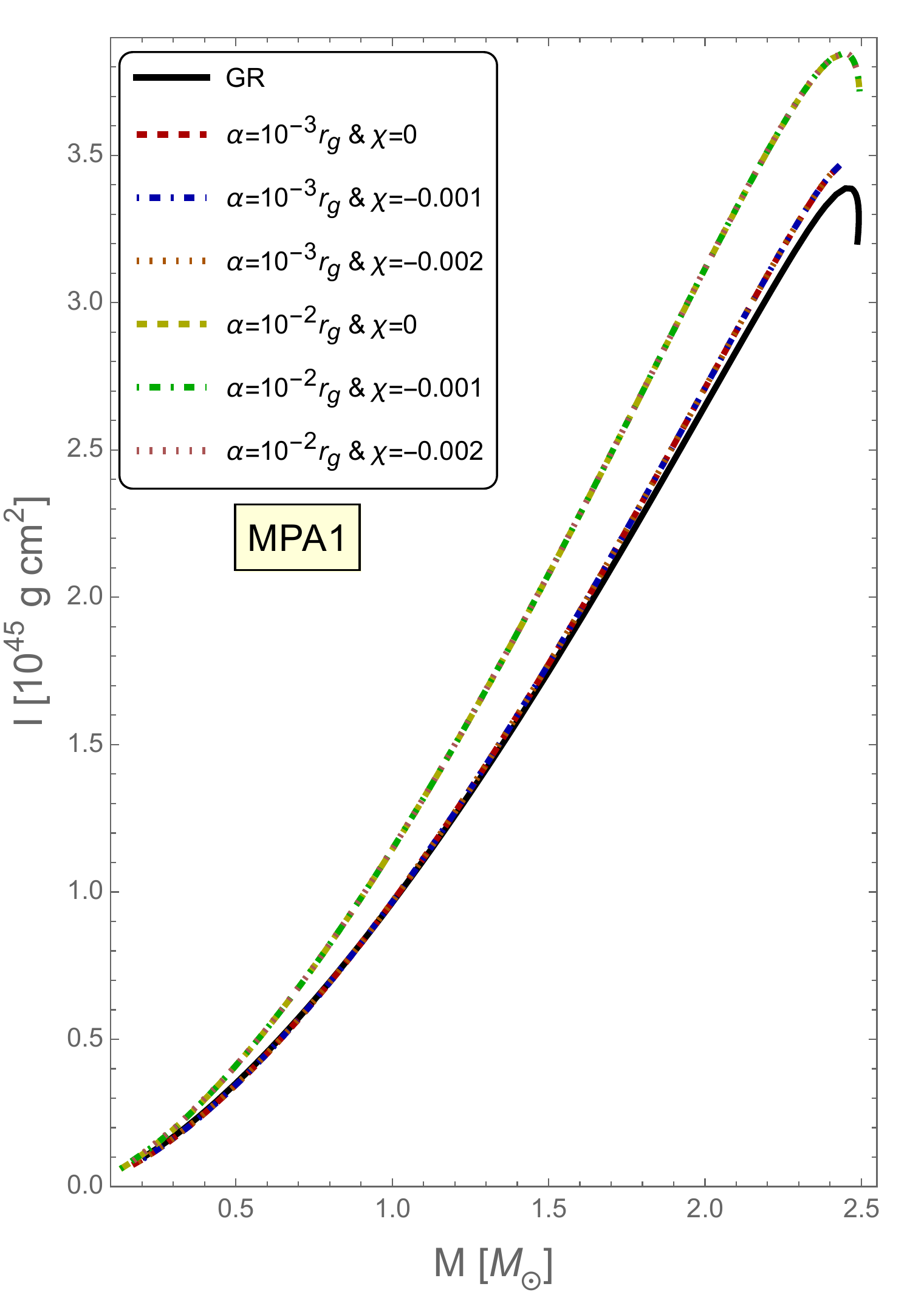}
 %   \caption{alf4}
    \label{fig:12e}
\end{subfigure}
\hfill
\begin{subfigure}{0.305\textwidth}
    \includegraphics[width=\textwidth]{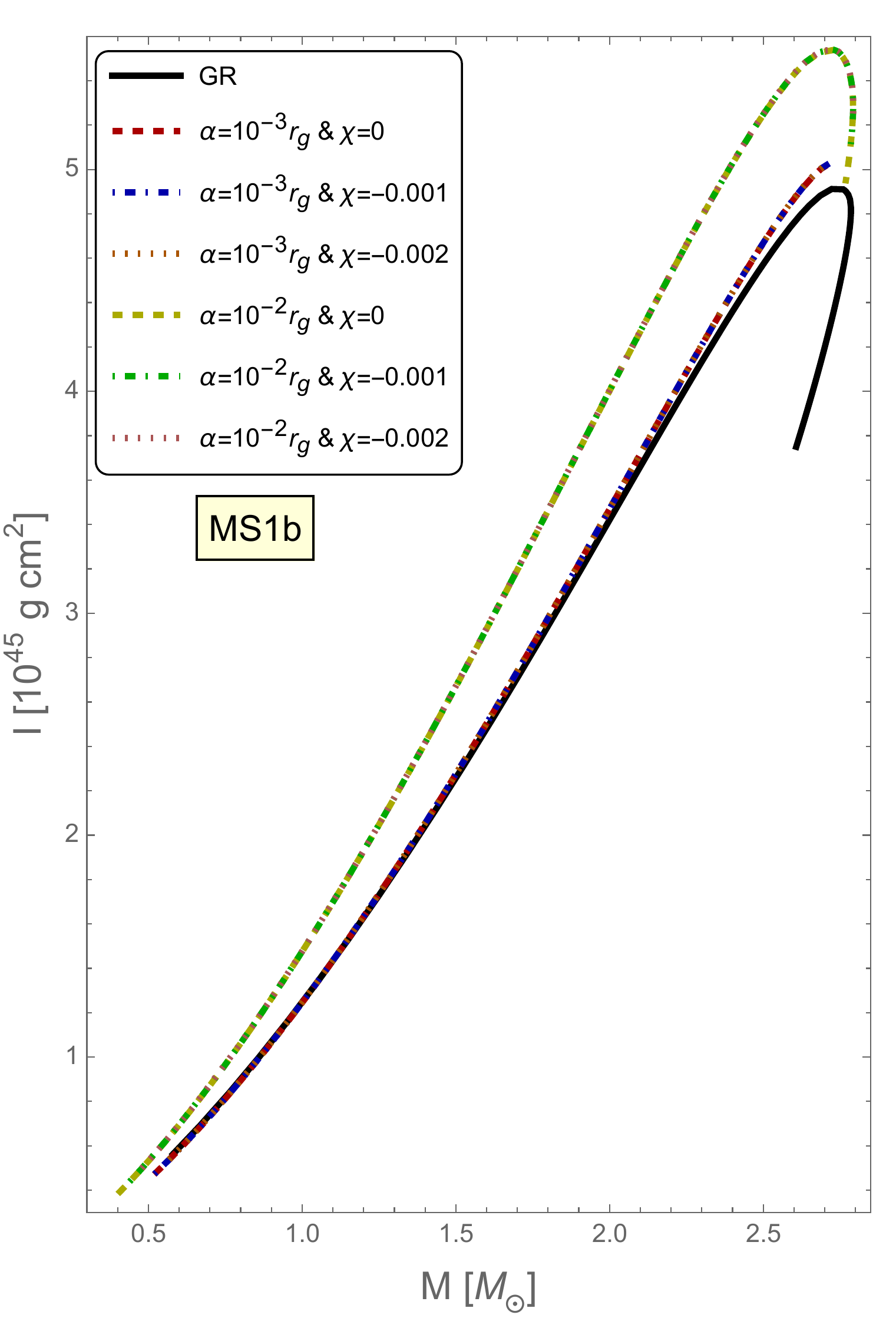}
%    \caption{pal6}
    \label{fig:12f}
\end{subfigure}
\caption{The moment of inertia as a function of the mass for $f(R,T)=R+\alpha R^{2}+2\chi T$ gravity.}
\label{fig:12}
\end{figure*}     
     
\textbf{\section{Equation of State (EoS)}     
  Neutron stars are ultra-high dense objects having densities above the nuclear saturation density, that is, $ \rho_{0}=2.7 \times 10^{14} gcm^{3}$, while their temperature is low as compared to their constituent particle's Fermi energy. A cold ultra-high dense neutron star's core can be made of exotic particles such as hyperons, pion or kaon condensates, and even strange quark matter other than nucleons—currently, several EoS candidates, generated using various methods such as relativistic mean field theory, quantum chromodynamics, and lattice QCD, represent a neutron star's core. We choose to use a piecewise polytropic approach \cite{Read:2008iy,PhysRevD.79.124033,PhysRevD.107.104039} for all of our EoSs in this paper. The piecewise polytropic EoS is constructed using phenomenological data on nuclear matter. We use six EoS data, which are WFF1, AP3, ENG, MPA1, MS1, and MS1b, for the highly dense neutron star's core and SLy \cite{Douchin:2001sv} for the lower density part describing the crust of the neutron star. Among them, WFF1 \cite{PhysRevC.38.1010} and AP3 \cite{PhysRevC.58.1804} are the variational methods, ENG \cite{Engvik:1995gn} and MPA1 \cite{Muther:1987xaa} are the relativistic Brueckner-Hartree-Fock EoSs, and MS1 and MS1b are the relativistic mean field theory EoSs \cite{Mueller:1996pm}. To construct parameterized the piecewise polytropic EoS, we rely on the Read et al. paper \cite{Read:2008iy}. The Sly crust is being matched with the lower-density part of each of the six highly dense EoSs to form a complete piecewise polytropic EoS. Hereafter, if we say, for example, WFF1 EoS, which means WFF1 EoS is constructed by adopting a piecewise polytropic approach by using WFF1 and SLy four parameterized data found in \cite{Read:2008iy} for neutron stars core and crust respectively.}

 \section{Results and Discussions}  
 \subsection{$f(R,T)=R+2\chi T$ Modified Gravity results}
All 11 equations (Eqs.\ref{EQ:14} to \ref{EQ:17}, Eqs.\ref{EQ:23} to \ref{EQ:24}, Eq.\ref{EQ:29} to \ref{EQ:32} and Eq.\ref{EQ:51}) are now numerically integrated from the centre to the surface of the stars and then from the surface of the stars to infinity. To improve the stability of numerical integration, we first make Taylor expansions of these equations around the centre and solve the expanded equations to get higher-order terms. The Eq.\ref{EQ:25} can be written as a total derivative, and its integration gives
\begin{equation}\label{EQ:52}
h_{0}=h_{0C}-\frac{p_{0} \left[(\kappa +3 \chi ) \gamma -\chi \right]}{(\kappa +2 \chi ) \gamma }+\frac{1}{3} r^2 e^{-\nu }
   \bar{\omega }^2
\end{equation}
where $h_{0C}$ is a constant, and its value is determined by the demand of continuity of the function $h_{0}$ across the star surface.

\textbf{According to Eq.\ref{EQ:16}, the value of matter-gravity coupling parameter $\chi$ should lie between $-4\pi $ and $0$. In addition, R. Lobato et al. \cite{Lobato2020} \& F. M. da Silva et al. \cite{daSilva2022} further restricted the value of $\chi$ to $-0.02$ based on the joint constraint provided by massive pulsars and GW 170817. The negative values of $\chi$ used in this study range from $0$ to $-0.02$. Our paper utilizes a piecewise polytropic equation of state. The neutron star's core can be one of six EoSs: WFF1, MS1, AP3, ENG, MPA1, or MS1b, and SLy EoS represents the crust. These six EoSs give rise to neutron stars having mass greater than $2M_{\odot}$.}

\textbf{We present stellar configuration in Fig-\ref{fig:1} for six different EoSs. The right and left insets of each subfigure of Fig-\ref{fig:1} show the mass-radius and mass-central density relations, respectively. According to this figure, the maximum mass of NSs for any EoS of our study does not vary much with coupling parameter $\chi$. In this $f(R, T)$ gravity theory, the maximum mass increment is 0.41\% for WFF1, 0.23\% for MS1 and MS1b, and 0.2\% for AP3, ENG and MPA1 for coupling parameter $\chi=-0.02$. The relationship between mass and central density is not affected by coupling parameter $\chi$. Fig-\ref{fig:2} shows that the effect of coupling parameters on the compactness of stars decreases with increasing mass. When $1M_{\odot}$ NSs are considered, compactness decreases approximately by a maximum of 24-25\% in MS1 and MS1b, by 19-20\% in MS1b and by a minimum of 16\% in WFF1, and at the maximum mass associated with any EoS, it decreases by about 2-6\%.}

\textbf{Fig-\ref{fig:3} shows the rotational correction in mass (left inset) and in binding energy (right inset) as a function of a stellar mass for all six EoSs. We observe that the change in mass and binding energy are more sensitive to $\chi$ at the lower masses than that of the higher masses. The maximum value of rotational mass correction in GR occurs at $0.98M_{\odot}$, $1.25M_{\odot}$, $1.08M_{\odot}$, $1.02M_{\odot}$, $1.1M_{\odot}$ and $1.23M_{\odot}$ for WFF1, MS1, AP3, ENG, MPA1 and MS1b EoSs respectively. The maximal value of rotational mass correction in this $f(R, T)$ modified gravity is reduced significantly, and the occurrence of maximization of mass correction is shifted towards higher masses. For $\chi=-0.02$, the maximum mass correction is reduced by 26-29\%, and maximization occurs at 50-56\% higher mass. The energy correction in GR gets maximized at $0.84M_{\odot}$, $1.07M_{\odot}$, $0.91M_{\odot}$, $0.9M_{\odot}$, $0.94M_{\odot}$ and $1.07M_{\odot}$ for WFF1, MS1, AP3, ENG, MPA1 and MS1b EoSs respectively. The energy correction peaks at 45-58\% higher mass, and its maximum value is downsized by 31-32\%. We depict the correction in radius due to the rotation as a function of stellar mass in the left inset of Fig-\ref{fig:4}. The right inset of Fig-\ref{fig:4} shows how the eccentricity of the star varies with the stellar mass for all six EoSs. The rotational correction in radius depends on the $\chi$, but this dependency on $\chi$ reduces as the mass of the star increases. In GR, as well as in this $f(R, T)=R+2\chi T$ modified gravity, the radius correction monotonically decreases as the mass increases. The rotational correction in radius depends on the $\chi$, but this dependency on $\chi$ reduces as the mass of the star increases. The eccentricity attains a maximum at $0.98M_{\odot}$, $1.25M_{\odot}$, $1.07M_{\odot}$, $1.04M_{\odot}$, $1.1M_{\odot}$ and $1.26M_{\odot}$ for WFF1, MS1, AP3, ENG, MPA1 and MS1b EoSs respectively in GR. In this $f(R, T)$ modified gravity, the eccentricity's maximum value is reduced, and its occurrence is also shifted towards the higher masses. For $\chi=-0.02$, the maximal value of eccentricity is lessened by 7-9\% and is shifted 31-41\% towards higher masses. The dependency of eccentricity on $\chi$ decreases as we approach the higher masses corresponding to any EoS from $1 M_{\odot}$.}

\textbf{The quadrupole moment of the stars depends on the coupling parameters $\chi$ as shown in Fig-\ref{fig:5}. In GR, as well as in $f(R, T)$, the value of the Quadrupole moment increases as the mass of NS increases up to the maximal value of the quadrupole moment, and then its value decreases as the mass of the NS further increases. As the coupling parameter $\chi$ decreases, the value of the quadrupole moment decreases. For $\chi=-0.02$, the maximal value of the quadrupole moment is lessened by 10-14\%, and its occurrence is shifted to 2-6\% higher masses. Fig-\ref{fig:6} indicates that the moment of inertia is not so sensitive to $\chi$. Its value slightly increases as $\chi$ decreases. Its value slightly increases as $\chi$ decreases. Its maximal value in $f(R, T)$, for $\chi=-0.02$, increases $>2\%$, and its occurrence of the maximal of the moment of inertia is shifted $.0.3\%$ as compared to its GR counterparts. Fig-\ref{fig:7} shows how tidal love numbers are varied with stellar masses. The love number increases as the mass of the NS increases until its maximum, and then it decreases as the mass increases. The maximal value of the tidal love number is increased, and its position is shifted towards lighter masses in this $f(R, T)$ as compared to its GR counterpart. For $\chi=-0.02$, its maximal value is increased by 3-4\% and is shifted 0.5-4\% towards lighter masses. In this $f(R, T)$, the love numbers of the heavier stars are not much different from those of the GR. }

\textbf{Fig-\ref{fig:8} shows that the relation between the dimensionless moment of inertia and the dimensionless love numbers is independent of the coupling parameter $\chi$. This $\bar{I}-\bar{\lambda}$ relation deviates a maximum of 0.3\% for all six EoSs. Similarly, the dimensionless quadrupole moments and the dimensionless love numbers relationship are too insensitive to $\chi$ as shown in Fig-\ref{fig:9}. The maximum deviation of $\bar{Q}-\bar{\lambda}$ relation in this $f(R, T)$ for $\chi=-0.02$ from GR counterparts is around 0.7\% for all six EoSs. The same I-Love-Q relation of GR holds in this $f(R, T)=R+2\chi T$ modified theory of gravity. In the left upper inset of Fig-\ref{fig:10}, we plotted the dimensionless moment of inertia and love number for different EoSs with fixed value $\chi$ (in Fig-\ref{fig:10a} to  Fig-\ref{fig:10d} the $\chi$ is fixed at $0$(GR), $-0.001$, $-0.01$ and $-0.02$ ). The left lower inset of this figure shows the deviations of $\bar{I}-\bar{\lambda}$ relations of different EoSs from the relation obtained from data fitting within $f(R, T)=R+2\chi T$ modified gravity. It shows that the deviation $\bar{I}-\bar{\lambda}$ relations for different EoSs is always less than 0.5\% for all the values of $\chi$. We show $\bar{Q}-\bar{\lambda}$ relation for six distinct EoSs in the right upper inset of Fig-\ref{fig:10}. And the right lower inset of Fig-\ref{fig:10} shows the deviation of $\bar{Q}-\bar{\lambda}$ from the data fitted  $\bar{Q}-\bar{\lambda}$ relation. Again, we see the deviations of $\bar{Q}-\bar{\lambda}$ relations of different EoSs are always less than 0.5\% from the data fitted $\bar{Q}-\bar{\lambda}$ relation. In $f(R, T)$, we see the degeneracy in the I-Love-Q relation holds for different EoSs. }

\subsection{$f(R,T)=R+\alpha R^{2}+2\chi T$ Modified Gravity results}
\textbf{In $f(R, T)=R+\alpha R^{2}+2\chi T$ modified gravity theory, we investigate the slow rotating of neutron stars up to first order in their spin. To obtain the M-R and M-I relation in this modified gravity, we solve the system of equations (i.e., Eq.\ref{EQ:4A}-\ref{EQ:7A} and Eq.\ref{EQ:9A}) using six EoSs for the six pairs of $\alpha$ and $\chi$ values. For $\alpha=10^{-3}r_{g}$, we consider three different values of $\chi$: $\chi=0$, $\chi=-0.001$ and $\chi=-0.002$. Similarly, for $\alpha=10^{-2}r_{g}$, we take $\chi$: $\chi=0$, $\chi=-0.001$ and $\chi=-0.002$. In their study, Pretel et al. established a constraint on the value of $\alpha$ that guarantees stellar stability. Specifically, the value of $\alpha$ must be less than or equal to $10^{-2}r_{g}$ to ensure that the star remains stable \cite{Pretel:2020rqx}. We present the M-R relation for six EoSs in the left inset of Fig.\ref{fig:11}. NSs with mass less than $1.5M_{\odot}$ become more compact as alpha increases, while the opposite happens for heavier NSs across all six EoSs. The radius of NSs increases as $\chi$ decreases for a fixed value of $\alpha$. The mass-central density relation, as shown in the right inset of each panel of Fig.\ref{fig:11}, demonstrates that the mass of neutron stars decreases as the value of $\alpha$ increases, given a certain lower central density. However, above this certain lower central density, the mass of neutron stars increases as the value of $\alpha$ increases.The same findings were discovered in two other studies, namely \cite{Capozziello:2015yza} and \cite{Yazadjiev_2014}. On the other hand, the mass of neutron stars seems to be insensitive to the gravity coupling parameter $\chi$. The central density is restricted to a certain limit by the requirement of having a positive Ricci scalar value at the centre. The relationship between mass and moment of inertia, as offered in Fig.\ref{fig:12}, shows that the moment of inertia increases as the $\alpha$ increases for fixed $\chi$. Like $f(R, T)=R+2\chi T$ modified gravity, the moment of inertia in this modified gravity is not very sensitive to $\chi$.}

\section{Effect of $2\chi T$ term on the Neutron stars}
\textbf{In this paper, we study NS in $f(R, T)=R+2\chi T$ and $f(R, T)=R+\alpha R^{2}+2\chi T$ modified gravity. In this section, we will discuss the effect of the $2\chi T$ term on the NS compared to its GR counterparts. This kind of comparative study was done in the ref-\cite{Astashenok:2013vza,Astashenok:2014nua,Astashenok:2014dja,Astashenok:2017dpo} for $f(R)$ modified gravity where they took different forms of $f(R)$ and showed that NSs are very much dependent on the form of $f(R)$. Odinstov et al. did a similar comparative study for several inflationary models using different EoS \cite{PhysRevD.107.104039}. They found among all the EoS they considered, only MPA1 EoS is compatible with all the constraints produced by all the inflationary models they studied. In our study, we find that the $2\chi T$ term has a substantial effect on the radius of NS below a certain central density value in both $f(R, T)=R+2\chi T$ and $f(R, T)=R+\alpha R^{2}+2\chi T$ modified gravity. In contrast, the mass of the NS does not depend on the $2\chi T$ for a given central density. Hence, below a certain central density value, the compactness is greatly influenced by this $2\chi T$ term. Hence, below a certain central density value, the compactness is greatly influenced by this $2\chi T$ term. Such kind of results were found in ref-\cite{Lobato2020,Pretel2022} for $f(R, T)=R+2\chi T$. Because of the rotation, there is a correction in mass and binding energy. These corrections in mass and binding energy are also affected by the term $2\chi T$. The change in radius due to rotation and eccentricity also relies on the $2\chi T$ term below a certain NS mass. On the other side, the moment of inertia seems to be unaffected by the term $2\chi T$ for all the masses of NS. In the lower masses NS, the tidal love number and quadrupole moments deviate from GR because of the $2\chi T$ term of the modified gravity. Whereas, the I-Love-Q relationship does not differ from its GR peers. In the modified $f(R, T)=R+2\chi T$ gravity, the I-Love-Q relationship is independent of EoS, which we see in the case of GR. The results clearly indicate that the matter-geometric coupling parameter has a considerably greater impact on lighter neutron stars in both of these modified gravity models.}

\section{Conclusions}
\textbf{In this paper, we analyze the behavior of a slowly rotating neutron star in the context of $f(R, T)$ modified gravity. We calculate various relevant quantities to the second order in the angular velocity of the neutron star for $f(R, T)=R+2\chi T$ and to the first order in the angular velocity for $f(R, T)=R+\alpha R^{2}+2\chi T$. We compare all the results calculated in these two $f(R, T)$ modified gravity with their GR counterparts.}

\textbf{We find that the increment of the maximal mass in $f(R, T)=R+2\chi T$ for any EoS is below 0.4\%, and the compactness in $f(R, T)=R+2\chi T$ is less than that of their GR counterparts. For all six EoSs, the radius of NSs in the $f(R, T)=R+\alpha R^{2}+2\chi T$ theory is lower below 1.5$M_{\odot}$ and higher above 1.5$M_{\odot}$ than in their corresponding GR counterparts.}

\textbf{Our results show that the second-order correction, like radius and binding energy correction, are more sensitive to the coupling parameter $\chi$ of $f(R, T)$ in lighter NS than the heavier NS in $f(R, T)=R+2\chi T$. We observe that the quadrupole moment and tidal love number are less, whereas the moment of inertia is slightly higher than their corresponding GR counterparts in $f(R, T)=R+2\chi T$ modified gravity. Lobato et al. \cite{Lobato2020} found similar kinds of results in their study, where they showed that the increment of mass of NS in $f(R, T)$ is less than 1\%. Pretel \cite{Pretel2022} recently studied the slowly rotating anisotropic neutron stars in $f(R, T)=R+2\chi T$ modified gravity. They found that the gravitational mass and moment of inertia get slightly modified due to the influence of the effects generated by the minimal matter-gravity coupling. Whereas the anisotropic makes substantial changes in both the mass and the moment of inertia. F. M. da Silva found in their study of rotating NSs in $f(R, T)$ gravity that the moment of inertia at lower angular speed increases with the decreases of the coupling parameter $\chi$. Likewise, Our results show that the gravitational mass and the moment of inertia slightly increase as the coupling parameter $\chi$ decreases. The moment of inertia is not very sensitive to $\chi$ in $f(R, T)=R+\alpha R^{2}+2\chi T$ modified gravity, but it varies with $\alpha$.
We observe that the EoS independent I-Love-Q relation also holds in $f(R, T)=R+2\chi T $ gravity, and it coincides with the general relativistic ones within less than 0.5\% even for the maximum allowed coupling parameters. This I-Love-Q relationship also holds in the scalar-tensor theory, as shown by Pani and Berti \cite{Pani2014}.}

We calculate all the spin-dependent quantities for the Newtonian mass-shedding angular speed $\Omega^{*}=\sqrt{\frac{M}{R^{3}}}$. If desired, one can compute all the spin-dependent quantities of NS for any angular velocity $\Omega$ such that $\epsilon=\frac{\Omega}{\Omega^{*}}<<1$ by simply multiplying the quantities of the first order and second order in the spin with $\epsilon$ and $\epsilon^{2}$ respectively. da Silva et al. \cite{daSilva2022} studied Neutron stars with angular speeds of 300Hz and 716Hz  are very small compared to the Newtonian mass-shedding limit. Therefore, we can get similar results using the slow rotation approximation. Indeed, our findings are very close to the results found in da Silva et al.'s study \cite{daSilva2022}.

\textbf{We observe that all the quantities, more or less, have distinguishably diverse in $f(R, T)=R+2\chi T$ from general relativistic ones for the NSs having mass in the range $0.5M_{\odot}-1.5M_{\odot}$. Our galaxy is abundant in NSs having gravitational mass around $1.4M_{\odot}$, and most of these NSs rotate. Therefore, old neutron stars with a mass around $1.4M_{\odot}$ which rotate slowly relative to the newly born NS, are crucial in determining the coupling parameter of $f(R, T)$ gravity or could rule out the possibility of $f(R, T)$ gravity entirely. In this scenario, our study of slowly rotating NS in $f(R, T)$ gravity could be helpful.}

\section*{Acknowledgements}
MK is grateful to the Inter-University Centre for Astronomy and Astrophysics (IUCAA), Pune, India for providing Associateship programme under which a part of this work was carried out. 
\bibliographystyle{apsrev}
\bibliography{reference}% Produces the bibliography via BibTeX.

%apsrev4-2.bst 2015-08-30 from 4.21a (PWD, AO, DPC/HNN) hacked
\begin{thebibliography}{93}%
\makeatletter
\providecommand \@ifxundefined [1]{%
 \@ifx{#1\undefined}
}%
\providecommand \@ifnum [1]{%
 \ifnum #1\expandafter \@firstoftwo
 \else \expandafter \@secondoftwo
 \fi
}%
\providecommand \@ifx [1]{%
 \ifx #1\expandafter \@firstoftwo
 \else \expandafter \@secondoftwo
 \fi
}%
\providecommand \natexlab [1]{#1}%
\providecommand \enquote  [1]{``#1''}%
\providecommand \bibnamefont  [1]{#1}%
\providecommand \bibfnamefont [1]{#1}%
\providecommand \citenamefont [1]{#1}%
\providecommand \href@noop [0]{\@secondoftwo}%
\providecommand \href [0]{\begingroup \@sanitize@url \@href}%
\providecommand \@href[1]{\@@startlink{#1}\@@href}%
\providecommand \@@href[1]{\endgroup#1\@@endlink}%
\providecommand \@sanitize@url [0]{\catcode `\\12\catcode `\$12\catcode
  `\&12\catcode `\#12\catcode `\^12\catcode `\_12\catcode `\%12\relax}%
\providecommand \@@startlink[1]{}%
\providecommand \@@endlink[0]{}%
\providecommand \url  [0]{\begingroup\@sanitize@url \@url }%
\providecommand \@url [1]{\endgroup\@href {#1}{\urlprefix }}%
\providecommand \urlprefix  [0]{URL }%
\providecommand \Eprint [0]{\href }%
\providecommand \doibase [0]{http://dx.doi.org/}%
\providecommand \selectlanguage [0]{\@gobble}%
\providecommand \bibinfo  [0]{\@secondoftwo}%
\providecommand \bibfield  [0]{\@secondoftwo}%
\providecommand \translation [1]{[#1]}%
\providecommand \BibitemOpen [0]{}%
\providecommand \bibitemStop [0]{}%
\providecommand \bibitemNoStop [0]{.\EOS\space}%
\providecommand \EOS [0]{\spacefactor3000\relax}%
\providecommand \BibitemShut  [1]{\csname bibitem#1\endcsname}%
\let\auto@bib@innerbib\@empty
%</preamble>
\bibitem [{\citenamefont {Chrusciel}\ \emph {et~al.}(2012)\citenamefont
  {Chrusciel}, \citenamefont {Lopes~Costa},\ and\ \citenamefont
  {Heusler}}]{Chrusciel2012}%
  \BibitemOpen
  \bibfield  {author} {\bibinfo {author} {\bibfnamefont {P.~T.}\ \bibnamefont
  {Chrusciel}}, \bibinfo {author} {\bibfnamefont {J.}~\bibnamefont
  {Lopes~Costa}}, \ and\ \bibinfo {author} {\bibfnamefont {M.}~\bibnamefont
  {Heusler}},\ }\href {\doibase 10.12942/lrr-2012-7} {\bibfield  {journal}
  {\bibinfo  {journal} {Living Rev. Rel.}\ }\textbf {\bibinfo {volume} {15}},\
  \bibinfo {pages} {7} (\bibinfo {year} {2012})},\ \Eprint
  {http://arxiv.org/abs/1205.6112}{arXiv:1205.6112 [gr-qc]}\BibitemShut
  {NoStop}%
\bibitem [{\citenamefont {Lattimer}\ and\ \citenamefont
  {Prakash}(2004)}]{Lattimer&Prakash2004}%
  \BibitemOpen
  \bibfield  {author} {\bibinfo {author} {\bibfnamefont {J.~M.}\ \bibnamefont
  {Lattimer}}\ and\ \bibinfo {author} {\bibfnamefont {M.}~\bibnamefont
  {Prakash}},\ }\href {\doibase 10.1126/science.1090720} {\bibfield  {journal}
  {\bibinfo  {journal} {Science}\ }\textbf {\bibinfo {volume} {304}},\ \bibinfo
  {pages} {536} (\bibinfo {year} {2004})}\BibitemShut {NoStop}%
\bibitem [{\citenamefont {Lattimer}\ and\ \citenamefont
  {Schutz}(2005)}]{Lattimer2004}%
  \BibitemOpen
  \bibfield  {author} {\bibinfo {author} {\bibfnamefont {J.~M.}\ \bibnamefont
  {Lattimer}}\ and\ \bibinfo {author} {\bibfnamefont {B.~F.}\ \bibnamefont
  {Schutz}},\ }\href {\doibase 10.1086/431543} {\bibfield  {journal} {\bibinfo
  {journal} {Astrophys. J.}\ }\textbf {\bibinfo {volume} {629}},\ \bibinfo
  {pages} {979} (\bibinfo {year} {2005})},\ \Eprint
  {http://arxiv.org/abs/astro-ph/0411470}{arXiv:astro-ph/0411470}\BibitemShut
  {NoStop}%
\bibitem [{\citenamefont {Lattimer}\ and\ \citenamefont
  {Prakash}(2007)}]{Lattimer2007}%
  \BibitemOpen
  \bibfield  {author} {\bibinfo {author} {\bibfnamefont {J.~M.}\ \bibnamefont
  {Lattimer}}\ and\ \bibinfo {author} {\bibfnamefont {M.}~\bibnamefont
  {Prakash}},\ }\href {\doibase https://doi.org/10.1016/j.physrep.2007.02.003}
  {\bibfield  {journal} {\bibinfo  {journal} {Phys. Rep.}\ }\textbf {\bibinfo
  {volume} {442}},\ \bibinfo {pages} {109} (\bibinfo {year} {2007})},\ \bibinfo
  {note} {the Hans Bethe Centennial Volume 1906-2006}\BibitemShut {NoStop}%
\bibitem [{\citenamefont {\"Ozel}\ \emph {et~al.}(2010)\citenamefont {\"Ozel},
  \citenamefont {Baym},\ and\ \citenamefont {G\"uver}}]{Ozel2010}%
  \BibitemOpen
  \bibfield  {author} {\bibinfo {author} {\bibfnamefont {F.}~\bibnamefont
  {\"Ozel}}, \bibinfo {author} {\bibfnamefont {G.}~\bibnamefont {Baym}}, \ and\
  \bibinfo {author} {\bibfnamefont {T.}~\bibnamefont {G\"uver}},\ }\href
  {\doibase 10.1103/PhysRevD.82.101301} {\bibfield  {journal} {\bibinfo
  {journal} {Phys. Rev. D}\ }\textbf {\bibinfo {volume} {82}},\ \bibinfo
  {pages} {101301} (\bibinfo {year} {2010})}\BibitemShut {NoStop}%
\bibitem [{\citenamefont {{Steiner}}\ \emph {et~al.}(2010)\citenamefont
  {{Steiner}}, \citenamefont {{Lattimer}},\ and\ \citenamefont
  {{Brown}}}]{Steiner2010}%
  \BibitemOpen
  \bibfield  {author} {\bibinfo {author} {\bibfnamefont {A.~W.}\ \bibnamefont
  {{Steiner}}}, \bibinfo {author} {\bibfnamefont {J.~M.}\ \bibnamefont
  {{Lattimer}}}, \ and\ \bibinfo {author} {\bibfnamefont {E.~F.}\ \bibnamefont
  {{Brown}}},\ }\href {\doibase 10.1088/0004-637X/722/1/33} {\bibfield
  {journal} {\bibinfo  {journal} {Astrophys. J.}\ }\textbf {\bibinfo {volume}
  {722}},\ \bibinfo {pages} {33} (\bibinfo {year} {2010})},\ \Eprint
  {http://arxiv.org/abs/1005.0811}{arXiv:1005.0811 [astro-ph.HE]}\BibitemShut
  {NoStop}%
\bibitem [{\citenamefont {Psaltis}\ \emph {et~al.}(2014)\citenamefont
  {Psaltis}, \citenamefont {\"Ozel},\ and\ \citenamefont
  {Chakrabarty}}]{Psaltis2013}%
  \BibitemOpen
  \bibfield  {author} {\bibinfo {author} {\bibfnamefont {D.}~\bibnamefont
  {Psaltis}}, \bibinfo {author} {\bibfnamefont {F.}~\bibnamefont {\"Ozel}}, \
  and\ \bibinfo {author} {\bibfnamefont {D.}~\bibnamefont {Chakrabarty}},\
  }\href {\doibase 10.1088/0004-637X/787/2/136} {\bibfield  {journal} {\bibinfo
   {journal} {Astrophys. J.}\ }\textbf {\bibinfo {volume} {787}},\ \bibinfo
  {pages} {136} (\bibinfo {year} {2014})},\ \Eprint
  {http://arxiv.org/abs/1311.1571}{arXiv:1311.1571 [astro-ph.HE]}\BibitemShut
  {NoStop}%
\bibitem [{\citenamefont {Guver}\ and\ \citenamefont {Ozel}(2013)}]{Guver2013}%
  \BibitemOpen
  \bibfield  {author} {\bibinfo {author} {\bibfnamefont {T.}~\bibnamefont
  {Guver}}\ and\ \bibinfo {author} {\bibfnamefont {F.}~\bibnamefont {Ozel}},\
  }\href {\doibase 10.1088/2041-8205/765/1/L1} {\bibfield  {journal} {\bibinfo
  {journal} {Astrophys. J. Lett.}\ }\textbf {\bibinfo {volume} {765}},\
  \bibinfo {pages} {L1} (\bibinfo {year} {2013})},\ \Eprint
  {http://arxiv.org/abs/1301.0831}{arXiv:1301.0831 [astro-ph.HE]}\BibitemShut
  {NoStop}%
\bibitem [{\citenamefont {{Miller}}\ \emph {et~al.}(2020)\citenamefont
  {{Miller}}, \citenamefont {{Chirenti}},\ and\ \citenamefont
  {{Lamb}}}]{Miller2020}%
  \BibitemOpen
  \bibfield  {author} {\bibinfo {author} {\bibfnamefont {M.~C.}\ \bibnamefont
  {{Miller}}}, \bibinfo {author} {\bibfnamefont {C.}~\bibnamefont
  {{Chirenti}}}, \ and\ \bibinfo {author} {\bibfnamefont {F.~K.}\ \bibnamefont
  {{Lamb}}},\ }\href {\doibase 10.3847/1538-4357/ab4ef9} {\bibfield  {journal}
  {\bibinfo  {journal} {Astrophys. J.}\ }\textbf {\bibinfo {volume} {888}},\
  \bibinfo {eid} {12} (\bibinfo {year} {2020})},\ \Eprint
  {http://arxiv.org/abs/1904.08907}{arXiv:1904.08907 [astro-ph.HE]}\BibitemShut
  {NoStop}%
\bibitem [{\citenamefont {Hu}\ \emph {et~al.}(2020)\citenamefont {Hu},
  \citenamefont {Kramer}, \citenamefont {Wex}, \citenamefont {Champion},\ and\
  \citenamefont {Kehl}}]{Hu2020}%
  \BibitemOpen
  \bibfield  {author} {\bibinfo {author} {\bibfnamefont {H.}~\bibnamefont
  {Hu}}, \bibinfo {author} {\bibfnamefont {M.}~\bibnamefont {Kramer}}, \bibinfo
  {author} {\bibfnamefont {N.}~\bibnamefont {Wex}}, \bibinfo {author}
  {\bibfnamefont {D.~J.}\ \bibnamefont {Champion}}, \ and\ \bibinfo {author}
  {\bibfnamefont {M.~S.}\ \bibnamefont {Kehl}},\ }\href {\doibase
  10.1093/mnras/staa2107} {\bibfield  {journal} {\bibinfo  {journal} {Mon. Not.
  Roy. Astron. Soc.}\ }\textbf {\bibinfo {volume} {497}},\ \bibinfo {pages}
  {3118} (\bibinfo {year} {2020})},\ \Eprint
  {http://arxiv.org/abs/2007.07725}{arXiv:2007.07725 [astro-ph.SR]}\BibitemShut
  {NoStop}%
\bibitem [{\citenamefont {LIGO}()}]{LIGO}%
  \BibitemOpen
  \bibfield  {author} {\bibinfo {author} {\bibnamefont {LIGO}},\ }\href@noop {}
  {}\Eprint
  {http://arxiv.org/abs/https://www.ligo.caltech.edu}{https://www.ligo.caltech.edu}\BibitemShut
  {NoStop}%
\bibitem [{\citenamefont {VIRGO}()}]{VIRGO}%
  \BibitemOpen
  \bibfield  {author} {\bibinfo {author} {\bibnamefont {VIRGO}},\ }\href@noop
  {} {}\Eprint
  {http://arxiv.org/abs/https://www.virgo-gw.eu/}{https://www.virgo-gw.eu/}\BibitemShut
  {NoStop}%
\bibitem [{\citenamefont {KAGRA}()}]{KAGRA}%
  \BibitemOpen
  \bibfield  {author} {\bibinfo {author} {\bibnamefont {KAGRA}},\ }\href@noop
  {} {}\Eprint
  {http://arxiv.org/abs/https://gwcenter.icrr.u-tokyo.ac.jp/en/}{https://gwcenter.icrr.u-tokyo.ac.jp/en/}\BibitemShut
  {NoStop}%
\bibitem [{\citenamefont {Abbott}\ \emph {et~al.}(2017)\citenamefont {Abbott}
  \emph {et~al.}}]{Abbott2017}%
  \BibitemOpen
  \bibfield  {author} {\bibinfo {author} {\bibfnamefont {B.~P.}\ \bibnamefont
  {Abbott}} \emph {et~al.},\ }\href {\doibase 10.1103/PhysRevLett.119.161101}
  {\bibfield  {journal} {\bibinfo  {journal} {Phys. Rev. Lett.}\ }\textbf
  {\bibinfo {volume} {119}},\ \bibinfo {pages} {161101} (\bibinfo {year}
  {2017})},\ \Eprint {http://arxiv.org/abs/1710.05832}{arXiv:1710.05832
  [gr-qc]}\BibitemShut {NoStop}%
\bibitem [{\citenamefont {Will}(2006)}]{Will2005}%
  \BibitemOpen
  \bibfield  {author} {\bibinfo {author} {\bibfnamefont {C.~M.}\ \bibnamefont
  {Will}},\ }\href {\doibase 10.12942/lrr-2006-3} {\bibfield  {journal}
  {\bibinfo  {journal} {Living Rev. Rel.}\ }\textbf {\bibinfo {volume} {9}},\
  \bibinfo {pages} {3} (\bibinfo {year} {2006})},\ \Eprint
  {http://arxiv.org/abs/gr-qc/0510072}{arXiv:gr-qc/0510072}\BibitemShut
  {NoStop}%
\bibitem [{\citenamefont {{Hartle}}(1967)}]{Hartle1967}%
  \BibitemOpen
  \bibfield  {author} {\bibinfo {author} {\bibfnamefont {J.~B.}\ \bibnamefont
  {{Hartle}}},\ }\href {\doibase 10.1086/149400} {\bibfield  {journal}
  {\bibinfo  {journal} {\apj}\ }\textbf {\bibinfo {volume} {150}},\ \bibinfo
  {pages} {1005} (\bibinfo {year} {1967})}\BibitemShut {NoStop}%
\bibitem [{\citenamefont {{Hartle}}\ and\ \citenamefont
  {{Thorne}}(1968)}]{Hartle1968}%
  \BibitemOpen
  \bibfield  {author} {\bibinfo {author} {\bibfnamefont {J.~B.}\ \bibnamefont
  {{Hartle}}}\ and\ \bibinfo {author} {\bibfnamefont {K.~S.}\ \bibnamefont
  {{Thorne}}},\ }\href {\doibase 10.1086/149707} {\bibfield  {journal}
  {\bibinfo  {journal} {\apj}\ }\textbf {\bibinfo {volume} {153}},\ \bibinfo
  {pages} {807} (\bibinfo {year} {1968})}\BibitemShut {NoStop}%
\bibitem [{\citenamefont {Abramowicz}\ \emph {et~al.}(2003)\citenamefont
  {Abramowicz}, \citenamefont {Almergren}, \citenamefont {Kluzniak},\ and\
  \citenamefont {Thampan}}]{Abramowicz2003}%
  \BibitemOpen
  \bibfield  {author} {\bibinfo {author} {\bibfnamefont {M.~A.}\ \bibnamefont
  {Abramowicz}}, \bibinfo {author} {\bibfnamefont {G.~J.~E.}\ \bibnamefont
  {Almergren}}, \bibinfo {author} {\bibfnamefont {W.}~\bibnamefont {Kluzniak}},
  \ and\ \bibinfo {author} {\bibfnamefont {A.~V.}\ \bibnamefont {Thampan}},\
  }\href@noop {} {\  (\bibinfo {year} {2003})},\ \Eprint
  {http://arxiv.org/abs/gr-qc/0312070}{arXiv:gr-qc/0312070}\BibitemShut
  {NoStop}%
\bibitem [{\citenamefont {Manko}\ \emph
  {et~al.}(2000{\natexlab{a}})\citenamefont {Manko}, \citenamefont {Mielke},\
  and\ \citenamefont {Sanabria-Gomez}}]{Manko2000b}%
  \BibitemOpen
  \bibfield  {author} {\bibinfo {author} {\bibfnamefont {V.~S.}\ \bibnamefont
  {Manko}}, \bibinfo {author} {\bibfnamefont {E.~W.}\ \bibnamefont {Mielke}}, \
  and\ \bibinfo {author} {\bibfnamefont {J.~D.}\ \bibnamefont
  {Sanabria-Gomez}},\ }\href {\doibase 10.1103/PhysRevD.61.081501} {\bibfield
  {journal} {\bibinfo  {journal} {Phys. Rev. D}\ }\textbf {\bibinfo {volume}
  {61}},\ \bibinfo {pages} {081501} (\bibinfo {year} {2000}{\natexlab{a}})},\
  \Eprint {http://arxiv.org/abs/gr-qc/0001081}{arXiv:gr-qc/0001081}\BibitemShut
  {NoStop}%
\bibitem [{\citenamefont {Manko}\ \emph
  {et~al.}(2000{\natexlab{b}})\citenamefont {Manko}, \citenamefont
  {Sanabria-Gomez},\ and\ \citenamefont {Manko}}]{Manko2000a}%
  \BibitemOpen
  \bibfield  {author} {\bibinfo {author} {\bibfnamefont {V.~S.}\ \bibnamefont
  {Manko}}, \bibinfo {author} {\bibfnamefont {J.~D.}\ \bibnamefont
  {Sanabria-Gomez}}, \ and\ \bibinfo {author} {\bibfnamefont {O.~V.}\
  \bibnamefont {Manko}},\ }\href {\doibase 10.1103/PhysRevD.62.044048}
  {\bibfield  {journal} {\bibinfo  {journal} {Phys. Rev. D}\ }\textbf {\bibinfo
  {volume} {62}},\ \bibinfo {pages} {044048} (\bibinfo {year}
  {2000}{\natexlab{b}})}\BibitemShut {NoStop}%
\bibitem [{\citenamefont {{Sibgatullin}}\ and\ \citenamefont
  {{Queen}}(1991)}]{Sibgatullin1991}%
  \BibitemOpen
  \bibfield  {author} {\bibinfo {author} {\bibfnamefont {N.~R.}\ \bibnamefont
  {{Sibgatullin}}}\ and\ \bibinfo {author} {\bibfnamefont {N.~M.}\ \bibnamefont
  {{Queen}}},\ }\href@noop {} {\emph {\bibinfo {title} {{Oscillations and Waves
  in Strong Gravitational and Electromagnetic Fields}}}}\ (\bibinfo
  {publisher} {Springer-Verlag, Berlin},\ \bibinfo {year} {1991})\BibitemShut
  {NoStop}%
\bibitem [{\citenamefont {Berti}\ \emph {et~al.}(2005)\citenamefont {Berti},
  \citenamefont {White}, \citenamefont {Maniopoulou},\ and\ \citenamefont
  {Bruni}}]{Berti2005}%
  \BibitemOpen
  \bibfield  {author} {\bibinfo {author} {\bibfnamefont {E.}~\bibnamefont
  {Berti}}, \bibinfo {author} {\bibfnamefont {F.}~\bibnamefont {White}},
  \bibinfo {author} {\bibfnamefont {A.}~\bibnamefont {Maniopoulou}}, \ and\
  \bibinfo {author} {\bibfnamefont {M.}~\bibnamefont {Bruni}},\ }\href
  {\doibase 10.1111/j.1365-2966.2005.08812.x} {\bibfield  {journal} {\bibinfo
  {journal} {Mon. Not. Roy. Astron. Soc.}\ }\textbf {\bibinfo {volume} {358}},\
  \bibinfo {pages} {923} (\bibinfo {year} {2005})}\BibitemShut {NoStop}%
\bibitem [{\citenamefont {Hinderer}(2008)}]{Hinderer2008}%
  \BibitemOpen
  \bibfield  {author} {\bibinfo {author} {\bibfnamefont {T.}~\bibnamefont
  {Hinderer}},\ }\href {\doibase 10.1086/533487} {\bibfield  {journal}
  {\bibinfo  {journal} {Astrophys. J.}\ }\textbf {\bibinfo {volume} {677}},\
  \bibinfo {pages} {1216} (\bibinfo {year} {2008})}\BibitemShut {NoStop}%
\bibitem [{\citenamefont {Flanagan}\ and\ \citenamefont
  {Hinderer}(2008)}]{Flanagan2008}%
  \BibitemOpen
  \bibfield  {author} {\bibinfo {author} {\bibfnamefont {E.~E.}\ \bibnamefont
  {Flanagan}}\ and\ \bibinfo {author} {\bibfnamefont {T.}~\bibnamefont
  {Hinderer}},\ }\href {\doibase 10.1103/PhysRevD.77.021502} {\bibfield
  {journal} {\bibinfo  {journal} {Phys. Rev. D}\ }\textbf {\bibinfo {volume}
  {77}},\ \bibinfo {pages} {021502} (\bibinfo {year} {2008})}\BibitemShut
  {NoStop}%
\bibitem [{\citenamefont {Binnington}\ and\ \citenamefont
  {Poisson}(2009)}]{Binnington2009}%
  \BibitemOpen
  \bibfield  {author} {\bibinfo {author} {\bibfnamefont {T.}~\bibnamefont
  {Binnington}}\ and\ \bibinfo {author} {\bibfnamefont {E.}~\bibnamefont
  {Poisson}},\ }\href {\doibase 10.1103/PhysRevD.80.084018} {\bibfield
  {journal} {\bibinfo  {journal} {Phys. Rev. D}\ }\textbf {\bibinfo {volume}
  {80}},\ \bibinfo {pages} {084018} (\bibinfo {year} {2009})}\BibitemShut
  {NoStop}%
\bibitem [{\citenamefont {Yagi}\ and\ \citenamefont {Yunes}(2013)}]{Yagi2013}%
  \BibitemOpen
  \bibfield  {author} {\bibinfo {author} {\bibfnamefont {K.}~\bibnamefont
  {Yagi}}\ and\ \bibinfo {author} {\bibfnamefont {N.}~\bibnamefont {Yunes}},\
  }\href {\doibase 10.1103/PhysRevD.88.023009} {\bibfield  {journal} {\bibinfo
  {journal} {Phys. Rev. D}\ }\textbf {\bibinfo {volume} {88}},\ \bibinfo
  {pages} {023009} (\bibinfo {year} {2013})}\BibitemShut {NoStop}%
\bibitem [{\citenamefont {Clifton}\ \emph {et~al.}(2012)\citenamefont
  {Clifton}, \citenamefont {Ferreira}, \citenamefont {Padilla},\ and\
  \citenamefont {Skordis}}]{Clifton2011}%
  \BibitemOpen
  \bibfield  {author} {\bibinfo {author} {\bibfnamefont {T.}~\bibnamefont
  {Clifton}}, \bibinfo {author} {\bibfnamefont {P.~G.}\ \bibnamefont
  {Ferreira}}, \bibinfo {author} {\bibfnamefont {A.}~\bibnamefont {Padilla}}, \
  and\ \bibinfo {author} {\bibfnamefont {C.}~\bibnamefont {Skordis}},\ }\href
  {\doibase 10.1016/j.physrep.2012.01.001} {\bibfield  {journal} {\bibinfo
  {journal} {Phys. Rept.}\ }\textbf {\bibinfo {volume} {513}},\ \bibinfo
  {pages} {1} (\bibinfo {year} {2012})},\ \Eprint
  {http://arxiv.org/abs/1106.2476}{arXiv:1106.2476 [astro-ph.CO]}\BibitemShut
  {NoStop}%
\bibitem [{\citenamefont {Nojiri}\ and\ \citenamefont
  {Odintsov}(2011)}]{Nojiri:2010wj}%
  \BibitemOpen
  \bibfield  {author} {\bibinfo {author} {\bibfnamefont {S.}~\bibnamefont
  {Nojiri}}\ and\ \bibinfo {author} {\bibfnamefont {S.~D.}\ \bibnamefont
  {Odintsov}},\ }\href {\doibase 10.1016/j.physrep.2011.04.001} {\bibfield
  {journal} {\bibinfo  {journal} {Phys. Rept.}\ }\textbf {\bibinfo {volume}
  {505}},\ \bibinfo {pages} {59} (\bibinfo {year} {2011})},\ \Eprint
  {http://arxiv.org/abs/1011.0544}{arXiv:1011.0544 [gr-qc]}\BibitemShut
  {NoStop}%
\bibitem [{\citenamefont {Nojiri}\ \emph {et~al.}(2017)\citenamefont {Nojiri},
  \citenamefont {Odintsov},\ and\ \citenamefont {Oikonomou}}]{Nojiri2017}%
  \BibitemOpen
  \bibfield  {author} {\bibinfo {author} {\bibfnamefont {S.}~\bibnamefont
  {Nojiri}}, \bibinfo {author} {\bibfnamefont {S.~D.}\ \bibnamefont
  {Odintsov}}, \ and\ \bibinfo {author} {\bibfnamefont {V.~K.}\ \bibnamefont
  {Oikonomou}},\ }\href {\doibase 10.1016/j.physrep.2017.06.001} {\bibfield
  {journal} {\bibinfo  {journal} {Phys. Rept.}\ }\textbf {\bibinfo {volume}
  {692}},\ \bibinfo {pages} {1} (\bibinfo {year} {2017})},\ \Eprint
  {http://arxiv.org/abs/1705.11098}{arXiv:1705.11098 [gr-qc]}\BibitemShut
  {NoStop}%
\bibitem [{\citenamefont {Berti}\ \emph {et~al.}(2015)\citenamefont {Berti}
  \emph {et~al.}}]{Berti2015}%
  \BibitemOpen
  \bibfield  {author} {\bibinfo {author} {\bibfnamefont {E.}~\bibnamefont
  {Berti}} \emph {et~al.},\ }\href {\doibase 10.1088/0264-9381/32/24/243001}
  {\bibfield  {journal} {\bibinfo  {journal} {Class. Quant. Grav.}\ }\textbf
  {\bibinfo {volume} {32}},\ \bibinfo {pages} {243001} (\bibinfo {year}
  {2015})},\ \Eprint {http://arxiv.org/abs/1501.07274}{arXiv:1501.07274
  [gr-qc]}\BibitemShut {NoStop}%
\bibitem [{\citenamefont {Silva}\ \emph {et~al.}(2016)\citenamefont {Silva},
  \citenamefont {Maselli}, \citenamefont {Minamitsuji},\ and\ \citenamefont
  {Berti}}]{Silva2016}%
  \BibitemOpen
  \bibfield  {author} {\bibinfo {author} {\bibfnamefont {H.~O.}\ \bibnamefont
  {Silva}}, \bibinfo {author} {\bibfnamefont {A.}~\bibnamefont {Maselli}},
  \bibinfo {author} {\bibfnamefont {M.}~\bibnamefont {Minamitsuji}}, \ and\
  \bibinfo {author} {\bibfnamefont {E.}~\bibnamefont {Berti}},\ }\href
  {\doibase 10.1142/S0218271816410066} {\bibfield  {journal} {\bibinfo
  {journal} {Int. J. Mod. Phys. D}\ }\textbf {\bibinfo {volume} {25}},\
  \bibinfo {pages} {1641006} (\bibinfo {year} {2016})},\ \Eprint
  {http://arxiv.org/abs/1602.05997}{arXiv:1602.05997 [gr-qc]}\BibitemShut
  {NoStop}%
\bibitem [{\citenamefont {Wu}\ \emph {et~al.}(2018)\citenamefont {Wu},
  \citenamefont {Li}, \citenamefont {Harko},\ and\ \citenamefont
  {Liang}}]{Wu2018}%
  \BibitemOpen
  \bibfield  {author} {\bibinfo {author} {\bibfnamefont {J.}~\bibnamefont
  {Wu}}, \bibinfo {author} {\bibfnamefont {G.}~\bibnamefont {Li}}, \bibinfo
  {author} {\bibfnamefont {T.}~\bibnamefont {Harko}}, \ and\ \bibinfo {author}
  {\bibfnamefont {S.-D.}\ \bibnamefont {Liang}},\ }\href {\doibase
  10.1140/epjc/s10052-018-5923-9} {\bibfield  {journal} {\bibinfo  {journal}
  {Eur. Phys. J. C}\ }\textbf {\bibinfo {volume} {78}},\ \bibinfo {pages} {430}
  (\bibinfo {year} {2018})},\ \Eprint
  {http://arxiv.org/abs/1805.07419}{arXiv:1805.07419 [gr-qc]}\BibitemShut
  {NoStop}%
\bibitem [{\citenamefont {De~Felice}\ and\ \citenamefont
  {Tsujikawa}(2010)}]{DeFelice2010}%
  \BibitemOpen
  \bibfield  {author} {\bibinfo {author} {\bibfnamefont {A.}~\bibnamefont
  {De~Felice}}\ and\ \bibinfo {author} {\bibfnamefont {S.}~\bibnamefont
  {Tsujikawa}},\ }\href {\doibase 10.12942/lrr-2010-3} {\bibfield  {journal}
  {\bibinfo  {journal} {Living Rev. Rel.}\ }\textbf {\bibinfo {volume} {13}},\
  \bibinfo {pages} {3} (\bibinfo {year} {2010})},\ \Eprint
  {http://arxiv.org/abs/1002.4928}{arXiv:1002.4928 [gr-qc]}\BibitemShut
  {NoStop}%
\bibitem [{\citenamefont {Sotiriou}\ and\ \citenamefont
  {Faraoni}(2010)}]{Sotiriou2008}%
  \BibitemOpen
  \bibfield  {author} {\bibinfo {author} {\bibfnamefont {T.~P.}\ \bibnamefont
  {Sotiriou}}\ and\ \bibinfo {author} {\bibfnamefont {V.}~\bibnamefont
  {Faraoni}},\ }\href {\doibase 10.1103/RevModPhys.82.451} {\bibfield
  {journal} {\bibinfo  {journal} {Rev. Mod. Phys.}\ }\textbf {\bibinfo {volume}
  {82}},\ \bibinfo {pages} {451} (\bibinfo {year} {2010})},\ \Eprint
  {http://arxiv.org/abs/0805.1726}{arXiv:0805.1726 [gr-qc]}\BibitemShut
  {NoStop}%
\bibitem [{\citenamefont {Lobo}(2008)}]{Lobo2008}%
  \BibitemOpen
  \bibfield  {author} {\bibinfo {author} {\bibfnamefont {F.~S.~N.}\
  \bibnamefont {Lobo}},\ }\href@noop {} {\  (\bibinfo {year} {2008})},\ \Eprint
  {http://arxiv.org/abs/0807.1640}{arXiv:0807.1640 [gr-qc]}\BibitemShut
  {NoStop}%
\bibitem [{\citenamefont {Astashenok}\ \emph {et~al.}(2013)\citenamefont
  {Astashenok}, \citenamefont {Capozziello},\ and\ \citenamefont
  {Odintsov}}]{Astashenok:2013vza}%
  \BibitemOpen
  \bibfield  {author} {\bibinfo {author} {\bibfnamefont {A.~V.}\ \bibnamefont
  {Astashenok}}, \bibinfo {author} {\bibfnamefont {S.}~\bibnamefont
  {Capozziello}}, \ and\ \bibinfo {author} {\bibfnamefont {S.~D.}\ \bibnamefont
  {Odintsov}},\ }\href {\doibase 10.1088/1475-7516/2013/12/040} {\bibfield
  {journal} {\bibinfo  {journal} {J. Cosmol. Astropart. Phys.}\ }\textbf
  {\bibinfo {volume} {12}},\ \bibinfo {pages} {040} (\bibinfo {year} {2013})},\
  \Eprint {http://arxiv.org/abs/1309.1978}{arXiv:1309.1978 [gr-qc]}\BibitemShut
  {NoStop}%
\bibitem [{\citenamefont {Astashenok}\ \emph
  {et~al.}(2015{\natexlab{a}})\citenamefont {Astashenok}, \citenamefont
  {Capozziello},\ and\ \citenamefont {Odintsov}}]{Astashenok:2014nua}%
  \BibitemOpen
  \bibfield  {author} {\bibinfo {author} {\bibfnamefont {A.~V.}\ \bibnamefont
  {Astashenok}}, \bibinfo {author} {\bibfnamefont {S.}~\bibnamefont
  {Capozziello}}, \ and\ \bibinfo {author} {\bibfnamefont {S.~D.}\ \bibnamefont
  {Odintsov}},\ }\href {\doibase 10.1088/1475-7516/2015/01/001} {\bibfield
  {journal} {\bibinfo  {journal} {J. Cosmol. Astropart. Phys.}\ }\textbf
  {\bibinfo {volume} {01}},\ \bibinfo {pages} {001} (\bibinfo {year}
  {2015}{\natexlab{a}})},\ \Eprint
  {http://arxiv.org/abs/1408.3856}{arXiv:1408.3856 [gr-qc]}\BibitemShut
  {NoStop}%
\bibitem [{\citenamefont {Astashenok}\ \emph
  {et~al.}(2015{\natexlab{b}})\citenamefont {Astashenok}, \citenamefont
  {Capozziello},\ and\ \citenamefont {Odintsov}}]{Astashenok:2014dja}%
  \BibitemOpen
  \bibfield  {author} {\bibinfo {author} {\bibfnamefont {A.~V.}\ \bibnamefont
  {Astashenok}}, \bibinfo {author} {\bibfnamefont {S.}~\bibnamefont
  {Capozziello}}, \ and\ \bibinfo {author} {\bibfnamefont {S.~D.}\ \bibnamefont
  {Odintsov}},\ }\href {\doibase 10.1016/j.physletb.2015.01.030} {\bibfield
  {journal} {\bibinfo  {journal} {Phys. Lett. B}\ }\textbf {\bibinfo {volume}
  {742}},\ \bibinfo {pages} {160} (\bibinfo {year} {2015}{\natexlab{b}})},\
  \Eprint {http://arxiv.org/abs/1412.5453}{arXiv:1412.5453 [gr-qc]}\BibitemShut
  {NoStop}%
\bibitem [{\citenamefont {Astashenok}\ \emph {et~al.}(2017)\citenamefont
  {Astashenok}, \citenamefont {Odintsov},\ and\ \citenamefont {de~la
  Cruz-Dombriz}}]{Astashenok:2017dpo}%
  \BibitemOpen
  \bibfield  {author} {\bibinfo {author} {\bibfnamefont {A.~V.}\ \bibnamefont
  {Astashenok}}, \bibinfo {author} {\bibfnamefont {S.~D.}\ \bibnamefont
  {Odintsov}}, \ and\ \bibinfo {author} {\bibfnamefont {A.}~\bibnamefont {de~la
  Cruz-Dombriz}},\ }\href {\doibase 10.1088/1361-6382/aa8971} {\bibfield
  {journal} {\bibinfo  {journal} {Class. Quant. Grav.}\ }\textbf {\bibinfo
  {volume} {34}},\ \bibinfo {pages} {205008} (\bibinfo {year} {2017})},\
  \Eprint {http://arxiv.org/abs/1704.08311}{arXiv:1704.08311
  [gr-qc]}\BibitemShut {NoStop}%
\bibitem [{\citenamefont {Astashenok}\ and\ \citenamefont
  {Odintsov}(2020{\natexlab{a}})}]{Astashenok:2020cfv}%
  \BibitemOpen
  \bibfield  {author} {\bibinfo {author} {\bibfnamefont {A.~V.}\ \bibnamefont
  {Astashenok}}\ and\ \bibinfo {author} {\bibfnamefont {S.~D.}\ \bibnamefont
  {Odintsov}},\ }\href {\doibase 10.1093/mnras/staa214} {\bibfield  {journal}
  {\bibinfo  {journal} {Mon. Not. Roy. Astron. Soc.}\ }\textbf {\bibinfo
  {volume} {493}},\ \bibinfo {pages} {78} (\bibinfo {year}
  {2020}{\natexlab{a}})},\ \Eprint
  {http://arxiv.org/abs/2001.08504}{arXiv:2001.08504 [gr-qc]}\BibitemShut
  {NoStop}%
\bibitem [{\citenamefont {Astashenok}\ and\ \citenamefont
  {Odintsov}(2020{\natexlab{b}})}]{Astashenok:2020cqq}%
  \BibitemOpen
  \bibfield  {author} {\bibinfo {author} {\bibfnamefont {A.~V.}\ \bibnamefont
  {Astashenok}}\ and\ \bibinfo {author} {\bibfnamefont {S.~D.}\ \bibnamefont
  {Odintsov}},\ }\href {\doibase 10.1093/mnras/staa2630} {\bibfield  {journal}
  {\bibinfo  {journal} {Mon. Not. Roy. Astron. Soc.}\ }\textbf {\bibinfo
  {volume} {498}},\ \bibinfo {pages} {3616} (\bibinfo {year}
  {2020}{\natexlab{b}})},\ \Eprint
  {http://arxiv.org/abs/2008.11271}{arXiv:2008.11271 [gr-qc]}\BibitemShut
  {NoStop}%
\bibitem [{\citenamefont {Astashenok}\ \emph {et~al.}(2020)\citenamefont
  {Astashenok}, \citenamefont {Capozziello}, \citenamefont {Odintsov},\ and\
  \citenamefont {Oikonomou}}]{Astashenok:2020qds}%
  \BibitemOpen
  \bibfield  {author} {\bibinfo {author} {\bibfnamefont {A.~V.}\ \bibnamefont
  {Astashenok}}, \bibinfo {author} {\bibfnamefont {S.}~\bibnamefont
  {Capozziello}}, \bibinfo {author} {\bibfnamefont {S.~D.}\ \bibnamefont
  {Odintsov}}, \ and\ \bibinfo {author} {\bibfnamefont {V.~K.}\ \bibnamefont
  {Oikonomou}},\ }\href {\doibase 10.1016/j.physletb.2020.135910} {\bibfield
  {journal} {\bibinfo  {journal} {Phys. Lett. B}\ }\textbf {\bibinfo {volume}
  {811}},\ \bibinfo {pages} {135910} (\bibinfo {year} {2020})},\ \Eprint
  {http://arxiv.org/abs/2008.10884}{arXiv:2008.10884 [gr-qc]}\BibitemShut
  {NoStop}%
\bibitem [{\citenamefont {Astashenok}\ \emph
  {et~al.}(2021{\natexlab{a}})\citenamefont {Astashenok}, \citenamefont
  {Capozziello}, \citenamefont {Odintsov},\ and\ \citenamefont
  {Oikonomou}}]{Astashenok:2021xpm}%
  \BibitemOpen
  \bibfield  {author} {\bibinfo {author} {\bibfnamefont {A.~V.}\ \bibnamefont
  {Astashenok}}, \bibinfo {author} {\bibfnamefont {S.}~\bibnamefont
  {Capozziello}}, \bibinfo {author} {\bibfnamefont {S.~D.}\ \bibnamefont
  {Odintsov}}, \ and\ \bibinfo {author} {\bibfnamefont {V.~K.}\ \bibnamefont
  {Oikonomou}},\ }\href {\doibase 10.1209/0295-5075/134/59001} {\bibfield
  {journal} {\bibinfo  {journal} {EPL}\ }\textbf {\bibinfo {volume} {134}},\
  \bibinfo {pages} {59001} (\bibinfo {year} {2021}{\natexlab{a}})},\ \Eprint
  {http://arxiv.org/abs/2106.01234}{arXiv:2106.01234 [gr-qc]}\BibitemShut
  {NoStop}%
\bibitem [{\citenamefont {Astashenok}\ \emph
  {et~al.}(2021{\natexlab{b}})\citenamefont {Astashenok}, \citenamefont
  {Capozziello}, \citenamefont {Odintsov},\ and\ \citenamefont
  {Oikonomou}}]{Astashenok:2021btj}%
  \BibitemOpen
  \bibfield  {author} {\bibinfo {author} {\bibfnamefont {A.~V.}\ \bibnamefont
  {Astashenok}}, \bibinfo {author} {\bibfnamefont {S.}~\bibnamefont
  {Capozziello}}, \bibinfo {author} {\bibfnamefont {S.~D.}\ \bibnamefont
  {Odintsov}}, \ and\ \bibinfo {author} {\bibfnamefont {V.~K.}\ \bibnamefont
  {Oikonomou}},\ }\href {\doibase 10.1209/0295-5075/ac3d6c} {\bibfield
  {journal} {\bibinfo  {journal} {EPL}\ }\textbf {\bibinfo {volume} {136}},\
  \bibinfo {pages} {59001} (\bibinfo {year} {2021}{\natexlab{b}})},\ \Eprint
  {http://arxiv.org/abs/2111.14179}{arXiv:2111.14179 [gr-qc]}\BibitemShut
  {NoStop}%
\bibitem [{\citenamefont {Astashenok}\ \emph
  {et~al.}(2021{\natexlab{c}})\citenamefont {Astashenok}, \citenamefont
  {Capozziello}, \citenamefont {Odintsov},\ and\ \citenamefont
  {Oikonomou}}]{Astashenok:2021peo}%
  \BibitemOpen
  \bibfield  {author} {\bibinfo {author} {\bibfnamefont {A.~V.}\ \bibnamefont
  {Astashenok}}, \bibinfo {author} {\bibfnamefont {S.}~\bibnamefont
  {Capozziello}}, \bibinfo {author} {\bibfnamefont {S.~D.}\ \bibnamefont
  {Odintsov}}, \ and\ \bibinfo {author} {\bibfnamefont {V.~K.}\ \bibnamefont
  {Oikonomou}},\ }\href {\doibase 10.1016/j.physletb.2021.136222} {\bibfield
  {journal} {\bibinfo  {journal} {Phys. Lett. B}\ }\textbf {\bibinfo {volume}
  {816}},\ \bibinfo {pages} {136222} (\bibinfo {year} {2021}{\natexlab{c}})},\
  \Eprint {http://arxiv.org/abs/2103.04144}{arXiv:2103.04144
  [gr-qc]}\BibitemShut {NoStop}%
\bibitem [{\citenamefont {Harko}\ \emph {et~al.}(2011)\citenamefont {Harko},
  \citenamefont {Lobo}, \citenamefont {Nojiri},\ and\ \citenamefont
  {Odintsov}}]{Harko2011}%
  \BibitemOpen
  \bibfield  {author} {\bibinfo {author} {\bibfnamefont {T.}~\bibnamefont
  {Harko}}, \bibinfo {author} {\bibfnamefont {F.~S.~N.}\ \bibnamefont {Lobo}},
  \bibinfo {author} {\bibfnamefont {S.}~\bibnamefont {Nojiri}}, \ and\ \bibinfo
  {author} {\bibfnamefont {S.~D.}\ \bibnamefont {Odintsov}},\ }\href {\doibase
  10.1103/PhysRevD.84.024020} {\bibfield  {journal} {\bibinfo  {journal} {Phys.
  Rev. D}\ }\textbf {\bibinfo {volume} {84}},\ \bibinfo {pages} {024020}
  (\bibinfo {year} {2011})}\BibitemShut {NoStop}%
\bibitem [{\citenamefont {Debnath}(2018)}]{Debnath:2018wct}%
  \BibitemOpen
  \bibfield  {author} {\bibinfo {author} {\bibfnamefont {P.~S.}\ \bibnamefont
  {Debnath}},\ }\href {\doibase 10.1142/S0219887819500051} {\bibfield
  {journal} {\bibinfo  {journal} {Int. J. Geom. Meth. Mod. Phys.}\ }\textbf
  {\bibinfo {volume} {16}},\ \bibinfo {pages} {1950005} (\bibinfo {year}
  {2018})},\ \Eprint {http://arxiv.org/abs/1907.02238}{arXiv:1907.02238
  [gr-qc]}\BibitemShut {NoStop}%
\bibitem [{\citenamefont {Bhattacharjee}\ and\ \citenamefont
  {Sahoo}(2020)}]{Bhattacharjee:2020eec}%
  \BibitemOpen
  \bibfield  {author} {\bibinfo {author} {\bibfnamefont {S.}~\bibnamefont
  {Bhattacharjee}}\ and\ \bibinfo {author} {\bibfnamefont {P.~K.}\ \bibnamefont
  {Sahoo}},\ }\href {\doibase 10.1016/j.dark.2020.100537} {\bibfield  {journal}
  {\bibinfo  {journal} {Phys. Dark Univ.}\ }\textbf {\bibinfo {volume} {28}},\
  \bibinfo {pages} {100537} (\bibinfo {year} {2020})},\ \Eprint
  {http://arxiv.org/abs/2003.14211}{arXiv:2003.14211 [gr-qc]}\BibitemShut
  {NoStop}%
\bibitem [{\citenamefont {Bhattacharjee}\ \emph {et~al.}(2020)\citenamefont
  {Bhattacharjee}, \citenamefont {Santos}, \citenamefont {Moraes},\ and\
  \citenamefont {Sahoo}}]{Bhattacharjee:2020jsf}%
  \BibitemOpen
  \bibfield  {author} {\bibinfo {author} {\bibfnamefont {S.}~\bibnamefont
  {Bhattacharjee}}, \bibinfo {author} {\bibfnamefont {J.~R.~L.}\ \bibnamefont
  {Santos}}, \bibinfo {author} {\bibfnamefont {P.~H. R.~S.}\ \bibnamefont
  {Moraes}}, \ and\ \bibinfo {author} {\bibfnamefont {P.~K.}\ \bibnamefont
  {Sahoo}},\ }\href {\doibase 10.1140/epjp/s13360-020-00583-6} {\bibfield
  {journal} {\bibinfo  {journal} {Eur. Phys. J. Plus}\ }\textbf {\bibinfo
  {volume} {135}},\ \bibinfo {pages} {576} (\bibinfo {year} {2020})},\ \Eprint
  {http://arxiv.org/abs/2006.04336}{arXiv:2006.04336 [gr-qc]}\BibitemShut
  {NoStop}%
\bibitem [{\citenamefont {Gamonal}(2021)}]{Gamonal:2020itt}%
  \BibitemOpen
  \bibfield  {author} {\bibinfo {author} {\bibfnamefont {M.}~\bibnamefont
  {Gamonal}},\ }\href {\doibase 10.1016/j.dark.2020.100768} {\bibfield
  {journal} {\bibinfo  {journal} {Phys. Dark Univ.}\ }\textbf {\bibinfo
  {volume} {31}},\ \bibinfo {pages} {100768} (\bibinfo {year} {2021})},\
  \Eprint {http://arxiv.org/abs/2010.03861}{arXiv:2010.03861
  [gr-qc]}\BibitemShut {NoStop}%
\bibitem [{\citenamefont {Shabani}\ and\ \citenamefont
  {Ziaie}(2018)}]{Shabani:2017kis}%
  \BibitemOpen
  \bibfield  {author} {\bibinfo {author} {\bibfnamefont {H.}~\bibnamefont
  {Shabani}}\ and\ \bibinfo {author} {\bibfnamefont {A.~H.}\ \bibnamefont
  {Ziaie}},\ }\href {\doibase 10.1140/epjc/s10052-018-5886-x} {\bibfield
  {journal} {\bibinfo  {journal} {Eur. Phys. J. C}\ }\textbf {\bibinfo {volume}
  {78}},\ \bibinfo {pages} {397} (\bibinfo {year} {2018})},\ \Eprint
  {http://arxiv.org/abs/1708.07874}{arXiv:1708.07874 [gr-qc]}\BibitemShut
  {NoStop}%
\bibitem [{\citenamefont {Moraes}\ \emph {et~al.}(2016)\citenamefont {Moraes},
  \citenamefont {Arba\~nil},\ and\ \citenamefont {Malheiro}}]{Moraes:2015uxq}%
  \BibitemOpen
  \bibfield  {author} {\bibinfo {author} {\bibfnamefont {P.~H. R.~S.}\
  \bibnamefont {Moraes}}, \bibinfo {author} {\bibfnamefont {J.~D.~V.}\
  \bibnamefont {Arba\~nil}}, \ and\ \bibinfo {author} {\bibfnamefont
  {M.}~\bibnamefont {Malheiro}},\ }\href {\doibase
  10.1088/1475-7516/2016/06/005} {\bibfield  {journal} {\bibinfo  {journal}
  {JCAP}\ }\textbf {\bibinfo {volume} {06}},\ \bibinfo {pages} {005} (\bibinfo
  {year} {2016})},\ \Eprint {http://arxiv.org/abs/1511.06282}{arXiv:1511.06282
  [gr-qc]}\BibitemShut {NoStop}%
\bibitem [{\citenamefont {Lobato}\ \emph {et~al.}(2020)\citenamefont {Lobato},
  \citenamefont {Lourenço}, \citenamefont {Moraes}, \citenamefont {Lenzi},
  \citenamefont {de~Avellar}, \citenamefont {de~Paula}, \citenamefont {Dutra},\
  and\ \citenamefont {Malheiro}}]{Lobato2020}%
  \BibitemOpen
  \bibfield  {author} {\bibinfo {author} {\bibfnamefont {R.}~\bibnamefont
  {Lobato}}, \bibinfo {author} {\bibfnamefont {O.}~\bibnamefont {Lourenço}},
  \bibinfo {author} {\bibfnamefont {P.}~\bibnamefont {Moraes}}, \bibinfo
  {author} {\bibfnamefont {C.}~\bibnamefont {Lenzi}}, \bibinfo {author}
  {\bibfnamefont {M.}~\bibnamefont {de~Avellar}}, \bibinfo {author}
  {\bibfnamefont {W.}~\bibnamefont {de~Paula}}, \bibinfo {author}
  {\bibfnamefont {M.}~\bibnamefont {Dutra}}, \ and\ \bibinfo {author}
  {\bibfnamefont {M.}~\bibnamefont {Malheiro}},\ }\href {\doibase
  10.1088/1475-7516/2020/12/039} {\bibfield  {journal} {\bibinfo  {journal} {J.
  Cosmol. Astropart. Phys.}\ }\textbf {\bibinfo {volume} {2020}},\ \bibinfo
  {pages} {039} (\bibinfo {year} {2020})}\BibitemShut {NoStop}%
\bibitem [{\citenamefont {da~Silva}\ \emph {et~al.}(2023)\citenamefont
  {da~Silva}, \citenamefont {Santos}, \citenamefont {Mota}, \citenamefont
  {da~Costa},\ and\ \citenamefont {Fabris}}]{daSilva2022}%
  \BibitemOpen
  \bibfield  {author} {\bibinfo {author} {\bibfnamefont {F.~M.}\ \bibnamefont
  {da~Silva}}, \bibinfo {author} {\bibfnamefont {L.~C.~N.}\ \bibnamefont
  {Santos}}, \bibinfo {author} {\bibfnamefont {C.~E.}\ \bibnamefont {Mota}},
  \bibinfo {author} {\bibfnamefont {T.~O.~F.}\ \bibnamefont {da~Costa}}, \ and\
  \bibinfo {author} {\bibfnamefont {J.~C.}\ \bibnamefont {Fabris}},\ }\href
  {\doibase 10.1140/epjc/s10052-023-11466-2} {\bibfield  {journal} {\bibinfo
  {journal} {Eur. Phys. J. C}\ }\textbf {\bibinfo {volume} {83}},\ \bibinfo
  {pages} {295} (\bibinfo {year} {2023})},\ \Eprint
  {http://arxiv.org/abs/2206.08469}{arXiv:2206.08469 [gr-qc]}\BibitemShut
  {NoStop}%
\bibitem [{\citenamefont {Pretel}\ \emph {et~al.}(2021)\citenamefont {Pretel},
  \citenamefont {Jor\'as}, \citenamefont {Reis},\ and\ \citenamefont
  {Arba\~nil}}]{Pretel:2020oae}%
  \BibitemOpen
  \bibfield  {author} {\bibinfo {author} {\bibfnamefont {J.~M.~Z.}\
  \bibnamefont {Pretel}}, \bibinfo {author} {\bibfnamefont {S.~E.}\
  \bibnamefont {Jor\'as}}, \bibinfo {author} {\bibfnamefont {R.~R.~R.}\
  \bibnamefont {Reis}}, \ and\ \bibinfo {author} {\bibfnamefont {J.~D.~V.}\
  \bibnamefont {Arba\~nil}},\ }\href {\doibase 10.1088/1475-7516/2021/04/064}
  {\bibfield  {journal} {\bibinfo  {journal} {JCAP}\ }\textbf {\bibinfo
  {volume} {04}},\ \bibinfo {pages} {064} (\bibinfo {year} {2021})},\ \Eprint
  {http://arxiv.org/abs/2012.03342}{arXiv:2012.03342 [gr-qc]}\BibitemShut
  {NoStop}%
\bibitem [{\citenamefont {Pretel}\ \emph {et~al.}(2022)\citenamefont {Pretel},
  \citenamefont {Tangphati}, \citenamefont {Banerjee},\ and\ \citenamefont
  {Pradhan}}]{Pretel:2022dbx}%
  \BibitemOpen
  \bibfield  {author} {\bibinfo {author} {\bibfnamefont {J.~M.~Z.}\
  \bibnamefont {Pretel}}, \bibinfo {author} {\bibfnamefont {T.}~\bibnamefont
  {Tangphati}}, \bibinfo {author} {\bibfnamefont {A.}~\bibnamefont {Banerjee}},
  \ and\ \bibinfo {author} {\bibfnamefont {A.}~\bibnamefont {Pradhan}},\ }\href
  {\doibase 10.1088/1674-1137/ac84cb} {\bibfield  {journal} {\bibinfo
  {journal} {Chin. Phys. C}\ }\textbf {\bibinfo {volume} {46}},\ \bibinfo
  {pages} {115103} (\bibinfo {year} {2022})},\ \Eprint
  {http://arxiv.org/abs/2207.12947}{arXiv:2207.12947 [gr-qc]}\BibitemShut
  {NoStop}%
\bibitem [{\citenamefont {Bora}\ and\ \citenamefont
  {Goswami}(2022)}]{Bora:2022dnu}%
  \BibitemOpen
  \bibfield  {author} {\bibinfo {author} {\bibfnamefont {J.}~\bibnamefont
  {Bora}}\ and\ \bibinfo {author} {\bibfnamefont {U.~D.}\ \bibnamefont
  {Goswami}},\ }\href {\doibase 10.1016/j.dark.2022.101132} {\bibfield
  {journal} {\bibinfo  {journal} {Phys. Dark Univ.}\ }\textbf {\bibinfo
  {volume} {38}},\ \bibinfo {pages} {101132} (\bibinfo {year} {2022})},\
  \Eprint {http://arxiv.org/abs/2207.12847}{arXiv:2207.12847
  [gr-qc]}\BibitemShut {NoStop}%
\bibitem [{\citenamefont {Das}\ \emph {et~al.}(2016)\citenamefont {Das},
  \citenamefont {Rahaman}, \citenamefont {Guha},\ and\ \citenamefont
  {Ray}}]{Das:2016mxq}%
  \BibitemOpen
  \bibfield  {author} {\bibinfo {author} {\bibfnamefont {A.}~\bibnamefont
  {Das}}, \bibinfo {author} {\bibfnamefont {F.}~\bibnamefont {Rahaman}},
  \bibinfo {author} {\bibfnamefont {B.~K.}\ \bibnamefont {Guha}}, \ and\
  \bibinfo {author} {\bibfnamefont {S.}~\bibnamefont {Ray}},\ }\href {\doibase
  10.1140/epjc/s10052-016-4503-0} {\bibfield  {journal} {\bibinfo  {journal}
  {Eur. Phys. J. C}\ }\textbf {\bibinfo {volume} {76}},\ \bibinfo {pages} {654}
  (\bibinfo {year} {2016})},\ \Eprint
  {http://arxiv.org/abs/1608.00566}{arXiv:1608.00566 [gr-qc]}\BibitemShut
  {NoStop}%
\bibitem [{\citenamefont {Rej}\ and\ \citenamefont {Bhar}(2021)}]{Rej:2021ngp}%
  \BibitemOpen
  \bibfield  {author} {\bibinfo {author} {\bibfnamefont {P.}~\bibnamefont
  {Rej}}\ and\ \bibinfo {author} {\bibfnamefont {P.}~\bibnamefont {Bhar}},\
  }\href {\doibase 10.1007/s10509-021-03943-5} {\bibfield  {journal} {\bibinfo
  {journal} {Astrophys. Space Sci.}\ }\textbf {\bibinfo {volume} {366}},\
  \bibinfo {pages} {35} (\bibinfo {year} {2021})},\ \Eprint
  {http://arxiv.org/abs/2105.12572}{arXiv:2105.12572 [gr-qc]}\BibitemShut
  {NoStop}%
\bibitem [{\citenamefont {Deb}\ \emph {et~al.}(2019{\natexlab{a}})\citenamefont
  {Deb}, \citenamefont {Ketov}, \citenamefont {Khlopov},\ and\ \citenamefont
  {Ray}}]{Deb:2018gzt}%
  \BibitemOpen
  \bibfield  {author} {\bibinfo {author} {\bibfnamefont {D.}~\bibnamefont
  {Deb}}, \bibinfo {author} {\bibfnamefont {S.~V.}\ \bibnamefont {Ketov}},
  \bibinfo {author} {\bibfnamefont {M.}~\bibnamefont {Khlopov}}, \ and\
  \bibinfo {author} {\bibfnamefont {S.}~\bibnamefont {Ray}},\ }\href {\doibase
  10.1088/1475-7516/2019/10/070} {\bibfield  {journal} {\bibinfo  {journal}
  {JCAP}\ }\textbf {\bibinfo {volume} {10}},\ \bibinfo {pages} {070} (\bibinfo
  {year} {2019}{\natexlab{a}})},\ \Eprint
  {http://arxiv.org/abs/1812.11736}{arXiv:1812.11736 [gr-qc]}\BibitemShut
  {NoStop}%
\bibitem [{\citenamefont {Deb}\ \emph {et~al.}(2019{\natexlab{b}})\citenamefont
  {Deb}, \citenamefont {Ketov}, \citenamefont {Maurya}, \citenamefont
  {Khlopov}, \citenamefont {Moraes},\ and\ \citenamefont {Ray}}]{Deb:2018sgt}%
  \BibitemOpen
  \bibfield  {author} {\bibinfo {author} {\bibfnamefont {D.}~\bibnamefont
  {Deb}}, \bibinfo {author} {\bibfnamefont {S.~V.}\ \bibnamefont {Ketov}},
  \bibinfo {author} {\bibfnamefont {S.~K.}\ \bibnamefont {Maurya}}, \bibinfo
  {author} {\bibfnamefont {M.}~\bibnamefont {Khlopov}}, \bibinfo {author}
  {\bibfnamefont {P.~H. R.~S.}\ \bibnamefont {Moraes}}, \ and\ \bibinfo
  {author} {\bibfnamefont {S.}~\bibnamefont {Ray}},\ }\href {\doibase
  10.1093/mnras/stz708} {\bibfield  {journal} {\bibinfo  {journal} {Mon. Not.
  Roy. Astron. Soc.}\ }\textbf {\bibinfo {volume} {485}},\ \bibinfo {pages}
  {5652} (\bibinfo {year} {2019}{\natexlab{b}})},\ \Eprint
  {http://arxiv.org/abs/1810.07678}{arXiv:1810.07678 [gr-qc]}\BibitemShut
  {NoStop}%
\bibitem [{\citenamefont {Maurya}\ \emph {et~al.}(2019)\citenamefont {Maurya},
  \citenamefont {Errehymy}, \citenamefont {Deb}, \citenamefont {Tello-Ortiz},\
  and\ \citenamefont {Daoud}}]{PhysRevD.100.044014}%
  \BibitemOpen
  \bibfield  {author} {\bibinfo {author} {\bibfnamefont {S.~K.}\ \bibnamefont
  {Maurya}}, \bibinfo {author} {\bibfnamefont {A.}~\bibnamefont {Errehymy}},
  \bibinfo {author} {\bibfnamefont {D.}~\bibnamefont {Deb}}, \bibinfo {author}
  {\bibfnamefont {F.}~\bibnamefont {Tello-Ortiz}}, \ and\ \bibinfo {author}
  {\bibfnamefont {M.}~\bibnamefont {Daoud}},\ }\href {\doibase
  10.1103/PhysRevD.100.044014} {\bibfield  {journal} {\bibinfo  {journal}
  {Phys. Rev. D}\ }\textbf {\bibinfo {volume} {100}},\ \bibinfo {pages}
  {044014} (\bibinfo {year} {2019})}\BibitemShut {NoStop}%
\bibitem [{\citenamefont {Biswas}\ \emph {et~al.}(2020)\citenamefont {Biswas},
  \citenamefont {Shee}, \citenamefont {Guha},\ and\ \citenamefont
  {Ray}}]{Biswas:2020gzd}%
  \BibitemOpen
  \bibfield  {author} {\bibinfo {author} {\bibfnamefont {S.}~\bibnamefont
  {Biswas}}, \bibinfo {author} {\bibfnamefont {D.}~\bibnamefont {Shee}},
  \bibinfo {author} {\bibfnamefont {B.~K.}\ \bibnamefont {Guha}}, \ and\
  \bibinfo {author} {\bibfnamefont {S.}~\bibnamefont {Ray}},\ }\href {\doibase
  10.1140/epjc/s10052-020-7725-0} {\bibfield  {journal} {\bibinfo  {journal}
  {Eur. Phys. J. C}\ }\textbf {\bibinfo {volume} {80}},\ \bibinfo {pages} {175}
  (\bibinfo {year} {2020})},\ \Eprint
  {http://arxiv.org/abs/2006.01619}{arXiv:2006.01619 [gr-qc]}\BibitemShut
  {NoStop}%
\bibitem [{\citenamefont {Maurya}\ and\ \citenamefont
  {Tello-Ortiz}(2020)}]{Maurya:2019iup}%
  \BibitemOpen
  \bibfield  {author} {\bibinfo {author} {\bibfnamefont {S.~K.}\ \bibnamefont
  {Maurya}}\ and\ \bibinfo {author} {\bibfnamefont {F.}~\bibnamefont
  {Tello-Ortiz}},\ }\href {\doibase 10.1016/j.aop.2020.168070} {\bibfield
  {journal} {\bibinfo  {journal} {Annals Phys.}\ }\textbf {\bibinfo {volume}
  {414}},\ \bibinfo {pages} {168070} (\bibinfo {year} {2020})},\ \Eprint
  {http://arxiv.org/abs/1906.11756}{arXiv:1906.11756 [gr-qc]}\BibitemShut
  {NoStop}%
\bibitem [{\citenamefont {Biswas}\ \emph {et~al.}(2021)\citenamefont {Biswas},
  \citenamefont {Deb}, \citenamefont {Ray},\ and\ \citenamefont
  {Guha}}]{Biswas:2021wfn}%
  \BibitemOpen
  \bibfield  {author} {\bibinfo {author} {\bibfnamefont {S.}~\bibnamefont
  {Biswas}}, \bibinfo {author} {\bibfnamefont {D.}~\bibnamefont {Deb}},
  \bibinfo {author} {\bibfnamefont {S.}~\bibnamefont {Ray}}, \ and\ \bibinfo
  {author} {\bibfnamefont {B.~K.}\ \bibnamefont {Guha}},\ }\href {\doibase
  10.1016/j.aop.2021.168429} {\bibfield  {journal} {\bibinfo  {journal} {Annals
  Phys.}\ }\textbf {\bibinfo {volume} {428}},\ \bibinfo {pages} {168429}
  (\bibinfo {year} {2021})}\BibitemShut {NoStop}%
\bibitem [{\citenamefont {Stairs}(2003)}]{Stairs2003}%
  \BibitemOpen
  \bibfield  {author} {\bibinfo {author} {\bibfnamefont {I.~H.}\ \bibnamefont
  {Stairs}},\ }\href {\doibase 10.12942/lrr-2003-5} {\bibfield  {journal}
  {\bibinfo  {journal} {Living Rev. Rel.}\ }\textbf {\bibinfo {volume} {6}},\
  \bibinfo {pages} {5} (\bibinfo {year} {2003})},\ \Eprint
  {http://arxiv.org/abs/astro-ph/0307536}{arXiv:astro-ph/0307536}\BibitemShut
  {NoStop}%
\bibitem [{\citenamefont {Yagi}\ and\ \citenamefont {Yunes}(2017)}]{Yagi2017}%
  \BibitemOpen
  \bibfield  {author} {\bibinfo {author} {\bibfnamefont {K.}~\bibnamefont
  {Yagi}}\ and\ \bibinfo {author} {\bibfnamefont {N.}~\bibnamefont {Yunes}},\
  }\href {\doibase 10.1016/j.physrep.2017.03.002} {\bibfield  {journal}
  {\bibinfo  {journal} {Phys. Rept.}\ }\textbf {\bibinfo {volume} {681}},\
  \bibinfo {pages} {1} (\bibinfo {year} {2017})},\ \Eprint
  {http://arxiv.org/abs/1608.02582}{arXiv:1608.02582 [gr-qc]}\BibitemShut
  {NoStop}%
\bibitem [{\citenamefont {Pappas}\ and\ \citenamefont
  {Apostolatos}(2014)}]{Pappas2014}%
  \BibitemOpen
  \bibfield  {author} {\bibinfo {author} {\bibfnamefont {G.}~\bibnamefont
  {Pappas}}\ and\ \bibinfo {author} {\bibfnamefont {T.~A.}\ \bibnamefont
  {Apostolatos}},\ }\href {\doibase 10.1103/PhysRevLett.112.121101} {\bibfield
  {journal} {\bibinfo  {journal} {Phys. Rev. Lett.}\ }\textbf {\bibinfo
  {volume} {112}},\ \bibinfo {pages} {121101} (\bibinfo {year}
  {2014})}\BibitemShut {NoStop}%
\bibitem [{\citenamefont {Pani}\ and\ \citenamefont {Berti}(2014)}]{Pani2014}%
  \BibitemOpen
  \bibfield  {author} {\bibinfo {author} {\bibfnamefont {P.}~\bibnamefont
  {Pani}}\ and\ \bibinfo {author} {\bibfnamefont {E.}~\bibnamefont {Berti}},\
  }\href {\doibase 10.1103/PhysRevD.90.024025} {\bibfield  {journal} {\bibinfo
  {journal} {Phys. Rev. D}\ }\textbf {\bibinfo {volume} {90}},\ \bibinfo
  {pages} {024025} (\bibinfo {year} {2014})}\BibitemShut {NoStop}%
\bibitem [{\citenamefont {Sham}\ \emph {et~al.}(2014)\citenamefont {Sham},
  \citenamefont {Lin},\ and\ \citenamefont {Leung}}]{Sham2013}%
  \BibitemOpen
  \bibfield  {author} {\bibinfo {author} {\bibfnamefont {Y.~H.}\ \bibnamefont
  {Sham}}, \bibinfo {author} {\bibfnamefont {L.~M.}\ \bibnamefont {Lin}}, \
  and\ \bibinfo {author} {\bibfnamefont {P.~T.}\ \bibnamefont {Leung}},\ }\href
  {\doibase 10.1088/0004-637X/781/2/66} {\bibfield  {journal} {\bibinfo
  {journal} {Astrophys. J.}\ }\textbf {\bibinfo {volume} {781}},\ \bibinfo
  {pages} {66} (\bibinfo {year} {2014})},\ \Eprint
  {http://arxiv.org/abs/1312.1011}{arXiv:1312.1011 [gr-qc]}\BibitemShut
  {NoStop}%
\bibitem [{\citenamefont {Yagi}\ \emph {et~al.}(2013)\citenamefont {Yagi},
  \citenamefont {Stein}, \citenamefont {Yunes},\ and\ \citenamefont
  {Tanaka}}]{Yagi2013b}%
  \BibitemOpen
  \bibfield  {author} {\bibinfo {author} {\bibfnamefont {K.}~\bibnamefont
  {Yagi}}, \bibinfo {author} {\bibfnamefont {L.~C.}\ \bibnamefont {Stein}},
  \bibinfo {author} {\bibfnamefont {N.}~\bibnamefont {Yunes}}, \ and\ \bibinfo
  {author} {\bibfnamefont {T.}~\bibnamefont {Tanaka}},\ }\href {\doibase
  10.1103/PhysRevD.87.084058} {\bibfield  {journal} {\bibinfo  {journal} {Phys.
  Rev. D}\ }\textbf {\bibinfo {volume} {87}},\ \bibinfo {pages} {084058}
  (\bibinfo {year} {2013})}\BibitemShut {NoStop}%
\bibitem [{\citenamefont {Kleihaus}\ \emph {et~al.}(2014)\citenamefont
  {Kleihaus}, \citenamefont {Kunz},\ and\ \citenamefont
  {Mojica}}]{Kleihaus2014}%
  \BibitemOpen
  \bibfield  {author} {\bibinfo {author} {\bibfnamefont {B.}~\bibnamefont
  {Kleihaus}}, \bibinfo {author} {\bibfnamefont {J.}~\bibnamefont {Kunz}}, \
  and\ \bibinfo {author} {\bibfnamefont {S.}~\bibnamefont {Mojica}},\ }\href
  {\doibase 10.1103/PhysRevD.90.061501} {\bibfield  {journal} {\bibinfo
  {journal} {Phys. Rev. D}\ }\textbf {\bibinfo {volume} {90}},\ \bibinfo
  {pages} {061501} (\bibinfo {year} {2014})}\BibitemShut {NoStop}%
\bibitem [{\citenamefont {Pani}\ \emph {et~al.}(2011)\citenamefont {Pani},
  \citenamefont {Berti}, \citenamefont {Cardoso},\ and\ \citenamefont
  {Read}}]{pani2011}%
  \BibitemOpen
  \bibfield  {author} {\bibinfo {author} {\bibfnamefont {P.}~\bibnamefont
  {Pani}}, \bibinfo {author} {\bibfnamefont {E.}~\bibnamefont {Berti}},
  \bibinfo {author} {\bibfnamefont {V.}~\bibnamefont {Cardoso}}, \ and\
  \bibinfo {author} {\bibfnamefont {J.}~\bibnamefont {Read}},\ }\href {\doibase
  10.1103/PhysRevD.84.104035} {\bibfield  {journal} {\bibinfo  {journal} {Phys.
  Rev. D}\ }\textbf {\bibinfo {volume} {84}},\ \bibinfo {pages} {104035}
  (\bibinfo {year} {2011})},\ \Eprint
  {http://arxiv.org/abs/1109.0928}{arXiv:1109.0928 [gr-qc]}\BibitemShut
  {NoStop}%
\bibitem [{\citenamefont {Yazadjiev}\ \emph {et~al.}(2016)\citenamefont
  {Yazadjiev}, \citenamefont {Doneva},\ and\ \citenamefont
  {Popchev}}]{Yazadjiev2016}%
  \BibitemOpen
  \bibfield  {author} {\bibinfo {author} {\bibfnamefont {S.~S.}\ \bibnamefont
  {Yazadjiev}}, \bibinfo {author} {\bibfnamefont {D.~D.}\ \bibnamefont
  {Doneva}}, \ and\ \bibinfo {author} {\bibfnamefont {D.}~\bibnamefont
  {Popchev}},\ }\href {\doibase 10.1103/PhysRevD.93.084038} {\bibfield
  {journal} {\bibinfo  {journal} {Phys. Rev. D}\ }\textbf {\bibinfo {volume}
  {93}},\ \bibinfo {pages} {084038} (\bibinfo {year} {2016})},\ \Eprint
  {http://arxiv.org/abs/1602.04766}{arXiv:1602.04766 [gr-qc]}\BibitemShut
  {NoStop}%
\bibitem [{\citenamefont {Boumaza}(2021)}]{Boumaza2021}%
  \BibitemOpen
  \bibfield  {author} {\bibinfo {author} {\bibfnamefont {H.}~\bibnamefont
  {Boumaza}},\ }\href {\doibase 10.1140/epjc/s10052-021-09222-5} {\bibfield
  {journal} {\bibinfo  {journal} {Eur. Phys. J. C}\ }\textbf {\bibinfo {volume}
  {81}},\ \bibinfo {pages} {448} (\bibinfo {year} {2021})}\BibitemShut
  {NoStop}%
\bibitem [{\citenamefont {Staykov}\ \emph {et~al.}(2014)\citenamefont
  {Staykov}, \citenamefont {Doneva}, \citenamefont {Yazadjiev},\ and\
  \citenamefont {Kokkotas}}]{Staykov2014}%
  \BibitemOpen
  \bibfield  {author} {\bibinfo {author} {\bibfnamefont {K.~V.}\ \bibnamefont
  {Staykov}}, \bibinfo {author} {\bibfnamefont {D.~D.}\ \bibnamefont {Doneva}},
  \bibinfo {author} {\bibfnamefont {S.~S.}\ \bibnamefont {Yazadjiev}}, \ and\
  \bibinfo {author} {\bibfnamefont {K.~D.}\ \bibnamefont {Kokkotas}},\ }\href
  {\doibase 10.1088/1475-7516/2014/10/006} {\bibfield  {journal} {\bibinfo
  {journal} {J. Cosmol. Astropart. Phys.}\ }\textbf {\bibinfo {volume} {10}},\
  \bibinfo {pages} {006} (\bibinfo {year} {2014})},\ \Eprint
  {http://arxiv.org/abs/1407.2180}{arXiv:1407.2180 [gr-qc]}\BibitemShut
  {NoStop}%
\bibitem [{\citenamefont {Pattersons}\ and\ \citenamefont
  {Sulaksono}(2021)}]{Pattersons2021}%
  \BibitemOpen
  \bibfield  {author} {\bibinfo {author} {\bibfnamefont {M.~L.}\ \bibnamefont
  {Pattersons}}\ and\ \bibinfo {author} {\bibfnamefont {A.}~\bibnamefont
  {Sulaksono}},\ }\href {\doibase 10.1140/epjc/s10052-021-09481-2} {\bibfield
  {journal} {\bibinfo  {journal} {Eur. Phys. J. C}\ }\textbf {\bibinfo {volume}
  {81}},\ \bibinfo {pages} {698} (\bibinfo {year} {2021})}\BibitemShut
  {NoStop}%
\bibitem [{\citenamefont {Pappas}\ \emph {et~al.}(2022)\citenamefont {Pappas},
  \citenamefont {Posada},\ and\ \citenamefont
  {Stuchl\'\i{}k}}]{Pappas:2022gtt}%
  \BibitemOpen
  \bibfield  {author} {\bibinfo {author} {\bibfnamefont {T.~D.}\ \bibnamefont
  {Pappas}}, \bibinfo {author} {\bibfnamefont {C.}~\bibnamefont {Posada}}, \
  and\ \bibinfo {author} {\bibfnamefont {Z.}~\bibnamefont {Stuchl\'\i{}k}},\
  }\href {\doibase 10.1103/PhysRevD.106.124014} {\bibfield  {journal} {\bibinfo
   {journal} {Phys. Rev. D}\ }\textbf {\bibinfo {volume} {106}},\ \bibinfo
  {pages} {124014} (\bibinfo {year} {2022})},\ \Eprint
  {http://arxiv.org/abs/2210.15597}{arXiv:2210.15597 [gr-qc]}\BibitemShut
  {NoStop}%
\bibitem [{\citenamefont {Pretel}(2022)}]{Pretel2022}%
  \BibitemOpen
  \bibfield  {author} {\bibinfo {author} {\bibfnamefont {J.~M.~Z.}\
  \bibnamefont {Pretel}},\ }\href {\doibase 10.1142/S0217732322501887}
  {\bibfield  {journal} {\bibinfo  {journal} {Mod. Phys. Lett. A}\ }\textbf
  {\bibinfo {volume} {37}},\ \bibinfo {pages} {2250188} (\bibinfo {year}
  {2022})},\ \Eprint {http://arxiv.org/abs/2301.02881}{arXiv:2301.02881
  [gr-qc]}\BibitemShut {NoStop}%
\bibitem [{\citenamefont {Barrientos~O.}\ and\ \citenamefont
  {Rubilar}(2014)}]{Barrientos2014}%
  \BibitemOpen
  \bibfield  {author} {\bibinfo {author} {\bibfnamefont {J.}~\bibnamefont
  {Barrientos~O.}}\ and\ \bibinfo {author} {\bibfnamefont {G.~F.}\ \bibnamefont
  {Rubilar}},\ }\href {\doibase 10.1103/PhysRevD.90.028501} {\bibfield
  {journal} {\bibinfo  {journal} {Phys. Rev. D}\ }\textbf {\bibinfo {volume}
  {90}},\ \bibinfo {pages} {028501} (\bibinfo {year} {2014})}\BibitemShut
  {NoStop}%
\bibitem [{\citenamefont {Regge}\ and\ \citenamefont
  {Wheeler}(1957)}]{Regge1957}%
  \BibitemOpen
  \bibfield  {author} {\bibinfo {author} {\bibfnamefont {T.}~\bibnamefont
  {Regge}}\ and\ \bibinfo {author} {\bibfnamefont {J.~A.}\ \bibnamefont
  {Wheeler}},\ }\href {\doibase 10.1103/PhysRev.108.1063} {\bibfield  {journal}
  {\bibinfo  {journal} {Phys. Rev.}\ }\textbf {\bibinfo {volume} {108}},\
  \bibinfo {pages} {1063} (\bibinfo {year} {1957})}\BibitemShut {NoStop}%
\bibitem [{\citenamefont {Read}\ \emph
  {et~al.}(2009{\natexlab{a}})\citenamefont {Read}, \citenamefont {Lackey},
  \citenamefont {Owen},\ and\ \citenamefont {Friedman}}]{Read:2008iy}%
  \BibitemOpen
  \bibfield  {author} {\bibinfo {author} {\bibfnamefont {J.~S.}\ \bibnamefont
  {Read}}, \bibinfo {author} {\bibfnamefont {B.~D.}\ \bibnamefont {Lackey}},
  \bibinfo {author} {\bibfnamefont {B.~J.}\ \bibnamefont {Owen}}, \ and\
  \bibinfo {author} {\bibfnamefont {J.~L.}\ \bibnamefont {Friedman}},\ }\href
  {\doibase 10.1103/PhysRevD.79.124032} {\bibfield  {journal} {\bibinfo
  {journal} {Phys. Rev. D}\ }\textbf {\bibinfo {volume} {79}},\ \bibinfo
  {pages} {124032} (\bibinfo {year} {2009}{\natexlab{a}})},\ \Eprint
  {http://arxiv.org/abs/0812.2163}{arXiv:0812.2163 [astro-ph]}\BibitemShut
  {NoStop}%
\bibitem [{\citenamefont {Read}\ \emph
  {et~al.}(2009{\natexlab{b}})\citenamefont {Read}, \citenamefont {Markakis},
  \citenamefont {Shibata}, \citenamefont {Ury\ifmmode~\bar{u}\else \={u}\fi{}},
  \citenamefont {Creighton},\ and\ \citenamefont
  {Friedman}}]{PhysRevD.79.124033}%
  \BibitemOpen
  \bibfield  {author} {\bibinfo {author} {\bibfnamefont {J.~S.}\ \bibnamefont
  {Read}}, \bibinfo {author} {\bibfnamefont {C.}~\bibnamefont {Markakis}},
  \bibinfo {author} {\bibfnamefont {M.}~\bibnamefont {Shibata}}, \bibinfo
  {author} {\bibfnamefont {K.~b.~o.}\ \bibnamefont {Ury\ifmmode~\bar{u}\else
  \={u}\fi{}}}, \bibinfo {author} {\bibfnamefont {J.~D.~E.}\ \bibnamefont
  {Creighton}}, \ and\ \bibinfo {author} {\bibfnamefont {J.~L.}\ \bibnamefont
  {Friedman}},\ }\href {\doibase 10.1103/PhysRevD.79.124033} {\bibfield
  {journal} {\bibinfo  {journal} {Phys. Rev. D}\ }\textbf {\bibinfo {volume}
  {79}},\ \bibinfo {pages} {124033} (\bibinfo {year}
  {2009}{\natexlab{b}})}\BibitemShut {NoStop}%
\bibitem [{\citenamefont {Odintsov}\ and\ \citenamefont
  {Oikonomou}(2023)}]{PhysRevD.107.104039}%
  \BibitemOpen
  \bibfield  {author} {\bibinfo {author} {\bibfnamefont {S.~D.}\ \bibnamefont
  {Odintsov}}\ and\ \bibinfo {author} {\bibfnamefont {V.~K.}\ \bibnamefont
  {Oikonomou}},\ }\href {\doibase 10.1103/PhysRevD.107.104039} {\bibfield
  {journal} {\bibinfo  {journal} {Phys. Rev. D}\ }\textbf {\bibinfo {volume}
  {107}},\ \bibinfo {pages} {104039} (\bibinfo {year} {2023})}\BibitemShut
  {NoStop}%
\bibitem [{\citenamefont {Douchin}\ and\ \citenamefont
  {Haensel}(2001)}]{Douchin:2001sv}%
  \BibitemOpen
  \bibfield  {author} {\bibinfo {author} {\bibfnamefont {F.}~\bibnamefont
  {Douchin}}\ and\ \bibinfo {author} {\bibfnamefont {P.}~\bibnamefont
  {Haensel}},\ }\href {\doibase 10.1051/0004-6361:20011402} {\bibfield
  {journal} {\bibinfo  {journal} {Astron. Astrophys.}\ }\textbf {\bibinfo
  {volume} {380}},\ \bibinfo {pages} {151} (\bibinfo {year} {2001})},\ \Eprint
  {http://arxiv.org/abs/astro-ph/0111092}{arXiv:astro-ph/0111092}\BibitemShut
  {NoStop}%
\bibitem [{\citenamefont {Wiringa}\ \emph {et~al.}(1988)\citenamefont
  {Wiringa}, \citenamefont {Fiks},\ and\ \citenamefont
  {Fabrocini}}]{PhysRevC.38.1010}%
  \BibitemOpen
  \bibfield  {author} {\bibinfo {author} {\bibfnamefont {R.~B.}\ \bibnamefont
  {Wiringa}}, \bibinfo {author} {\bibfnamefont {V.}~\bibnamefont {Fiks}}, \
  and\ \bibinfo {author} {\bibfnamefont {A.}~\bibnamefont {Fabrocini}},\ }\href
  {\doibase 10.1103/PhysRevC.38.1010} {\bibfield  {journal} {\bibinfo
  {journal} {Phys. Rev. C}\ }\textbf {\bibinfo {volume} {38}},\ \bibinfo
  {pages} {1010} (\bibinfo {year} {1988})}\BibitemShut {NoStop}%
\bibitem [{\citenamefont {Akmal}\ \emph {et~al.}(1998)\citenamefont {Akmal},
  \citenamefont {Pandharipande},\ and\ \citenamefont
  {Ravenhall}}]{PhysRevC.58.1804}%
  \BibitemOpen
  \bibfield  {author} {\bibinfo {author} {\bibfnamefont {A.}~\bibnamefont
  {Akmal}}, \bibinfo {author} {\bibfnamefont {V.~R.}\ \bibnamefont
  {Pandharipande}}, \ and\ \bibinfo {author} {\bibfnamefont {D.~G.}\
  \bibnamefont {Ravenhall}},\ }\href {\doibase 10.1103/PhysRevC.58.1804}
  {\bibfield  {journal} {\bibinfo  {journal} {Phys. Rev. C}\ }\textbf {\bibinfo
  {volume} {58}},\ \bibinfo {pages} {1804} (\bibinfo {year}
  {1998})}\BibitemShut {NoStop}%
\bibitem [{\citenamefont {Engvik}\ \emph {et~al.}(1996)\citenamefont {Engvik},
  \citenamefont {Bao}, \citenamefont {Hjorth-Jensen}, \citenamefont {Osnes},\
  and\ \citenamefont {Ostgaard}}]{Engvik:1995gn}%
  \BibitemOpen
  \bibfield  {author} {\bibinfo {author} {\bibfnamefont {L.}~\bibnamefont
  {Engvik}}, \bibinfo {author} {\bibfnamefont {G.}~\bibnamefont {Bao}},
  \bibinfo {author} {\bibfnamefont {M.}~\bibnamefont {Hjorth-Jensen}}, \bibinfo
  {author} {\bibfnamefont {E.}~\bibnamefont {Osnes}}, \ and\ \bibinfo {author}
  {\bibfnamefont {E.}~\bibnamefont {Ostgaard}},\ }\href {\doibase
  10.1086/177827} {\bibfield  {journal} {\bibinfo  {journal} {Astrophys. J.}\
  }\textbf {\bibinfo {volume} {469}},\ \bibinfo {pages} {794} (\bibinfo {year}
  {1996})},\ \Eprint
  {http://arxiv.org/abs/nucl-th/9509016}{arXiv:nucl-th/9509016}\BibitemShut
  {NoStop}%
\bibitem [{\citenamefont {M\"uther}\ \emph {et~al.}(1987)\citenamefont
  {M\"uther}, \citenamefont {Prakash},\ and\ \citenamefont
  {Ainsworth}}]{Muther:1987xaa}%
  \BibitemOpen
  \bibfield  {author} {\bibinfo {author} {\bibfnamefont {H.}~\bibnamefont
  {M\"uther}}, \bibinfo {author} {\bibfnamefont {M.}~\bibnamefont {Prakash}}, \
  and\ \bibinfo {author} {\bibfnamefont {T.~L.}\ \bibnamefont {Ainsworth}},\
  }\href {\doibase 10.1016/0370-2693(87)91611-X} {\bibfield  {journal}
  {\bibinfo  {journal} {Phys. Lett. B}\ }\textbf {\bibinfo {volume} {199}},\
  \bibinfo {pages} {469} (\bibinfo {year} {1987})}\BibitemShut {NoStop}%
\bibitem [{\citenamefont {Mueller}\ and\ \citenamefont
  {Serot}(1996)}]{Mueller:1996pm}%
  \BibitemOpen
  \bibfield  {author} {\bibinfo {author} {\bibfnamefont {H.}~\bibnamefont
  {Mueller}}\ and\ \bibinfo {author} {\bibfnamefont {B.~D.}\ \bibnamefont
  {Serot}},\ }\href {\doibase 10.1016/0375-9474(96)00187-X} {\bibfield
  {journal} {\bibinfo  {journal} {Nucl. Phys. A}\ }\textbf {\bibinfo {volume}
  {606}},\ \bibinfo {pages} {508} (\bibinfo {year} {1996})},\ \Eprint
  {http://arxiv.org/abs/nucl-th/9603037}{arXiv:nucl-th/9603037}\BibitemShut
  {NoStop}%
\bibitem [{\citenamefont {Pretel}\ \emph {et~al.}(2020)\citenamefont {Pretel},
  \citenamefont {Jor\'as},\ and\ \citenamefont {Reis}}]{Pretel:2020rqx}%
  \BibitemOpen
  \bibfield  {author} {\bibinfo {author} {\bibfnamefont {J.~M.~Z.}\
  \bibnamefont {Pretel}}, \bibinfo {author} {\bibfnamefont {S.~E.}\
  \bibnamefont {Jor\'as}}, \ and\ \bibinfo {author} {\bibfnamefont {R.~R.~R.}\
  \bibnamefont {Reis}},\ }\href {\doibase 10.1088/1475-7516/2020/11/048}
  {\bibfield  {journal} {\bibinfo  {journal} {JCAP}\ }\textbf {\bibinfo
  {volume} {11}},\ \bibinfo {pages} {048} (\bibinfo {year} {2020})},\ \Eprint
  {http://arxiv.org/abs/2008.00536}{arXiv:2008.00536 [gr-qc]}\BibitemShut
  {NoStop}%
\bibitem [{\citenamefont {Capozziello}\ \emph {et~al.}(2016)\citenamefont
  {Capozziello}, \citenamefont {De~Laurentis}, \citenamefont {Farinelli},\ and\
  \citenamefont {Odintsov}}]{Capozziello:2015yza}%
  \BibitemOpen
  \bibfield  {author} {\bibinfo {author} {\bibfnamefont {S.}~\bibnamefont
  {Capozziello}}, \bibinfo {author} {\bibfnamefont {M.}~\bibnamefont
  {De~Laurentis}}, \bibinfo {author} {\bibfnamefont {R.}~\bibnamefont
  {Farinelli}}, \ and\ \bibinfo {author} {\bibfnamefont {S.~D.}\ \bibnamefont
  {Odintsov}},\ }\href {\doibase 10.1103/PhysRevD.93.023501} {\bibfield
  {journal} {\bibinfo  {journal} {Phys. Rev. D}\ }\textbf {\bibinfo {volume}
  {93}},\ \bibinfo {pages} {023501} (\bibinfo {year} {2016})},\ \Eprint
  {http://arxiv.org/abs/1509.04163}{arXiv:1509.04163 [gr-qc]}\BibitemShut
  {NoStop}%
\bibitem [{\citenamefont {Yazadjiev}\ \emph {et~al.}(2014)\citenamefont
  {Yazadjiev}, \citenamefont {Doneva}, \citenamefont {Kokkotas},\ and\
  \citenamefont {Staykov}}]{Yazadjiev_2014}%
  \BibitemOpen
  \bibfield  {author} {\bibinfo {author} {\bibfnamefont {S.~S.}\ \bibnamefont
  {Yazadjiev}}, \bibinfo {author} {\bibfnamefont {D.~D.}\ \bibnamefont
  {Doneva}}, \bibinfo {author} {\bibfnamefont {K.~D.}\ \bibnamefont
  {Kokkotas}}, \ and\ \bibinfo {author} {\bibfnamefont {K.~V.}\ \bibnamefont
  {Staykov}},\ }\href {\doibase 10.1088/1475-7516/2014/06/003} {\bibfield
  {journal} {\bibinfo  {journal} {Journal of Cosmology and Astroparticle
  Physics}\ }\textbf {\bibinfo {volume} {2014}},\ \bibinfo {pages} {003}
  (\bibinfo {year} {2014})}\BibitemShut {NoStop}%
\end{thebibliography}%

\end{document}